\documentclass[11pt]{article}
\pdfoutput=1
\usepackage[LGR,T1]{fontenc}
\usepackage[greek,british]{babel}
\usepackage[latin9]{inputenc}
\usepackage[square,comma,numbers,sort&compress]{natbib}
\usepackage{bbold}
\usepackage[font=small]{caption}
\usepackage{bm}
\usepackage{bbm}
\usepackage{amsmath}
\usepackage{amsfonts}
\usepackage{amssymb}
\usepackage{wasysym}
\usepackage{graphicx, rotating}
\usepackage{epstopdf}
\usepackage{epsfig}
\usepackage{empheq}
\usepackage{latexsym}
\usepackage{multirow}
\usepackage{stmaryrd}
\usepackage{color}
\usepackage[colorlinks=true,linkcolor=black,anchorcolor=black,citecolor=blue,filecolor=cyan,menucolor=red,runcolor=filecolor,urlcolor=blue,bookmarks=true,bookmarksnumbered=true]{hyperref}%
\usepackage{color}
\usepackage{enumerate}

\definecolor{rosso}{cmyk}{0,1,1,0.4}
\definecolor{rossos}{cmyk}{0,1,1,0.55}
\definecolor{rossoc}{cmyk}{0,1,1,0.2}
\definecolor{blu}{cmyk}{1,1,0,0.3}
\definecolor{blus}{cmyk}{1,1,0,0.6}
\definecolor{bluc}{cmyk}{1,1,0,0.1}
\definecolor{verde}{cmyk}{0.92,0,0.59,0.25}
\definecolor{verdec}{cmyk}{0.92,0,0.59,0.15}
\definecolor{verdes}{cmyk}{0.92,0,0.59,0.4}
\definecolor{grigio}{cmyk}{0,0,0,0.07}
\definecolor{rosa}{cmyk}{0,0.1,0.1,0.02}
\definecolor{rosino}{cmyk}{0,0.05,0.05,0.02}
\definecolor{rosas}{cmyk}{0,0.3,0.25,0.05}
\definecolor{celeste}{cmyk}{0.1,0,0,0.02}
\definecolor{giallino}{cmyk}{0,0,0.4,0.02}
\definecolor{rosso}{cmyk}{0,1,1,0.4}
\definecolor{rossos}{cmyk}{0,1,1,0.55}
\definecolor{rossoc}{cmyk}{0,1,1,0.2}
\definecolor{blu}{cmyk}{1,1,0,0.3}
\definecolor{bluc}{cmyk}{1,1,0,0.1}
\definecolor{blucc}{cmyk}{0.7,0.5,0,0}
\definecolor{viola}{cmyk}{0,1,0,0.6}
\definecolor{viola2}{cmyk}{0,1,0.2,0.6}
\definecolor{verde}{cmyk}{0.92,0,0.59,0.25}
\definecolor{verdec}{cmyk}{0.92,0,0.59,0.15}
\definecolor{verdes}{cmyk}{0.92,0,0.59,0.4}
\definecolor{verdino}{cmyk}{0.12,0,0.09,0.05}
\definecolor{giallo}{cmyk}{0,0,1,0}
\definecolor{gialloverde}{cmyk}{0.44,0,0.74,0}
\definecolor{grey}{rgb}{0.6,0.6,0.6}
\definecolor{fuchsia}{rgb}{1,0,1}

\usepackage{chngcntr}
\counterwithin{figure}{section}
\counterwithin{table}{section}
\counterwithin{equation}{section}

\def\l{\label}
\def\La{\mathcal{L}}
\def\Amp{\mathcal{A}}
\def\({\left(}
\def\){\right)}
\def\f{\frac}
\def\be{\begin{equation}}
\def\ee{\end{equation}}
\def\bry{\begin{array}}
\def\ery{\end{array}}
\def\bes{\begin{subequations}}
\def\ees{\end{subequations}}
\def\bit{\begin{itemize}}
\def\eit{\end{itemize}}
\def\ben{\begin{enumerate}}
\def\een{\end{enumerate}}
\def\dst{\displaystyle}

\def\de{\partial}
\def\demub{\de_{\mu}}

\def\demua{\de^{\mu}}

\def\ovl{\overline}
\def\Tr{\text{Tr}}
\def\bes{\begin{subequations}}
\def\ees{\end{subequations}}
\def\bea{\begin{eqnarray}}
\def\eea{\end{eqnarray}}
\def\bry{\begin{array}}
\def\ery{\end{array}}
\def\bit{\begin{itemize}}
\def\eit{\end{itemize}}

\def\tr{\textrm{Tr}}

\def\dst{\displaystyle}
\def\f{\frac}

\def\i{\text{i}}
\def\di{\text{d}}
\def\e{\text{e}}
\def\PI{\text{\textrm{\greektext p}}}

\def\r{\mathbf{r}}

\newcommand{\nocontentsline}[3]{}
\newcommand{\tocless}[2]{\bgroup\let\addcontentsline=\nocontentsline#1{#2}\egroup}

\makeatother

\title{
\vspace{-1.5cm}
\normalsize{\hspace{9.9cm}DFPD-2014/TH/10, LPN14-072}
\vspace{1cm}
\vspace{0.0 cm}
{\huge
\bf Symmetries, Sum Rules and Constraints \vspace{2mm}\\
on Effective Field Theories
\vspace*{0.5cm} 
}}
\vspace{2cm}
\author{{\Large\text{Brando Bellazzini$^{1,2}$\,}\footnote{\href{brando.bellazzini@cea.fr}{brando.bellazzini@cea.fr}}\text{\,\,, Luca Martucci$^{2,3}$\,}\footnote{\href{luca.martucci@pd.infn.it}{luca.martucci@pd.infn.it}}\text{\,\, and Riccardo Torre$^{2,3,4}$\,}\footnote{\href{riccardo.torre@pd.infn.it}{riccardo.torre@pd.infn.it}}} \vspace{4mm}\\ 
{\small\emph{$^{1}$Institut de Physique Th\'eorique, CEA-Saclay, F-91191 Gif-sur-Yvette Cedex, France}}\\ \vspace{-15pt} \\
{\small\emph{$^{2}$Dipartimento di Fisica e Astronomia, Universit\'a di Padova, Via Marzolo 8, I-35131 Padova, Italy}} \\
{\small\emph{$^{3}$INFN Sezione di Padova, Via Marzolo 8, I-35131 Padova, Italy}}\\ \vspace{-15pt} \\
{\small\emph{$^{4}$SISSA, Via Bonomea 265, I-34136 Trieste, Italy}}
}

\date{}

\begin{document}
\baselineskip=16pt

\maketitle \thispagestyle{empty}
\begin{center}
{\Large Abstract}
\end{center}
Using unitarity, analyticity and crossing symmetry, we derive universal sum rules for scattering amplitudes in theories invariant under an arbitrary symmetry group. 
The sum rules relate the coefficients of the energy expansion of the scattering amplitudes in the IR to total cross sections integrated all the way up to the UV. Exploiting the group structure of the symmetry, we systematically determine all the independent sum rules and  positivity conditions on the expansion coefficients.
For effective field theories the amplitudes in the IR are calculable and hence the sum rules set constraints on the parameters of the effective Lagrangian. We clarify the impact of gauging on the sum rules for Goldstone bosons in spontaneously broken gauge theories.
We discuss explicit examples  that are relevant for $WW$-scattering, composite Higgs models, and chiral perturbation theory.  Certain sum rules based on custodial symmetry and its extensions provide constraints on the Higgs boson coupling to the electroweak gauge bosons.
  
\parbox[c]{12cm}{
\medskip
\noindent
}

\newpage
\tableofcontents

\section{Introduction}
\l{sec:introduction}

Effective field theories (EFTs) are the standard language that describes the dynamics of low-energy degrees of freedom in terms of a series of increasingly higher dimension operators, $\sum_{i, n} c^{(n)}_i/\Lambda^n \,\mathcal{O}^{(n)}_i$. Only a finite set of the (a priori unknown) low-energy coefficients (LECs) $c^{(n)}_i$ enters in the physical observables at any given order in $(p/\Lambda)^n$. When the theory is invariant under a symmetry group $H$, the LECs are further restricted  by demanding invariant $\mathcal{O}^{(n)}_i$. 
Chiral perturbation theory in QCD is the EFT prototype for the dynamics of pions at low-energies: the leading $O(p^2)$ Lagrangian is controlled by just two parameters, namely the pion decay constant and the pion mass. Beyond leading order, higher derivative operators become relevant and several other LECs need to be included.

While symmetry restrictions are crucial for the EFT to make sense and be predictive, they do not exhaust all physical conditions that the LECs must satisfy whenever the underlying ultraviolet (UV) theory has a Lorentz invariant, unitary, analytic, and crossing symmetric S-matrix. These requirements translate into dispersion relations that relate the LECs in the infrared (IR) to certain integrals over the energy of total cross-sections.  For example, the theory $ \mathcal{L}=(\partial_\mu\pi)^2/2+c /\Lambda^4 (\partial_\mu\pi\partial^\mu\pi)^2+\ldots $ for one Goldstone Boson (GB) $\pi$, invariant under a shift symmetry $\pi\rightarrow \pi+c$, admits sensible UV completions only for $c\geq 0$ because the forward elastic scattering amplitude $\Amp(s)$  satisfies \cite{Adams:2006ch}
\be
\Amp^{\prime\prime}(0)=\f{4}{\PI}\int_0^\infty \di s\, \f{\sigma^{\text{tot}}(s)}{s^{2}} \geq 0\,.
\ee
The left-hand side can be calculated within the EFT in terms of $c$, whereas the right-hand side is the total cross-section integrated all the way up to the UV where the EFT is not valid.
This \mbox{UV-IR} connection provides additional constraints on the LECs. The recent proof of the $a$-theorem \cite{Komargodski:2011vj} is actually based on such a  twice-subtracted dispersion relation for the dilaton elastic scattering where $c\sim a_{UV}-a_{IR}\geq0$.
Analogously, for the $SU(2)$ chiral Lagrangian one can derive dispersion relations that provide positivity constraints on the LECs $\ell_{4,5}$ \cite{Pham:1985cr,Adams:2006ch,Distler:2006if}. In fact, for  $\pi\pi$ scattering in QCD one can even go beyond the forward limit and implement unitarity, crossing symmetry, and analyticity in a set of twice-subtracted dispersion relations known as Roy equations \cite{Roy:1971tc},  see e.g.~refs.~\cite{Caprini:2011ky,Colangelo:2001df} for recent discussions. 
Similar twice-subtracted dispersion relations have been derived in the context of particle physics beyond the Standard Model (SM). For example, ref.~\cite{Distler:2006if} studied twice-subtracted dispersion relations for the scattering of longitudinally polarized Electroweak (EW) vector bosons $W$ and $Z$ in the EW chiral Lagrangian. 

All the examples above set constraints on LECs at $O(p^4)$. Indeed, the twice-subtracted dispersion relations ensure the UV convergence of the integral of the total cross-sections that cannot exceed the Froissart bound  $\sigma(s)\sim \log^2 s$ \cite{Froissart:1961ux}. However, as it was noticed in ref.~\cite{Falkowski:2012bu}, certain linear combinations of the scattering amplitudes may still be convergent with just one subtraction and thus give sum rules for the leading LECs at $O(p^2)$. In particular, inspired by the results of ref.~\cite{Low:2009di}, the authors of ref.~\cite{Falkowski:2012bu} derived the sum rule 
\be
1-a^2=\f{v^2}{6\PI}\int_0^\infty \f{\di s}{s}\left(2\sigma^{\text{tot}}_{I=0}+3\sigma^{\text{tot}}_{I=1}-5\sigma^{\text{tot}}_{I=2}\right)\,,
\ee
where $I$ is the weak isospin. This is a constraint for the $O(p^2)$ coupling constant $a$ of a Higgs-like singlet $h$ coupled to the GBs emerging from the spontaneous breaking $SU(2)_L\times SU(2)_R\to SU(2)_{V}$ with an interaction term $a\, h\, \partial_{\mu} \pi^i \partial^{\mu} \pi^i/v$. Within chiral perturbation theory in QCD, this equation with $a=0$ is known as Olsson sum-rule \cite{Olsson}, and it is convergent because the combination of cross-sections under the integral does not couple to the pomeron \cite{Ananthanarayan:2000ht}.  Indeed, the amplitudes that saturate the Froissart bound at high energies drop in that linear combination. More recently, ref.~\cite{Urbano:2013ty} derived a sum rule for the elastic forward scattering of $\mathbf{4}$-plets $\pi^a=(\pi^{1,2,3},h)$ of an approximate custodial $SO(4)$ in composite Higgs models, while ref.~\cite{Grinstein:2013fia} studied perturbative unitarity sum rules in weakly coupled models with several Higgs bosons.

In this paper we build on these previous results and consider the elastic forward $2\rightarrow 2$ scattering of an arbitrary real, unitary representation $\mathbf{r}$ of an internal symmetry group $H$. Using unitarity, analyticity, and crossing symmetry we derive universal sum rules for the scattering amplitudes that encompass and generalize all previous examples, including once-subtracted dispersion relations,  shedding light on the underlying general structure of the coefficients of the scattering amplitudes at any order, as well as on the LECs at $O(p^2)$. 
EFTs for GBs associated with a coset $G/H$ (where $G$ may or may not be compact) are the prototypes of theories where our sum rules apply. But in fact, our approach is also valid for arbitrary spins and masses. We discuss in detail the sum rules for the scattering of longitudinally polarized EW gauge bosons $W_{L}$'s, and carefully compare the results to the gauge-less limit with GBs.
We prove positivity constraints on the coefficients of the scattering amplitudes that generalize those found in ref.~\cite{Adams:2006ch} for the shift symmetry to arbitrary groups. In particular,
we show that the amplitude coefficients must lie within a convex polyhedral cone. We describe how to identify the cone edges, which  determine the `strongest' positivity constraints that,
linearly combined with positive coefficients, generate the entire cone.

\bigskip
\noindent{\bf \emph{Sneak preview and summary of the results}}
\vspace{2mm}

\noindent  In the remaining part of the introduction we outline the main ideas and part of the results of this paper while skipping most of the technical details related e.g.~to the massless limit, the IR convergence, the possible IR residues, and the good analytic behavior of the scattering amplitudes around, say, $s=0$. The main ideas and results presented here carry over the general case as we show in the bulk of the paper.

The sum rules for the $2\rightarrow 2$ elastic scattering are derived from dispersion relations that relate the low-energy forward ($t=0$) eigen-amplitudes $\Amp_I(s)$ within each irreducible representation (irrep) $\r_I$ found in $\r \otimes \r$
\be
\Amp_I(s) \sim  a^{(0)}_I +a^{(1)}_I s + a^{(2)}_I s^2+\dots\,,
\ee
to certain linear combinations of integrals of total cross-sections.  
For example, in an index-free matrix notation, the sum rules for massless states  for one and two subtractions are
\begin{subequations}
\label{eq:sumrule1}
\begin{align}
\label{eq:sumrule1a}
 P_{-}\,a^{(1)}= &\,\frac{2}{\PI}
 \int_{0}^{\infty} \frac{\di s}{s} P_{-}\,\sigma^{\text{tot}}(s)\,, \\
 \label{eq:sumrule1b}
  P_{+}\,a^{(2)}=&\frac{2}{\PI}
 \int_{0}^{\infty} \frac{\di s}{s^2} P_{+}\,\sigma^{\text{tot}}(s)\,,
 \end{align}
 \end{subequations}
where $P_{\pm}=(\mathbbm{1}\pm X)/2$ are the projection operators into the $\pm1$-eigenspace of the involutory crossing matrix $X$ that acts on the eigen-amplitudes by exchanging $s\leftrightarrow u$ channels as $\Amp_I(u)=\sum_J X_{IJ}\Amp_J(s)$. 
Moreover, the amplitudes are not all independent because satisfy the constraints
\begin{equation}
 \label{eq:sumrule1c}
  P_{-}\,a^{(2n)}=0\,,\qquad P_{+}\,a^{(2n+1)}=0\,,
\end{equation}
which do not rely on unitarity but  depend only on the symmetry structure of the theory and crossing symmetry.
The crossing matrix is completely independent of the dynamics and fully determined by the symmetry $H$. The left-hand side of the sum rules \eqref{eq:sumrule1} represents the IR side where the coefficients $a^{(1,2)}_I$ can possibly be calculated within an EFT in terms of the LECs, whereas the integrals over the total cross sections encode information from any energy scale up to the UV, where the EFT is no longer valid. The presence of $P_{-}$ in the sum rule with one subtraction in eq.~\eqref{eq:sumrule1} is crucial to project out the UV divergent contribution of the integral in eq.~\eqref{eq:sumrule1a}, making it thus convergent, analogously to the Olsson sum rule in QCD. 
We show that in absence of degeneracy the number of linearly independent sum rules with an even (odd) number of subtractions equals the number of (anti-)symmetric irreps in $\r\otimes\r$. Explicit expressions for these linearly independent sum rules can simply be obtained by diagonalizing the crossing matrix.

We also provide an algorithm to systematically construct the strongest positivity constraints on the scattering coefficients $a^{(n)}_I$ when $n$ is even. 
In particular, we show that the crossing matrix is unitary with respect to the positive definite  (diagonal) metric $\mathcal{G}_{IJ}=\mathrm{dim}\,\mathbf{r}_I\, \delta_{IJ}$ made of the dimensions $\mathrm{dim}\,\mathbf{r}_I$ of the irreps. Equation~\eqref{eq:sumrule1b} can thus be written in terms of a scalar product $\langle \mathbf{v},a^{(2)}\rangle= \sum_{IJ} v^*_{I} \mathcal{G}_{IJ}a^{(2)}_J$  that involves only positive quantities
\begin{equation}
\langle \mathbf{v},a^{(2)}\rangle=\frac{2}{\PI}\int_0^\infty \frac{\di s}{s^2}\langle \mathbf{v},\sigma^{\mathrm tot}(s)\rangle \quad \Longrightarrow \quad \langle \mathbf{v},a^{(2)}\rangle\geq 0\,,
\end{equation}
whenever the $+1$-eigenvector $\mathbf{v}$ of $X$ has positive real components $v_I$. In fact, we show that there always  exist $\mathrm{\dim}\, \mathcal{V}_{+}\neq 0$ linearly independent such positivity constraints (where $\mathcal{V}_{\pm}$ is the $\pm1$-eigenspace) provided by vectors $\mathbf{v}$ that live in a convex polyhedral cone whose edges are the intersection of $\mathcal{V}_{+}$ and the positive quadrant $\mathbb{R}^{m}_+$ where $m=\mathrm{dim}\, X$. The strongest positivity constraints $\langle \mathbf{v}_{\mathrm{edge}},a^{(2)}\rangle\geq 0$ on the $a^{(2)}_I$ are those associated with the scalar product along the edge generators $\mathbf{v}_{\mathrm{edge}}$ of the polyhedral cone.

For odd $n$, we show that no such general positivity constraints can be obtained.
Therefore, one cannot univocally determine the sign of the associated $O(p^2)$ LECs from eq.~\eqref{eq:sumrule1a}.
Nevertheless, it turns out that the sum rules often allow us to pin down the quantum numbers of the states that are needed to obtain specific signs for the LECs.

Let us briefly discuss a concrete example.
Taking e.g.~$H=SO(N\neq 4)$ and $\mathbf{r}= \mathbf{N}$, i.e.~the fundamental representation (for $N\geq 3$). The product decomposes as $\mathbf{N}\otimes \mathbf{N}=\mathbf{1}\oplus \mathbf{A}\oplus\mathbf{S}$, where \mbox{($\mathbf{A}$) $\mathbf{S}$} is the traceless (anti-)symmetric representation. The crossing matrix $X$ has one $-1$-eigenvalue and two $+1$-eigenvalues. Hence eq.~\eqref{eq:sumrule1a} gives one (once subtracted) sum rule
\be
\label{eq:example_SON}
 2a^{(1)}_{\mathbf{1}}+N a^{(1)}_{\mathbf{A}}-(N+2)a^{(1)}_{\mathbf{S}}
= \frac{2}{\PI}\int_0^\infty \frac{\di s}{s}\left[2\sigma^{\mathrm{tot}}_{\mathbf{1}}+N\sigma^{\mathrm{tot}}_{\mathbf{A}}-(N+2)\sigma^{\mathrm{tot}}_{\mathbf{S}}\right]\,,
\ee
eq.~\eqref{eq:sumrule1c} gives the constraints
\begin{equation}
\label{constexamplesoNintro}
a^{(1)}_{\mathbf{S}}=-a^{(1)}_{\mathbf{A}}=-\frac{1}{N-1}a^{(1)}_{\mathbf{1}}\,,\qquad 2a^{(2)}_{\mathbf{1}}+N a^{(2)}_{\mathbf{A}}-(N+2)a^{(2)}_{\mathbf{S}}=0\,,
\end{equation} 
while eq.~\eqref{eq:sumrule1b} gives two other (twice-subtracted) sum rules -- see the text for their explicit form -- leading to the two strongest positivity constraints
\begin{equation}
\label{ineqsonexample}
a^{(2)}_{\mathbf{A}}+a^{(2)}_{\mathbf{S}} \geq 0\,, \qquad a^{(2)}_{\mathbf{1}}+(N-1)a^{(2)}_{\mathbf{S}} \geq 0\,,
\end{equation}
which correspond to the conditions $\langle\mathbf{v}^{i}_{\mathrm{edge}}, a^{(2)}\rangle\geq 0$ where $\mathbf{v}^{1}_{\mathrm{edge}}=(0,N,N+2)^T$ and \sloppy\mbox{$\mathbf{v}^{2}_{\mathrm{edge}}=(N+2,2,0)^T$} are the edge generators of polyhedral convex cone in which the amplitude coefficients must lie.
Using the constraints \eqref{constexamplesoNintro}, the positivity constraints \eqref{ineqsonexample} imply $a^{(2)}_{\mathbf{S}}\geq0$.

The power of the sum rules emerges when one calculates the coefficients $a^{(n)}_{I}$ in terms of the LECs of an EFT. Let us take for example the theory of GBs coming from the symmetry breaking pattern  $SO(N+1)\rightarrow SO(N)$ (a sphere) or $SO(N,1)\rightarrow SO(N)$ (a hyperboloid), and let us add to this theory of GBs extra light \emph{Higgs-like} states $h\in\mathbf{1}$ and $h_{ab}\in\mathbf{S}$ coupled as $\left(a h \delta_{ab}+b h_{ab}\right)\partial_\mu\pi^a \partial^\mu\pi^b /f_\pi$. 
Equation~\eqref{eq:example_SON} therefore becomes a sum rule that constrains the LECs: 
\be
\label{eq:example2_SONhiggs}
\left(\pm 1-a^2+\frac{N+2}{2N}b^2\right)=\frac{f_\pi^2}{2\PI N}\int_0^\infty \frac{\di s}{s}\left[2\sigma^{\mathrm{tot}}_\mathbf{1}+N\sigma^{\mathrm{tot}}_{\mathbf{A}}-(N+2)\sigma^{\mathrm{tot}}_{\mathbf{S}}\right]\,.
\ee
The signs $+$ and $-$ correspond to the sphere and hyperboloid respectively.
For \sloppy\mbox{$SO(4)/SO(3)\sim SU(2)_L\times SU(2)_R/SU(2)_V$} one recovers the sum rule of ref.~\cite{Falkowski:2012bu}, and the original Olsson sum rule for $a=b=0$. 

The scattering of $\mathbf{4}$'s of $SO(4)$ is relevant in every custodially symmetric composite Higgs model. It is quite special because the anti-symmetric $\mathbf{6}\in SO(4)$ is further reducible into two anti-symmetric representations $(\mathbf{3},\mathbf{1})$ and $(\mathbf{1},\mathbf{3})$ of $SU(2)_{L}\times SU(2)_{R}$. In turn, this theory admits two sum rules for odd $n$ and other two for even $n$ as we show in detail in section~\ref{subsec:SO4}. In particular, we find a new once-subtracted sum rule in addition to the sum rule found in ref.~\cite{Urbano:2013ty}.

For $W_{L}W_{L}\to W_{L}W_{L}$ scattering, one would be tempted, by invoking the Equivalence Theorem (ET) in the custodial limit $g'=0$, to directly extrapolate the result \eqref{eq:example2_SONhiggs} obtained for GBs in $SO(4)/SO(3)$. 
However, when using the ET one has to carefully take into account the $t$-channel $W$-exchange diagram, since the squared mass $m_W^2$ cannot be discarded in the forward limit $t=0$ due to a pole $1/(t-m_{W}^{2})$. In fact, such a term gives a finite contribution $\sim g^2/(2m_{W}^{2})=2/v^{2}$ independent of the gauge coupling $g$ which thus affects the left-hand side of the sum rule \eqref{eq:example2_SONhiggs} that gets replaced as $(1-a^{2})\,\to\, (3-a^{2})$. Alternatively, one can work directly with $W_L$ as external states and reproduce, in the forward limit, the same result in agreement e.g.~with ref.~\cite{Espriu:2014jya}. However, as we show in section \ref{sectWWscattering}, the additional contribution to the left-hand side of the sum rule is exactly canceled by an additional finite contribution to the right-hand side, coming from the integral of the amplitude along a big circle at infinity in the complex $s$ plane. This subtle point boils down to identifying the correct analytic structure of the theory and has often been overlooked in previous works. 
Moreover, the sum rule has been previously derived with $g^\prime=0$, where no photon exchange in the $t$-channel occurs.\footnote{We thank Adam Falkowski for remarking this point.} In fact, the $t$-channel exchange of a massless spin-1 boson has a Coulomb singularity at $t=0$, and one may question the validity of the sum rule \eqref{eq:example2_SONhiggs} for the SM with a small but finite $g^\prime$. Nevertheless, even in this case, we show in subsection \ref{sec:photonincluded} that a cancellation between these extra gauge contributions on both sides of the sum rule (derived  departing from the strict forward limit) occurs, again thanks to analyticity. In light of these results, we are able to show that the sum rules obtained for GBs at vanishing gauge couplings do actually carry over to the full gauge theory in the approximation of small, but finite, $g^\prime\ll1$.

\vspace{5mm}
We suggest the reader interested in physical applications to go directly to section \ref{examples} where we provide a self-contained summary of the tools developed in the previous sections, as well as detailed examples thoroughly worked out.

 The paper is organized as follows. In section \ref{sumrules} we introduce our general approach, discuss on general grounds the UV and IR convergence of the sum rules, and describe the relation with EFTs. In section~\ref{sec:positivity} we derive the positivity constraints emerging from even-subtracted dispersion relations. In section~\ref{examples} we give several examples of the application of our general approach to particularly interesting physical cases. We study the scattering of fundamentals of $SO(N\neq 4)$ and of adjoints of $SU(N\geq 4)$ for every $N$. We analyze in detail the special cases of $SO(3)$ and $SO(4)$ which are relevant for the EW chiral Lagrangian and composite Higgs models, as well as $SU(2)$ and $SU(3)$ for chiral QCD. We finally devote section~\ref{sectWWscattering} to longitudinal $WW$ scattering in the EW chiral Lagrangian and show the cancellation of the contributions from $t$-channel gauge boson exchange. In section \ref{conclusions} we draw our conclusions and highlight possible interesting applications of our results. Appendix \ref{app:crossingM} contains an extensive discussion of the crossing matrix $X$ and its general properties. Appendix \ref{AnalyticLR} is devoted to a discussion of the analytic structure of the amplitude in the presence of light unstable resonances. In appendix~\ref{beyondforward} we go beyond the forward limit and discuss the sum rules at $t\neq0$.  Appendix \ref{AppProjectors} describes the construction of the crossing matrix for $SO(N)$ and $SU(N)$. Appendix \ref{WWscattering} reports the full expression of the $W_{L}W_{L}\to W_{L}W_{L}$ scattering amplitude at tree level.

\section{Sum rules}
\label{sumrules}

Let us focus on the $2\to2$ elastic scattering $\left|a\right>\left|b\right>\to \left|c\right>\left|d\right>$ with $a,b,c,d=1\,\ldots,\dim \r$ belonging to the real (non necessarily irreducible) representation $\r=\ovl{\r}$ of a symmetry group $H$. For concreteness we focus on real particles but the same arguments can be extended by properly including charge conjugation.
Two-particle  states can be decomposed into irreps ${\bf r}_{I(\xi)}$
\be
\r\otimes \r=\bigoplus_{I(\xi) } \r_{I(\xi)}
\label{CGseries}
\ee
where
$I$ is a (collective) index  that identifies inequivalent irreps, while $\xi$ labels possible degenerate identical irreps appearing in the decomposition. 
For example, in the scattering of triplets $\mathbf{3}$ under $SO(3)\sim SU(2)$, we have  $\mathbf{3}\otimes \mathbf{3}=\mathbf{1}\oplus\mathbf{3}\oplus\mathbf{5}$. 
For $SU(3)$, the scattering of adjoints $\mathbf{8}$ decomposes as $\mathbf{8}\otimes \mathbf{8}=\mathbf{1} \oplus \mathbf{8}_1 \oplus \mathbf{8}_2 \oplus \mathbf{10} \oplus \ovl{\mathbf{10}}  \oplus \mathbf{27}$ so that the $\mathbf{8}$ are degenerate because appear twice on the right-hand side of eq.~\eqref{CGseries}.
Equation \eqref{CGseries} allows us to decompose $\left|a\right>\left|b\right>\equiv |ab\rangle\in  \r\otimes \r$, $a,b=1\,\ldots,\dim \r$, as 
\be\label{CGcoeff}
|ab\rangle=\sum_{I(\xi),i}C^{ab}_{I(\xi)i} | I(\xi),i\rangle\,,
\ee
where $|I(\xi),i\rangle$ ($i=1,\ldots,\dim \r_{I}$) is a basis of $\r_{I(\xi)}$ and $C^{ab}_{I(\xi)i}$ denote the Clebsch-Gordan (CG) coefficients relating the two bases.

By the Wigner-Eckart theorem the scattering amplitudes among different irreps can be written just in terms of eigen-amplitudes $\Amp_{I(\xi\xi^\prime)}(s,t)$: 
\be
\label{WE}
\Amp_{I(\xi)i\rightarrow J(\xi')j} (s,t)=\delta_{ij}\delta_{IJ}\Amp_{I(\xi\xi^\prime)}(s,t)\, .
\ee
Here $s$, $t$ and $u$ are the standard Mandelstam variables $s=(p_a + p_b)^2$, $t=(p_a -p_c)^2$,  $u=(p_a -p_d)^2$ with $s+t+u=4m^2$. Notice that the {\it mixed} eigen-amplitues $\Amp_{I(\xi\xi^\prime)}(s,t)$ between degenerate irreps with $\xi\neq\xi'$, can be in principle non-vanishing unless other selection rules can be invoked. We come back to this point later on.

Hereafter, unless stated otherwise, we refer to forward scattering only
\be
\Amp_{I(\xi\xi^\prime)}(s)\equiv\Amp_{I(\xi\xi^\prime)}(s,t=0)\,.
\ee
Furthermore, we assume throughout this paper that the amplitudes obey the ordinary first principles of:
\begin{enumerate}
\item[(1)] \textit{Analyticity}, which allows us  to extend ${\cal A}_{I(\xi\xi')}(s)$ to an analytic function over the complex plane, with poles and branch cuts corresponding to the contributions of stable particles and of the continuum to the scattering process, as for instance in figure \ref{fig:path}. 

\item[(2)] \textit{Unitarity}, which gives the optical theorem
	\be\l{opticaltheorem}
		\mathrm{Im}\Amp_{I(\xi\xi)}(s)= s\sqrt{1-\f{4m^{2}}{s}} \sigma^{\mathrm{tot}}_{I(\xi\xi)}(s)
	\ee
	for $s$ on-shell and where $m$ is the mass of the particles $\r$ and also implies, via analytic continuation, that
	\be
		\label{eq:realitycondition}
		\Amp_{I(\xi\xi^\prime)}(s)^*=\Amp_{I(\xi^\prime\xi)}(s^*)\,,
	\ee
	which generalizes the Schwarz reflection principle.
\item[(3)] \textit{Crossing symmetry}, which relates e.g.~$s$- and $u$-channel amplitudes
	\be
		\label{eq:crossA}
		\Amp_{ab\rightarrow cd}(s) =\Amp_{a d\rightarrow cb}(u)\,,
	\ee
where
$u=4m^2-s$. In addition to $s\leftrightarrow u$ we can also completely exchange the initial and final states, $s\leftrightarrow s$, for which crossing symmetry implies
the following relations between the eigen-amplitudes
\be
\label{eq:Amp_real_irrep}
\Amp_{I(\xi\xi^\prime)}(s)=\Amp_{\bar{I}(\xi^\prime\xi)}(s)\,.
\ee
\end{enumerate}

\noindent We often adopt an index-free notation for the eigen-amplitudes:
\be\label{indfree}
\Amp(s)\equiv\left(\begin{array}{c}\vdots \\
\Amp_{I(\xi\xi')}(s)\\
\vdots
\end{array}\right)\,,
\ee
where the collective index $I(\xi\xi^\prime)$ is now restricted to {\em independent} eigen-amplitudes, i.e.~eigen-amplitudes which are unrelated by eq.~\eqref{eq:Amp_real_irrep}. For example, for the scattering $\mathbf{8}\otimes \mathbf{8}=\mathbf{1} \oplus \mathbf{8}_1 \oplus \mathbf{8}_2 \oplus \mathbf{10} \oplus \ovl{\mathbf{10}}  \oplus \mathbf{27}$ in chiral $SU(3)$ eq.~\eqref{eq:Amp_real_irrep} gives $\Amp_{\bf 8_{12}}=\Amp_{\bf 8_{21}}$ and $\Amp_{\ovl{{\bf 10}}}=\Amp_{\bf 10}$ and then only $\Amp_{\bf 8_{12}}$ (or $\Amp_{\bf 8_{21}}$) and \mbox{$\Amp_{{\bf 10}}$ (or $\Amp_{\ovl{{\bf 10}}}$)} appear in the index-free vector $\Amp(s)$.

In appendix  \ref{app:crossingM} it is shown that the $s\leftrightarrow u$ crossing symmetry acts on the eigen-amplitudes via a constant involutory \textit{crossing matrix} $X$ \footnote{More explicitly, the crossing matrix carries two collective indices $X_{I(\xi\xi^\prime) J(\zeta\zeta^\prime)}$ and eq.~\eqref{eq:crossEigenA} in components reads $\Amp_{I(\xi\xi^\prime)}(u) = \sum_{J\zeta\zeta^\prime} X_{I(\xi\xi^\prime)J(\zeta\zeta^\prime)} \Amp_{J(\zeta\zeta^\prime)}(s)$.
For scattering amplitudes of non-real particles, one should include in $X$ the charge conjugation operator.}
\be\label{eq:crossEigenA}
\Amp(u) =X \Amp(s)\,,\qquad\qquad X^2=\mathbbm{1}\,.
\ee
The entries $X_{I(\xi\xi^\prime)J(\zeta\zeta^\prime)}$ of the crossing matrix  are constructed in terms of the  CG  coefficients defined in eq.~\eqref{CGcoeff} and we refer to the appendix for their detailed description. Let us stress that, consistently with eq.~\eqref{eq:crossEigenA}, the indices of $X_{I(\xi\xi^\prime)J(\zeta\zeta^\prime)}$  must be restricted to those labeling the independent eigen-amplitudes which enter the index-free vector ${\cal A}$, as discussed after eq.~\eqref{indfree}. Also notice that, since $X^2=\mathbbm{1}$, all the eigenvalues of $X$ are either $+1$ or $-1$.

While we leave a detailed discussion of $X$ to appendix \ref{app:crossingM} and to the following sections,  we mention here one of its important properties. One of the $+1$-eigenvectors of $X$ is given by the vector $v$ with components
\be\label{onevector}
v_{I(\xi\xi')}=\left\{\begin{array}{l} 1\quad~~~ \text{if }\xi=\xi'\\
0\quad~~~ \text{if }\xi\neq\xi'\end{array}\right.\,.
\ee

Let us also anticipate that we encounter several cases in which $X$ is block-diagonal in non-mixed and mixed indices, i.e.
\be\label{blockdiag}
X=\left(\begin{array}{cc} \hat X &  \\   & X_{\rm mix} \end{array}\right)\,,
\ee
where $\hat X$ has only non-mixed entries $\hat X_{I(\xi\xi)J(\zeta\zeta)}$. Even though most of the following discussion is general and does not require eq.~\eqref{blockdiag}, in most physical applications it could be convenient to work with $\hat X$ instead of $X$. For instance this is the case for chiral $SU(3)$. We refer to $\hat X$ as the {\em reduced} crossing matrix. It clearly satisfies $\hat X^2=\mathbbm{1}$ and then it has eigenvalues $\pm 1$.
Furthermore, the vector in eq.~\eqref{onevector} restricts to a $+1$-eigenvector $\hat v$ of $\hat X$ with identical components $\hat v_{I(\xi\xi)}=1$.

\subsection{Dispersion relations}
\l{sec:dispersionrelations}

\begin{figure}[t!]
\begin{center}
\includegraphics[scale=0.5]{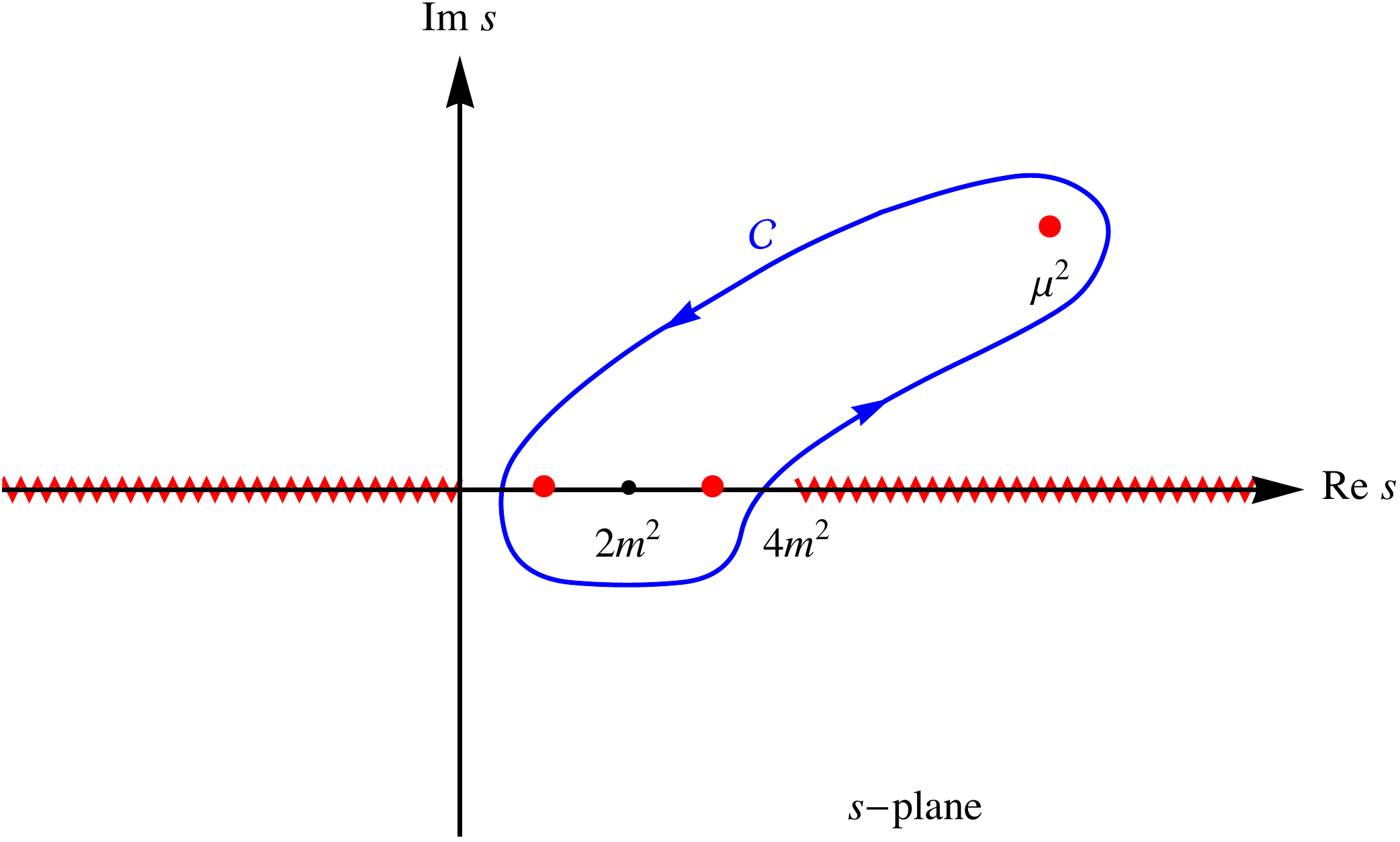}
\caption{Analytic structure of the amplitude $\Amp(s)$ in the $(\text{Re}\,s,\text{Im}\,s)$ plane. The contour $\mathcal{C}$ corresponding to the Cauchy integral formula \eqref{eq:Cauchy}, encloses the point $\mu^{2}$ around which the amplitude is expanded and the poles at $s=s_{i}$ (red points) corresponding to propagating particles with masses lighter than $4m^{2}$. The analytic structure of $\Amp(s)$ is symmetric under reflection around $2m^{2}$.}\label{fig:path}
\end{center}
\end{figure}

By analyticity, expanding the amplitude ${\cal A}(s)$ around a certain (complexified) scale $s=\mu^2$
\be
{\cal A}(s)=\sum_n {\cal A}^{(n)}(\mu^2)(s-\mu^2)^n\,,
\ee
one can use the Cauchy integral formula to express the coefficients ${\cal A}^{(n)}(\mu^2)$ as
\be\label{eq:Cauchy}
\Amp^{(n)}(\mu^2)+	\sum_{s_{i}}\mathrm{Res}\left[\frac{\Amp(s)}{(s-\mu^2)^{n+1}}\right]=\frac{1}{2\PI \i}\oint_\mathcal{C}\di s \frac{\Amp(s)}{(s-\mu^2)^{n+1}}\,,
\ee
where the left-hand side is the contribution from the residues at the poles $s=s_{i}$ (and their crossed) and $s=\mu^2$ enclosed by a contour $\mathcal{C}$ in the complex $s$-plane that does not cross any singular point, see figure \ref{fig:path}. For convenience we introduce the notation $a^{(n)}\equiv \Amp^{(n)}(\mu^{2}=0)$ or, more explicitly
\be
a^{(n)}_{I(\xi\xi')}\equiv \Amp^{(n)}_{I(\xi\xi')}(\mu^{2}=0)\,.
\ee

We consider the following analytic structure of the amplitude $\Amp(s)$: there is a branch cut running on the real axis from $s=4m^2$ (corresponding to the physical threshold of $\r$'s pair production) to $+\infty$. Crossing symmetry at $t=0$, i.e.~$s\to u=-s+4m^{2}$, enforces another cut from $-\infty$ to $s=0$. In addition, there may be mass poles for $s=s_i$ and, by crossing symmetry, for $s=4m^{2}-s_{i}$ on the real axis  below $4m^2$ associated to light propagating particles. Heavier resonances do not give poles in the physical Riemann sheet. Should the light poles at $s=s_i$ be unstable as well, they would move to another Riemann sheet hidden by a longer cut, see figure \ref{fig:path3}. The analytic structure in this case is discussed in appendix \ref{AnalyticLR}.

\begin{figure}[t!]
\begin{center}
\includegraphics[scale=0.5]{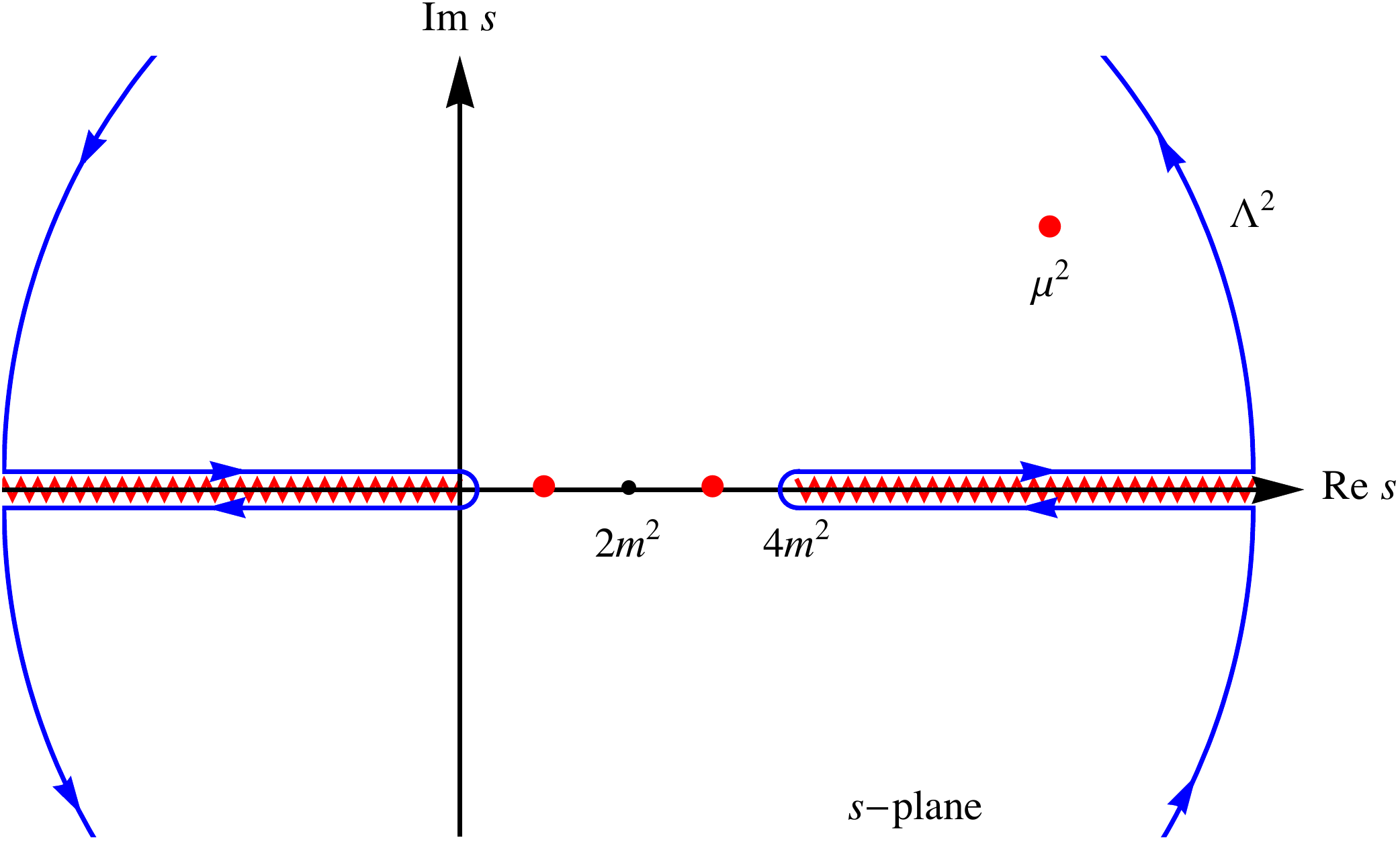}
\caption{The contour $\mathcal{C}$ of figure \ref{fig:path} deformed to a path along the branch cuts plus a big circle at $s=\Lambda^{2}$.}\label{fig:path2}
\end{center}
\end{figure}

We can now smoothly deform the integration contour $\mathcal{C}$ as in figure \ref{fig:path2}.
The right-hand side of eq.~\eqref{eq:Cauchy} can be written as the sum of two terms: one comes from the big circle of radius $\Lambda^2$ centered around $2m^2$
\be
\label{cinfty}
c^{\Lambda(n)}=\int_0^{2\PI} \frac{\di\theta}{2\PI} \frac{|s_{\Lambda}|\e^{\i\theta}\Amp(|s_{\Lambda}| \e^{\i\theta})}{\left(|s_{\Lambda}| \e^{\i\theta}-\mu^2\right)^{n+1}}\,,\qquad |s_{\Lambda}|=2m^2+\Lambda^2\,,
\ee
with $\Lambda$ eventually going to infinity, and the other from the integrals along the branch cuts:
\begin{align}\label{eq:dispersion0}
 \int_{4m^2}^{\Lambda^2+2m^2} \frac{\di s}{2\PI \i} \left[\frac{\Amp(s+\i\epsilon)-\Amp(s-\i\epsilon)}{(s-\mu^2)^{n+1}}+(-1)^n \frac{\Amp(-s+4m^2-\i\epsilon)-\Amp(-s+4m^2+\i\epsilon)}{(s-4m^2+\mu^2)^{n+1}} \right]\,.
\end{align} 
By the crossing symmetry \eqref{eq:crossEigenA}, we can rewrite eq.~\eqref{eq:Cauchy} as follows:
\be
\begin{aligned}
\label{eq:dispersion1}
 & \Amp^{(n)}(\mu^2)+ \sum_{\mathrm{s_i}}\mathrm{Res}\left[\frac{\Amp(s)}{(s-\mu^2)^{n+1}}\right]
=  \\
& c^{\Lambda\,(n)}+ \int_{4m^2}^{\Lambda^2+2m^2} \frac{\di s}{2\PI \i} \left[\frac{1}{(s-\mu^2)^{n+1}}+(-1)^n \frac{X}{(s-4m^2+\mu^2)^{n+1}} \right]\left[\Amp(s+\i\epsilon)-\Amp(s-\i\epsilon) \right]\,.
\end{aligned} 
\ee
In general, the condition \eqref{eq:realitycondition} implied by unitarity gives 
\be
\label{imaginaryandrealpart}
\Amp(s+\i\epsilon)-\Amp(s-\i\epsilon)=2\,\mathrm{Re}\,\Amp^{-}(s+\i\epsilon)+2\,\i\,\mathrm{Im}\,\Amp^{+}(s+\i\epsilon)\,,
\ee
where, in components, 
\be
\Amp^{\pm}_{I(\xi\xi^\prime)}(s)\equiv \frac{1}{2}\left[\Amp_{I(\xi\xi^\prime)}(s)\pm \Amp_{I(\xi'\xi)}(s)\right]
\ee
are the symmetric and anti-symmetric combinations with respect to the degeneration indices $\xi$ and $\xi^\prime$. 
In absence of degeneracy  we clearly have $\Amp^{+}= \Amp$ and $ \Amp^{-}=0$, but this can easily happen also in the degenerate case. For instance, the mixed amplitudes  may vanish by means of other selection rules which in fact remove the degeneracy\footnote{For example, a tiny mass splitting, angular momentum conservation or extra quantum numbers. For instance, this is the case for adjoints of $SU(3)$ where $\Amp_{\mathbf{8_{12}}}=0$ at $t=0$ for angular momentum conservation (see subsection \ref{SUN}).} so that  $\Amp_{I(\xi\xi^\prime)}=\Amp_{I}\delta_{\xi\xi^\prime}$ and then $\Amp^{+}= \Amp$.  Moreover, $\Amp^+=\Amp$ is granted whenever degenerate irreps are all real, as it must be the case for small enough real representations $\r$,  because of crossing symmetry  \eqref{eq:Amp_real_irrep}. In all these cases we can identify $\Amp^{+}$ with $\Amp$ and write
\be
\label{eq:dispersion2}
  \sum \mbox{ (residues)}^{(n)}
=   c^{\Lambda\,(n)}+ \int_{4m^2}^{\Lambda^2+2m^2} \frac{\di s}{\PI} \left[\frac{1}{(s-\mu^2)^{n+1}}+(-1)^n \frac{X}{(s-4m^2+\mu^2)^{n+1}} \right]\mathrm{Im}\,\Amp(s+\i\epsilon)\,,
\ee
where the left-hand side is a shorthand for
\be
\label{residuedef}
\sum \mbox{ (residues)}^{(n)}= \Amp^{(n)}(\mu^2)+\sum_{s_{i}}\mathrm{Res}\left[\frac{\Amp(s)}{(s-\mu^2)^{n+1}}\right]\,.
\ee
In the following we assume that ${\cal A}^-=0$ and therefore that eq.~\eqref{eq:dispersion2} holds, bearing in mind that whenever ${\cal A}^-$ can be non-vanishing one needs to use eq.~\eqref{eq:dispersion1} instead of eq.~\eqref{eq:dispersion2}.

\subsection{Convergence}
\label{sec:convergence}

Equation \eqref{eq:dispersion2} represents a set of general dispersion relations with $n$ subtractions. It is well known that for $n\geq2$, the integrals are convergent and $c^{\Lambda(n)}\rightarrow 0$  for $\Lambda\rightarrow\infty$, i.e.~when the radius of the big circle is sent to infinity, thanks to the Froissart bound $|\Amp(s)|\leq \text{contst}\times  s\log^2 s$ for $s\to \infty$ \cite{Froissart:1961ux}. For example, ref.~\cite{Manohar:2008tc} derived such a dispersion relation with two subtractions for the particular case of chiral $SU(2)$ in QCD. For $n=1$ one would instead expect no convergence when the amplitude saturates the Froissart bound.
However, only definite directions in the amplitude space may grow maximally fast so that even for $n=1$ one can find certain linear combinations which are convergent.
For $n=1$, the integral at large $s$ in eq.~\eqref{eq:dispersion2} is indeed dominated by
\begin{align}
\label{eq:asymptoticdispersion}
\frac{2}{\PI}\int^{\infty} \frac{\di s}{s^2}\, P_-\,\mathrm{Im}\Amp(s)\,,
\end{align}
where $P_{\pm}$ denote the projection operators associated with the $\pm1$-eigenspaces of the crossing matrix $X$:
\be\label{ppm}
P_{\pm}\equiv \frac{1}{2}\left(\mathbbm{1} \pm X\right)\,.
\ee 

As we will presently see, it follows from eq.~\eqref{eq:asymptoticdispersion} that in order to draw conclusions about the once-subtracted ($n=1$) dispersion relations it is sufficient to add to the three first principles listed above one further assumption:
\begin{enumerate}
\item[(4)]  {\it Universal asymptotic behavior} of the amplitude, that is the asymptotic scattering amplitude at large $s$ is the same for all irreps. More precisely we assume the following leading asymptotic behavior 
\be
\label{eq:Asymptotic}
\Amp_{I(\xi\xi^\prime)}(s) \sim \text{const}\times\,s\, \delta_{\xi\xi^\prime}\,, \qquad\mbox{for }s\rightarrow \infty \qquad (\mbox{modulo factors of }\log s )\,,
\ee
where the constant factor is independent of $I(\xi\xi^\prime)$.
\end{enumerate}
This condition refines the way an amplitude can saturate the Froissart bound: $\Amp(s)$ and hence $c^{\Lambda(1)}$ are allowed to grow maximally fast but in a universal way. Indeed, the condition \eqref{eq:Asymptotic}
is equivalent to demanding that $\Amp(s)$, and then $c^{\Lambda(1)}$,  is asymptotically proportional to the $+1$-eigenvector $v$ of the crossing matrix $X$ defined in eq.~\eqref{onevector}. This means that the leading asymptotic contribution to $\Amp(s)$ and $c^{\Lambda(1)}$ is annihilated by $P_-$.

We then see that eq.~\eqref{eq:Asymptotic} is sufficient for guaranteeing that eq.~\eqref{eq:asymptoticdispersion}, and hence the integral in eq.~\eqref{eq:dispersion2} for $n=1$, converges for $\Lambda\rightarrow \infty$. 
Moreover, if we project the entire sum rule \eqref{eq:dispersion2} with $n=1$ onto the $-1$-eigenspace of the crossing matrix $X$, we get rid of the big circle contribution since  $P_-c^{\infty(1)}=0$.
We then arrive at the expression
\be
\label{eq:dispersionn=1}
 P_- \sum \mbox{ (residues)}^{(1)}
=    \int_{4m^2}^{\infty} \frac{\di s}{\PI} \left[\frac{1}{(s-\mu^2)^{2}}+ \frac{1}{(s-4m^2+\mu^2)^{2}} \right]P_{-}\,\mathrm{Im}\,\Amp(s+\i\epsilon)\,,
\ee
which represents a set of once-subtracted dispersion relations involving only finite quantities. (Notice that the same argument can be repeated starting from the more general dispersion relations in eq.~\eqref{eq:dispersion1}.) In fact, by crossing symmetry and analyticity alone we know that \mbox{$\Amp(2m^2+s)=X\Amp(2m^2-s)$}, and thus $P_{+} c^{\infty(1)}=0$ too. Therefore, under the assumption of universal asymptotic growth \eqref{eq:Asymptotic} of the $\Amp_I$, the integral contribution along the big circle averages to $c^{\infty(1)}=0$ and we can write another dispersion relation,
\be
\label{sr1subtrconstr}
P_+ \sum \mbox{ (residues)}^{(1)}
=    \int_{4m^2}^{\infty} \frac{\di s}{\PI} \left[\frac{1}{(s-\mu^2)^{2}}- \frac{1}{(s-4m^2+\mu^2)^{2}} \right]P_{+}\,\mathrm{Im}\,\Amp(s+\i\epsilon)\,.
\ee
For $\mu^2=2m^2$, this equation represents just the constraints $P_+ \sum \mbox{ (residues)}^{(1)}=0$ imposed by crossing symmetry rather than a genuine once-subtracted sum rule as opposed to the eq.~\eqref{eq:dispersionn=1}.

Let us discuss now the validity of the condition \eqref{eq:Asymptotic}.
Strongly coupled theories can have amplitudes which saturate the Froissart bound and could in principle violate the condition \eqref{eq:Asymptotic}. However, whenever the fastest growth  is reached by exchanging an $H$-singlet object in the  $t$-channel, the corresponding eigen-amplitude does satisfy  the condition \eqref{eq:Asymptotic} because of unitarity of the CG coefficients
\be
\label{example_asymtotic}
\Amp_{ab\rightarrow cd}\sim s\,\delta^{ac}\delta^{bd} \quad \Longrightarrow\quad \Amp_{I(\xi\xi^\prime)}\sim s \,\delta_{\xi\xi^\prime}\,.
\ee
QCD, for example, satisfies the condition \eqref{eq:Asymptotic} because the  Froissart bound is indeed saturated by the exchange of the pomeron, a completely neutral composite object with the quantum numbers of the vacuum. It is in fact the universality expressed by eq.~\eqref{eq:Asymptotic} that gives rise to the Pomeranchuk's theorem \cite{Collins:1971hv}. Reference \cite{Foldy:1963zz} has indeed formally shown that, whenever the imaginary part of the amplitude is independent of the quantum numbers of the scattering states, the $H$-singlet exchange alone dominates the amplitude.   
Moreover, any model that respects the  Regge theory is also satisfying the condition \eqref{eq:Asymptotic} since the leading Regge trajectory is again due to a neutral object exchanged in the $t$-channel giving the behavior in eq.~\eqref{example_asymtotic} \cite{Collins:1971hv}.
The condition \eqref{eq:Asymptotic} is very general and, to the best of our knowledge, there exists no strongly coupled counter-example that violates it.    

Weakly coupled theories in the UV require more care. On the one hand, one would expect scattering amplitudes to fall with energy or, at most, become constant or admit perhaps a mild logarithmic growth.  For those, $\mbox{const}=0$ in eq.~\eqref{eq:Asymptotic} (meaning that the amplitude does not saturate the Froissart bound) and the convergence of the dispersion relations with one subtraction holds.
On the other hand, the amplitudes involving propagating massive spin-1 states in the UV may grow faster than  $\log s$.
Indeed, spontaneously broken gauge theories contain spin-1 bosons with masses $m_V$ that, propagating in the $t$-channel, contribute to the real forward scattering amplitudes  with a term $\delta \Amp\sim c_*^2 s$: even though the integral over the imaginary amplitudes (that is total cross-sections) remains finite, $\delta \Amp$ gives a finite contribution, coming from the big circle in the UV (see eq.~\eqref{cinfty}), $\delta c^{\infty(1)}=c_*^2$,  which is not necessarily projected out by $P_{-}$,  as stressed e.g.~in ref.~\cite{Falkowski:2012bu}.   In principle, one should therefore add this finite contribution $P_{-} \delta c^{\infty(1)}$  to the right-hand side of eq.~\eqref{eq:dispersionn=1}. Nevertheless, despite appearances, such an extra $P_{-} \delta c^{\infty(1)}$  from $t$-channel exchange is actually harmless when the massive gauge degrees of freedom in the IR and the UV are the same. The massive gauge bosons  contribute indeed to the left-hand side (the IR-side) of the sum rules too, and by the very same amount $\delta\Amp^{(1)} =c_*^2$ (see section \ref{sectWWscattering} for an explicit example). This is the case whenever the extra contribution to the amplitude is the same in the UV and in the IR, so that trivially, by analyticity, its integral is the same along the contours $\cal{C}$ and the big circle at $\Lambda^{2}$ (see figures \ref{fig:path} and \ref{fig:path2}).
This reasoning is not spoilt by the running of the gauge coupling or higher loops contributions because the exchanged momentum $t$ is zero while $s=\Lambda^2\rightarrow\infty$. In this regime the eikonal approximation becomes exact enforcing the above cancellation. 

The net contribution from the massive gauge bosons propagating in the $t$-channel, if they are stable, is thus only through the IR residues at $s=m_W^2$ and its crossed point, which is however negligible when $\mu^2\gg m_W^2$, see section~\ref{sumruleEFTsect} for details.

In summary,  the once-subtracted dispersion relations \eqref{eq:dispersionn=1} are theoretically on a firm ground. There is however a last important caveat: we have always assumed that it is possible to take the forward limit $t=0$.  When massless spin-1 states are propagating in the $t$-channel this may not be the case and one should add an IR regulator that provides a mass gap, or alternatively move away from the strict forward limit as done in the Roy equations, that exploit the partial wave expansion \cite{Roy:1971tc,Caprini:2011ky,Colangelo:2001df}. We come back to this point when discussing the sum rules for gauge theories in section~\ref{sectWWscattering} where we make use of dispersion relations at finite $t$ (presented in appendix~\ref{beyondforward}) that are needed to avoid the Coulomb singularity from the photon exchange. Anticipating the final result, by analyticity, a cancellation similar to that of massive gauge bosons discussed above takes place. In fact, the extra contribution from the massless vectors cancels between the two sides of the dispersion relation before taking the limit $t\rightarrow 0$.

\subsection{The sum rules}
\l{sec:sumrules}

Let us come back to the general dispersion relations \eqref{eq:dispersion2}. For $n\geq 2$, by using the projectors $P_\pm$ introduced in eq.~\eqref{ppm}, they can be projected onto the  $\pm 1$-eigenspaces of $X$ as
\be \label{exact_sumrulen1}
P_{\pm}\sum \mbox{ (residues)}^{(n)}=
 \int_{4m^2}^{\infty} \frac{\di s}{\PI} \left[\frac{1}{(s-\mu^2)^{n+1}}\pm \frac{(-1)^n}{(s-4m^2+\mu^2)^{n+1}} \right]P_{\pm}\,\mathrm{Im}\Amp(s+\i\epsilon)\,.
\ee
Under the conditions discussed in subsection \ref{sec:convergence}, this equation holds for $n=1$ too (and in fact, projected with $P_{-}$ only, it holds at the crossing symmetric point $\mu^{2}=2m^{2}$ for $n=0$ as well).

These dispersion relations involve integrals over  the physical region $s\geq 4m^2$. We want now to link them to physical observables such as the total cross-section. Indeed, unitarity implies the optical theorem \eqref{opticaltheorem} for the elastic forward scattering.  
Notice, however, that we cannot always focus just on the elastic forward amplitudes $\Amp_{I(\xi\xi)}$, since the crossing matrix $X$, and hence the projectors $P_\pm$, may bring non-elastic terms from mixed amplitudes $\Amp_{I(\xi\xi^\prime)}$ with $\xi\neq\xi'$ into the game.  In such a case, we cannot write
 \be\l{opttheor2}
 \mathrm{Im}\Amp(s+\i \epsilon)=s\sqrt{1-\f{4m^{2}}{s} } \sigma^{\text{tot}}(s)
 \ee
in eq.~\eqref{exact_sumrulen1}.
 
However, such a problem is often absent or can easily be circumvented.  For example, whenever non-trivial mixed amplitudes are absent, $\Amp_{I(\xi\xi^\prime)}=\Amp_{I(\xi\xi)} \delta_{\xi\xi^\prime}$,  
as for instance in the presence of additional selection rules, eq.~\eqref{opttheor2} holds and then from eq.~\eqref{exact_sumrulen1} one gets the sum rules
  \begin{align}
 \label{exact_sumrulen2}
 P_{\pm}\sum \mbox{ (residues)}^{(n)}= &
 \int_{4m^2}^{\infty} \frac{\di s}{\PI} \left[\frac{s}{(s-\mu^2)^{n+1}}\pm \frac{(-1)^{n} s}{(s-4m^2+\mu^2)^{n+1}} \right] \sqrt{1-\f{4m^{2}}{s}} P_{\pm}\sigma^{\text{tot}}(s)\,.
 \end{align}
Clearly, the same sum rules hold when the crossing matrix has the block-diagonal structure \eqref{blockdiag}, up to restricting to the non-mixed amplitudes and replacing $X\to\hat{X}$. 

Furthermore, even when $X$ does not have the form \eqref{blockdiag} and  the mixed amplitudes are non-vanishing, one could still obtain sum rules involving physical observables.  
Indeed, we can  consider the elastic scattering amplitude between mixed states $1/\sqrt{2}(| I(\xi)\rangle+|I(\xi^\prime)\rangle)$
\begin{align}
\Amp_{I(\xi)+I(\xi^\prime)}= \frac{1}{2}\Amp_{I(\xi)}+\frac{1}{2}\Amp_{I(\xi^\prime)}+\frac{1}{2}[\Amp_{I(\xi\xi^\prime)}+\Amp_{I(\xi^\prime\xi)}]=\frac{1}{2}\Amp_{I(\xi)}+\frac{1}{2}\Amp_{I(\xi^\prime)}+\Amp_{I(\xi\xi^\prime)}
\end{align}
and define $\sigma^{\text{tot}}_{I(\xi\xi^\prime)}$ for $\xi\neq \xi^\prime$ as 
\be\label{mixedsigma}
\sigma^{\text{tot}}_{I(\xi\xi^\prime)}\equiv \frac{\mathrm{Im}\Amp_{I(\xi\xi^\prime)}}{s  \sqrt{1-\f{4m^{2}}{s}}}=\sigma^{\text{tot}}_{I(\xi)+I(\xi^\prime)}-\frac{1}{2}\left(\sigma^{\text{tot}}_{I(\xi)}+\sigma^{\text{tot}}_{I(\xi^\prime)}\right)\,,
\ee
so that the sum rules \eqref{exact_sumrulen2} still hold and involve only physical cross-sections.

%
\subsection{Sum rules and EFT}
\label{sumruleEFTsect}

So far we have not used the freedom of choosing $\mu^2$.
Apart from where the singularities are located in the $s$-plane we can choose $\mu^2$ in eq.~\eqref{eq:Cauchy} as we like, although some choices may  be more useful than others. 
There are two choices that recommend themselves. 

The first choice corresponds to take the crossing symmetric point $\mu^2=2m^2$ that allows one to nicely disentangle, in Eq.~\eqref{exact_sumrulen2}, the actual sum rules
\begin{subequations}
\label{symmpointsumruleforpositivity}
 \begin{align}
 \label{symmpointsumruleforpositivityn=1}
 P_{-}\sum \mbox{ (residues)}^{(2k+1)} &= \frac{2}{\PI}
 \int_{4m^2}^{\infty} \di s \frac{s}{(s-2m^2)^{2k+2}}\sqrt{1-\frac{4m^{2}}{s}}P_{-}\sigma^{\text{tot}}(s)\,, &&k\geq 0\,, \\
 P_{+}\sum \mbox{ (residues)}^{(2k)} &= \frac{2}{\PI}
 \int_{4m^2}^{\infty} \di s \frac{s}{(s-2m^2)^{2k+1}}\sqrt{1-\frac{4m^{2}}{s}}P_{+}\sigma^{\text{tot}}(s)\,,&&k\geq1\,,\
  \end{align}
 \end{subequations}
 from the constraints 
\begin{subequations}
\label{constraintforpositivity}
 \begin{align}
 P_{-}\sum \mbox{ (residues)}^{(2k)} &= 0\,,&&k\geq0\,,\\
 \label{constraintforpositivityb}
 P_{+}\sum \mbox{ (residues)}^{(2k+1)} &= 0\,,&&k\geq0\,.
 \end{align}
 \end{subequations}
 Actually, these constraints \eqref{constraintforpositivity}  follow directly from the definition \eqref{residuedef} and eq.~\eqref{eq:crossEigenA}. 
 Notice that eq.~\eqref{constraintforpositivityb} for $k=0$ implies $P_{+}c^{\infty(1)}=0$ and therefore, should the assumption \eqref{eq:Asymptotic} be satisfied, $c^{\infty(1)}=0$.
   
The sum rules have two crucial properties:  they are IR and UV convergent,\footnote{Let us recall that only for one subtraction, $k=0$ in eq.~\eqref{symmpointsumruleforpositivityn=1}, the UV convergence of the  integral in eq.~\eqref{symmpointsumruleforpositivityn=1} is not automatically guaranteed and that there could be an additional constant $c^{(1)\infty}$ appearing on the right-hand side of eq.~\eqref{symmpointsumruleforpositivityn=1}, as discussed in subsection \ref{sec:convergence}.} and all the quantities on the right-hand side except for the projectors are real and have positive definite sign. 
The definite sign turns out to be crucial to derive positivity constraints that we discuss in the next section.

The other useful choice corresponds to $\mu^2$ much bigger than all IR mass scales, namely \mbox{$\mathrm{Re}\mu^2\sim \mathrm{Im}\mu^2\gg m_{\mathrm{IR}}^2 \approx m^2,s_i^2$}.
With such a choice, the dispersion relations \eqref{eq:dispersion2} and the sum rules (\ref{exact_sumrulen2}) take a simpler form by dropping all the IR structures.  In particular, we do not need to keep track of the IR residues. For instance  \eqref{eq:dispersion2} can be approximately written as
\be
\label{eq:dispersion3}
\Amp^{(n)}(\mu^2)=
c^{\Lambda\,(n)}+ \int_{4m^2}^{\Lambda^2} \frac{\di s}{\PI} \left[\frac{1}{(s-\mu^2)^{n+1}}+(-1)^n \frac{X}{(s+\mu^2)^{n+1}} \right]\mathrm{Im}\Amp(s+\i\epsilon),
\ee 
which holds up to small corrections of $O(m^2/\mu^2, m^2_i/\mu^2, m^2/\Lambda^2_{\mathrm{IR}})$ where $\Lambda_{\mathrm{IR}}$ is the cutoff of the EFT.  
Note, however, that unless the IR masses are really small this is possible only for the first few subtractions, i.e.~for  $n=1,2$ or so. Indeed, if we want to be able to calculate the left-hand side within the EFT, $|\mu|^2$ is bounded from above by the IR cutoff $\Lambda^2_{\mathrm{IR}}$, while the coefficients $\Amp^{(n)}(\mu^2)$ are generically suppressed by higher powers of $\Lambda_{\mathrm{IR}}^2$. For example, the choice $|\mu|^2\lesssim \Lambda_{\mathrm{IR}}^{2}/4\pi$ represents a compromise that works reasonably well for $n=1,2$.  In any case, as long as the IR side of the sum rule is calculable within the EFT one can always check whether this approximation is valid. If it is not, then one should keep the residues  on the left-hand side.

The scale $\mu^2$ acts  essentially as the scale where we probe the scattering process \cite{Adams:2006ch}.   
 By truncating the EFT at $O(p^{2n})$  we are tolerating errors of $O((\mu^2/\Lambda^2_{\mathrm{IR}})^{n+1})$ in our calculations. 
For example,  in a theory of GBs in the IR, the left-hand side calculated with the $O(p^2)$ Lagrangian is practically $\mu^2$ independent, whereas the $\mu^2$-dependence on the right-hand side accounts only for higher order terms (such as the  neglected $O(p^4)$ which includes loops and the logarithmic running of the $O(p^2)$ LECs) and/or the small IR deformations that enter e.g.~as $m_i^2/\mu^2$.

The approximate dispersion relations \eqref{eq:dispersion3} assume a neater form by projecting them with $P_\pm$ as done above. 
For instance, the once-subtracted dispersion relations become 
\bes
\begin{align}
\label{sumruleonce}
P_{-}\Amp^{(1)}(\mu^2)=\frac{2}{\PI}
\int_{4m^2}^{\infty} \di s\,  \frac{(s^2+\mu^4)}{(s^2-\mu^4)^2}  \,P_{-}\mathrm{Im}\Amp(s+\i\epsilon)\,,\\
P_{+}\Amp^{(1)}(\mu^2)=\frac{2}{\PI}
\int_{4m^2}^{\infty} \di s\,  \frac{2\mu^2 s}{(s^2-\mu^4)^2}  \,P_{+}\mathrm{Im}\Amp(s+\i\epsilon)\,,
\end{align}
\ees  
where again we are neglecting the masses with respect to $\mu^2$.
Moreover, should the masses be very small or vanishing, we can take $\mu^2$ small as well, or effectively vanishing, while keeping $m_i^2/\mu^2\ll 1$. Take for instance the sum rules (\ref{exact_sumrulen2}). In this limit they simplify even further in actual sum rules
\begin{subequations}
\label{zeromassrules}
\begin{align}
\label{zeromassrulesa}
  P_{-}\,a^{(2k+1)}& = \frac{2}{\PI}
 \int_{0}^{\infty} \frac{\di s}{s^{2k+1}} P_{-}\,\sigma^{\text{tot}}(s)\,,   &&k\geq 0\,,\\
 \label{zeromassrulesb}
 P_{+}\,a^{(2k)} &=\frac{2}{\PI}
 \int_{0}^{\infty} \frac{\di s}{s^{2k}} P_{+}\,\sigma^{\text{tot}}(s)\,,  &&k\geq 1\,,
 \end{align}
 \end{subequations}
 and constraints
 \begin{subequations}
\label{zeromassconstr}
\begin{align}
\label{zeromassconstra}
 P_{+}\,a^{(2k+1)}=0\,, && k\geq 0\,,\\
 \label{zeromassconstrb}
P_{-}\,a^{(2k)}=0\,, && k\geq 0\,,
 \end{align}
 \end{subequations}
assuming they are IR convergent for the given integer $k$. This is again a condition that one can explicitly verify with the EFT at hand. 
For example, a generic theory of GBs from a non-linear sigma model  gives an IR convergent once-subtracted sum rule in the limit $\mu^2\rightarrow0$. Theories with a shift-symmetry $\pi^a\rightarrow \pi^a+c^a$ give convergent twice-subtracted sum rules for $\mu^2\rightarrow 0$. 

If instead $\mu^2\rightarrow 0$ is a singular limit, one can not only resort to the regular expressions with finite $\mu^2$, but can actually try to isolate all the sources of IR divergence on the same side, and then take the limit $\mu\rightarrow 0$ at the end: since one side of the sum rule is convergent by construction the other must be so too.

A more explicit version of the sum rules \eqref{symmpointsumruleforpositivity} and \eqref{zeromassrules} is obtained by expressing them in a basis adapted to the $\pm 1$-eigenspaces ${\cal V}_\pm$ of the matrix $X$.  Let us denote by $m$ the rank of the matrix $X$, so that $m=m_++m_-$, where
$m_\pm=\dim {\cal V}_\pm$. One can then construct a matrix $M$ which diagonalizes the matrix $X$. In particular we can choose $M$ such that, if we split $\{I(\xi\xi')\}=\{\alpha, a\}$, with $\alpha=1,\ldots,m_-$ and $a=1,\ldots,m_+$, the projectors $P_\pm$ take the block-diagonal form
\be
M P_{-} M^{-1}=\left(\begin{array}{cc} \mathbbm{1}_{m_-} & 0 \\ 0 & 0 \end{array} \right)\,,\qquad M P_{+} M^{-1}=\left(\begin{array}{ll} 0 & 0 \\ 0 & \mathbbm{1}_{m_+} \end{array} \right)\,.
\ee
Then eqs.~\eqref{zeromassrules} give the following explicit set of sum rules
\begin{subequations}
\label{zeromassrules1}
\begin{align}
[Ma^{(2k+1)}]_\alpha &= \frac{2}{\PI}
 \int_{0}^{\infty} \frac{\di s}{s^{2k+1}} [M\sigma^{\text{tot}}(s)]_\alpha\,, && \hspace{-0.5cm}\alpha=1,\ldots,m_-\, ,\hspace{0.5cm}&& k\geq 0\,,
  \label{zeromassrules1b}\\
 [M a^{(2k)}]_a &=\frac{2}{\PI}
 \int_{0}^{\infty} \frac{\di s}{s^{2k}} [M\sigma^{\text{tot}}(s)]_a\,, && \hspace{-0.5cm}a=1,\ldots,m_+\, ,&& k\geq 1\,,
 \label{zeromassrules1a}
 \end{align}
 \end{subequations}
 while the constraints \eqref{zeromassconstr} take the form
 \begin{subequations}\label{zerorules}
\begin{align}
\label{zerorulesa}
[Ma^{(2k+1)}]_a &= 0 && \hspace{-0.5cm}a=1,\ldots,m_+\, ,&& k\geq 0 \,,\\
\label{zerorulesb}
 [M a^{(2k)}]_\alpha &= 0 && \hspace{-0.5cm}\alpha=1,\ldots,m_-\, ,\hspace{0.5cm}&& k\geq 0\, .
 \end{align}
 \end{subequations}
 Analogously, eqs.~\eqref{symmpointsumruleforpositivity}  and \eqref{constraintforpositivity} provide very similar sum rules and constraints for the choice \mbox{$\mu^2=2m^2$} up to the replacing
 \begin{equation}
 \label{substitutionmusymmetric}
 \frac{1}{s^{n}}\rightarrow \frac{s}{(s-2m^2)^{n+1}}\sqrt{1-\frac{4m^2}{s}}\,,\qquad a^{(n)}\rightarrow \Amp^{(n)}(\mu^2=2m^2)
 \end{equation}
 into the eqs.~\eqref{zeromassrules1} and \eqref{zerorules}, integrating from $4m^2$, and retaining all the residues on the left-hand side.


\section{Positivity constraints}
\l{sec:positivity}

In order to derive positivity constraints from our sum rules \eqref{exact_sumrulen2}, one needs to choose a real $\mu^2$. 
In practice any point on the real axis below the branch cut threshold at $s=4m^2$ could be a good choice. 
To be definite, let us take the crossing symmetric point $\mu^2=2m^2$ that has been considered to set analogous positivity constraints on the LECs $\ell_{4,5}$ of the QCD chiral Lagrangian \cite{Pham:1985cr}. We can then use the sum rules \eqref{symmpointsumruleforpositivity}.
 Let us also assume for simplicity that there are no poles below the cut. Therefore, the sum rules \eqref{symmpointsumruleforpositivity} and the constraints \eqref{constraintforpositivity} for even $n=2k$  take the form 
 \begin{subequations}
 \label{massivesumruleforpositivity}
  \begin{align}
 P_{+}\Amp^{(2k)}(2m^2) &=\frac{2}{\PI}
 \int_{4m^2}^{\infty} \di s \frac{s}{(s-2m^2)^{2k+1}}\sqrt{1-\frac{4m^{2}}{s}}P_{+}\sigma^{\text{tot}}(s)\,,
 \label{masspositivityb}\\
   P_{-}\Amp^{(2k)}(2m^2) &= 0\, .\label{masspositivitya}
 \end{align}
 \end{subequations}
Analogous expressions for odd $n=2k+1$ are obtained simply by replacing $P_{\pm}\rightarrow P_{\mp}$. By using the positivity of the cross sections and  
the properties of the projectors we are able to systematically analyze the existence of positivity constraints on linear combinations of the coefficients ${\cal A}^{(n)}(2m^2)$ that can be related to the LECs of the EFT one is interested in. 
Should the limit $m,\mu\rightarrow0$ be regular, we can even remove all the mass scales, so that the above some rules and constraints reduce to the form \eqref{zeromassrules} and \eqref{zeromassconstr}, and we can thus study  $a^{(n)}=\Amp^{(n)}(0)$ as it is done e.g.~for $n=2$ in the theory of GBs with a shift symmetry \cite{Adams:2006ch}, as well as for the dilaton in the proof of the $a$-theorem  \cite{Komargodski:2011vj}. In fact, should the cut actually extend all the way down to $s=0$, the limit $m, \mu\rightarrow0$, whenever it exists, would be the only sensible choice to discuss positivity constraints on the amplitude coefficients.

In the following, for notational convenience, we take $\mu=2m^2\rightarrow 0$  and use eqs.~\eqref{zeromassrules} and \eqref{zeromassconstr}, bearing in mind that the exact same arguments can be repeated by working with finite masses and eq.~\eqref{massivesumruleforpositivity} or its odd-$n$ counterpart.

\subsection{Positivity for even $n$}
\label{sec:evenpos}

Let us first restrict to even $n=2k$. 
In order to simplify the discussion, in this subsection we  assume that the crossing matrix $X$ has the block-diagonal structure \eqref{blockdiag}. 
Then we can focus on the real reduced matrix $\hat X$. We can correspondingly project all vectors appearing in eqs.~\eqref{zeromassrules} and \eqref{zeromassconstr} to the non-mixed components, adding a hat to  distinguish them. The (real) $\pm1$-eigenspaces of $\hat X$ are then denoted by $\hat{\cal V}_\pm$ and have dimensions $\hat m_\pm$, 
while $\hat m=\hat m_++\hat m_-$ gives the rank of $\hat X$.
It follows from the discussion of appendix \ref{app:crossingM} that the matrix $\hat X$ is orthogonal with respect to a metric $\hat {\cal G}$, which can be obtained by reducing to non-mixed indices the metric ${\cal G}$,  see eqs.~\eqref{Gmetric} and \eqref{Gunitary}. This reduction is important in the following.

Let us first consider the eq.~\eqref{zeromassconstrb}. This just says that the vector $\hat a^{(2k)}$ is constrained to lie in complexified $\hat{\cal V}_+$. Furthermore,
 the reality of the right-hand side of  the (reduced) eq.~\eqref{zeromassrulesb} implies that $\hat a^{(2k)}$ is actually real. This is clear if we reinstate finite masses and use $\mu^2=2m^2$, since the non-mixed amplitudes are real below the branch cut.   Hence we can conclude that
\be\label{zeroconst}
\hat a^{(2k)}\in \hat{\cal V}_+\,.
\ee
On the other hand, eq.~\eqref{zeromassrulesb} explicitly relates $\hat a^{(2k)}$ to the projected cross section vector $P_{+}\sigma^{\text{tot}}(s)$. In order study its implications,
let us denote by $\langle\cdot,\cdot\rangle$ the inner product associated with $\hat{\cal G}$, for instance $\langle \hat{\bf v}^1,\hat{\bf v}^2\rangle =\hat{\bf v}^{1\,\rm T}\hat{\cal G}\hat{\bf v}^2$.\footnote{More explicitly, $\hat{\bf v}^{1\,\rm T}\hat{\cal G}\hat{\bf v}^2$ clearly stands for $\sum_{I(\zeta\zeta')}\sum_{J(\xi\xi')}(\hat{\bf v}^1)_{I(\zeta\zeta')} \hat{\cal G}_{I(\zeta\zeta')J(\xi\xi')} (\hat{\bf v}^2)_{J(\xi\xi')}(s)$.} Then eq.~\eqref{zeromassrulesb} can be written as
\be\label{zeroruleseven}
\langle \hat{\bf v},\hat{a}^{(2k)}\rangle= \,\frac{2}{\PI}
 \int_{0}^{\infty} \frac{\di s}{s^{2k}}  \langle \hat{\bf v},\hat\sigma^{\text{tot}}(s)\rangle\,, \qquad \forall \hat{\bf v}\in\hat{\cal V}_+\, .
\ee
Remember that the vector $\hat\sigma(s)$ has all non-negative entries and
suppose that $\hat{\bf v}$ has real non-negative entries too. Then
\be\label{posv}
\langle \hat{\bf v},\hat\sigma^{\text{tot}}(s)\rangle\geq 0\,.
\ee

In more geometrical terms, which are useful for later generalizations, the vector $\hat{\sigma}(s)$  takes values in a convex polyhedral cone ${\cal C}\simeq \mathbb{R}^{\hat m}_+$ and 
requiring that $\hat{\bf v}$ has only non-negative entries is equivalent to requiring that $\hat{\bf v}$ lies in the dual cone ${\cal C}^*\simeq \mathbb{R}^{\hat m}_+$. 
Notice that the $+1$-eigenvector $\hat v$ 
 introduced below eq.~\eqref{blockdiag} lies inside ${\cal C}^*$. Hence, $\hat{\cal V}_+\cap {\cal C}^*$ is a non-empty $\hat m_+$-dimensional convex polyhedral cone.

From eq.~\eqref{zeroruleseven} we immediately get the following positivity constraints 
\be\label{poscons}
\langle \hat{\bf v},\hat a^{(2k)}\rangle \geq 0\quad~~~~\text{for}\quad \hat{\bf v}\in \hat{\cal V}_+\cap {\cal C}^*\,,
\ee
which must be accompanied by eq.~\eqref{zeroconst}, which in fact reduces the number of possible independent (even) amplitude coefficients to $\hat m_+$. Then eq.~\eqref{poscons} identifies an $\hat m_+$-dimensional cone  $\hat{\cal V}_+\cap {\cal C}^*$ of positivity constraints for such $\hat m_+$ independent amplitude coefficients.

\begin{figure}[htbp]
\vspace{2mm}\begin{center}
\includegraphics[scale=0.35]{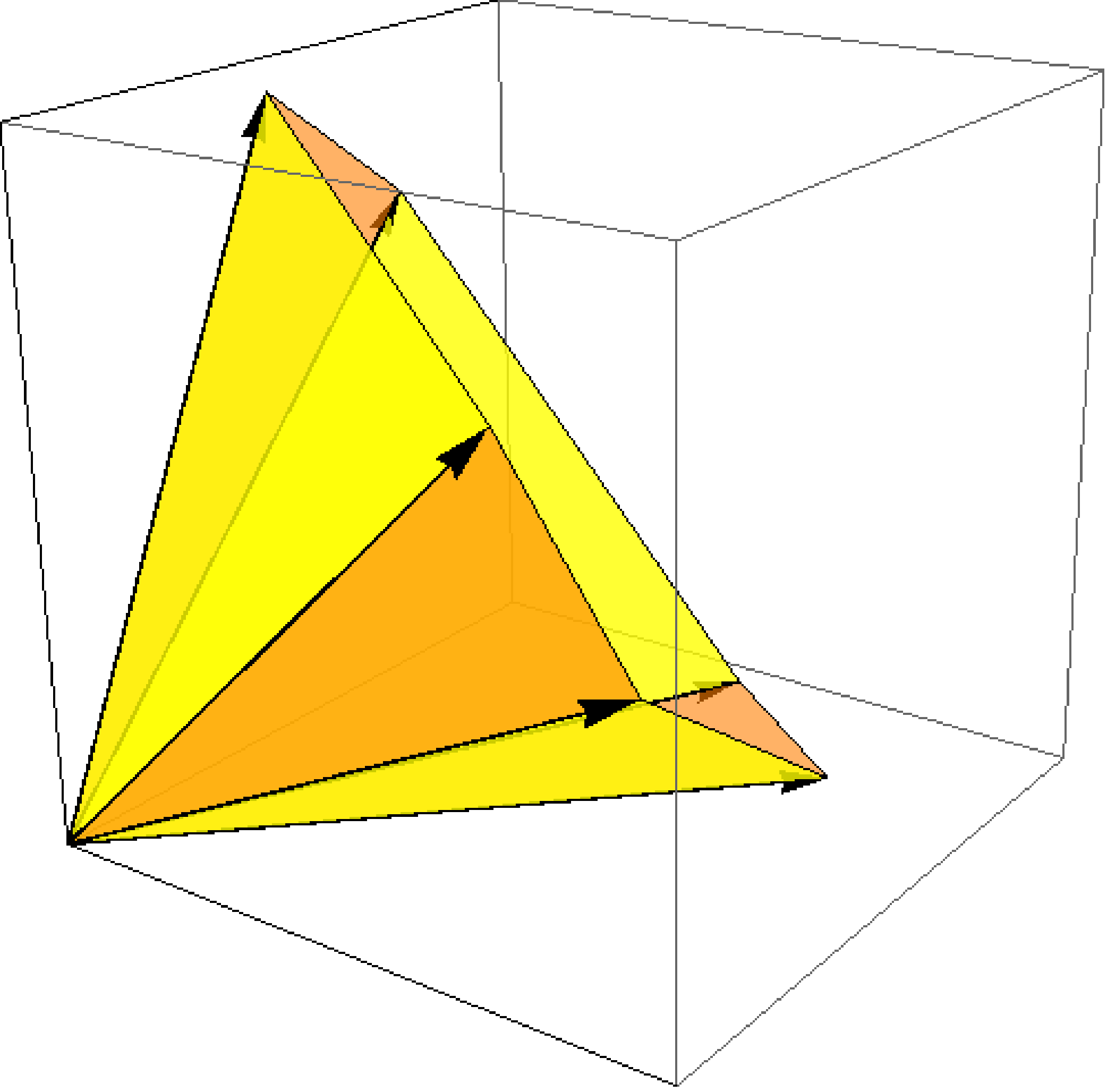}
\caption{Example of a convex polyhedral cone with six faces in three dimensions. All the edges of the cone and the orange faces lie on one of the faces of the first quadrant $\mathbb{R}^{3}_+$, while the yellow faces are internal to the quadrant. When the ambient space is generalized to $\hat{m}$ dimensions and the convex polyhedral cone to an $\hat{m}_{+}<\hat{m}$ dimensional one with $q\geq \hat{m}_{+}$ edges, the $q$ edge vectors lying on the faces of the $\mathbb{R}^{\hat m}_+$ space represent the unique choice of basis vectors that can generate the entire cone through linear combinations with only positive coefficients. We call the positivity constraints represented by these vectors the strongest positivity constraints.}
\label{polycone}
\end{center}
\end{figure}

To make these positivity constraints \eqref{poscons} more explicit, we can select  a set of vectors $\hat {\bf v}^{A}$, with $A=1,\ldots,q$ and $q\geq \hat m_+$,
which generate the edges of the polyhedral cone $\hat{\cal V}_+\cap {\cal C}^*$. Such a convex polyhedral cone with all the ``generating'' vectors lying on the faces of the $\mathbb{R}^{\hat m}_+$ space is shown in a 3-dimensional cartoon in figure \ref{polycone}. In practice  $\hat {\bf v}^{A}$ are  
generators with all non-negative entries of the one-dimensional subspaces resulting from the  intersection of $\hat{\cal V}_+$ with all the $\hat m-\hat m_++1$ planes obtained
by setting $\hat m_+-1$ components of $\mathbb{R}^{\hat m}$ equal to zero. In other words, they are identified by the equation $P_-\hat {\bf v}^{A}=0$ together with the vanishing of all the possible subsets of $\hat m_+-1$ components.
With such a choice, $\hat {\bf v}\in  \hat{\cal V}_+\cap {\cal C}^*$ if and only if
$\hat {\bf v}=\rho_A\hat{\bf v}^A$ with $\rho_A\geq 0$. Hence, once we have constructed this particular set of vectors, 
we can rewrite eq.~\eqref{poscons} as the  $q\geq \hat m_+$ positivity constraints
\be\label{exposs}
\langle \hat{\bf v}^A,\hat{a}^{(2k)}\rangle \geq 0\quad~~~~ A=1,\ldots, q\, .
\ee

One can also obtain this set of positivity constraints  directly from the sum rules written in the form \eqref{zeromassrules1a}, restricted to the non-mixed sector. The prescription is the following. Take the generic linear combination $ \tau^a[\hat M\hat{a}^{(2k)}]_a$, with $\hat m_+$ parameters $\tau^a$. This gives a linear combination of the $\hat m$ components of $\hat{a}^{(2k)}$. Choose $\hat m_+-1$ out of these $\hat m$  components  and impose that their coefficients in $\tau^a[\hat M\hat{a}^{(2k)}]_a$ are vanishing. This gives $\hat m_+-1$ equations which fix the $\hat m_+$ parameters $\tau^a$ in terms of a single one. 
If we can choose these constrained $\tau^a$'s so  that $\tau^a[\hat M\hat{a}^{(2k)}]_a$ has all positive coefficients, then it gives one of the combinations appearing on the left-hand side of the positivity constraints \eqref{exposs}. Otherwise we discard it. Then, to obtain all the other positivity constraints in eq.~\eqref{exposs}, one should repeat the procedure for all the other subsets of $\hat m_+-1$  out of the $\hat m$  components  of $\hat{a}^{(2k)}$. Finally, recall that these conditions are supplemented by \eqref{zeroconst}, which is more explicitly given by the set of equations   \eqref{zerorulesb} restricted to the non-mixed sector.

In the following sections we show how this prescription practically works in several examples.

\subsection{Non-positivity for odd $n$}

One can now wonder whether the above procedure outlined for even $n=2k$ could be mimicked for isolating some positivity constraints for odd $n=2k+1$ as well.
Unfortunately, it is easy to see that this is never possible. 

Let us again assume eq.~\eqref{blockdiag} and restrict to the non-mixed sector. First of all, by repeating the above argument starting from eq.~\eqref{zeromassconstra}, one would be lead to the conclusion that $\hat a^{(2k+1)}\in \hat{\cal V}_-$ and $\langle \hat{\bf w},\hat a^{(2k+1)}\rangle \geq 0$ for any  $\hat{\bf w}\in\hat{\cal V}_-\cap {\cal C}^*$ (with $k\geq1$).  Now the key point is that, being the matrix $\hat X$ $\hat{\cal G}$-orthogonal, the eigenspaces $\hat{\cal V}_+$ and $\hat{\cal V}_-$ are $\hat{\cal G}$-orthogonal in the sense that $\langle \hat{\bf v},\hat{\bf w}\rangle=0$ for any $\hat{\bf v}\in\hat{\cal V}_+$  and $\hat{\bf w}\in\hat{\cal V}_-$. In particular, we know that the vector $\hat v$  introduced below eq.~\eqref{blockdiag} belongs to $\hat{\cal V}_+$. Then $\langle \hat v,\hat{\bf w}\rangle=0$ for all $\hat{\bf w}\in\hat{\cal V}_-$. But, being $\hat{\cal G}$ diagonal (and with positive definite entries), $\langle \hat v,\hat{\bf w}\rangle$ is a linear combination of the components of $\hat{\bf w}$ with just positive coefficients. Hence, $\langle \hat v,\hat{\bf w}\rangle=0$ implies that at least one component of $\hat{\bf w}$ is negative and then $\hat{\bf w}$ cannot belong to ${\cal C}^*$. Therefore the set $\hat{\cal V}_-\cap {\cal C}^*$ is empty and there are no positivity constraints coming from the same argument used for even $n$.

\subsection{Inclusion of the mixed sector}

The above derivation of the positivity constraints for even $n$ can be extended to the case in which $X$ does not take the form of eq.~\eqref{blockdiag} and mixed amplitudes are included. We just briefly outline the general idea without spelling out the details. 

First, analogously to the case discussed above, eq.~\eqref{zeromassconstrb} says that
\be
a^{(2k)}\in {\cal V}_+\,,
\ee
while eq.~\eqref{zeromassrulesb} can be rewritten as
\be\label{zeroruleseven2}
\langle {\bf v},a^{(2k)}\rangle= \,\frac{2}{\PI}
 \int_{0}^{\infty} \frac{\di s}{s^{2k}}  \langle {\bf v},\sigma^{\text{tot}}(s)\rangle\,, \qquad \forall \mathbf{v}\in \mathcal{V}_{+}\,.
\ee
Here we have to take into account that ${\bf v}\in{\cal V}_+$ cannot be  generically restricted to be real and the pairing $\langle \cdot,\cdot\rangle$ corresponds to the complete metric ${\cal G}$, for instance  $\langle {\bf v}^1,{\bf v}^2\rangle ={\bf v}^{1\,\dagger}{\cal G}{\bf v}^2$.

We observe that now the vector $\sigma(s)$ can be seen as a linear combination with positive coefficients of the form
\be
\sigma^{\text{tot}}(s)=\sum_{I\xi}\sigma^{\text{tot}}_{I(\xi)}(s){\bf u}_{I(\xi\xi)}+\sum^{\text{tot}}_{I\xi\neq\xi'}\sigma_{I(\xi)+I(\xi')}(s){\bf u}_{I(\xi\xi')}\,.
\ee 
As one can easily check, the vectors ${\bf u}_{I(\xi\xi)}$ have one component equal to 1, a number (given by the degeneration of ${\bf r}_{I(\xi)}$) of components equal to $-\frac12$, and the other entries equal to zero.  On the other hand, the vectors  ${\bf u}_{I(\xi\xi')}$, with $\xi\neq\xi'$,  have one component equal to 1 and the others vanishing. Since $\sigma^{\text{tot}}_{I(\xi)}(s)$ and $\sigma^{\text{tot}}_{I(\xi)+I(\xi')}(s)$  are positive, we see that $\sigma^{\text{tot}}(s)$ lie in the convex polyhedral cone ${\cal C}\subset \mathbb{R}^m$ whose edges are generated by the vectors ${\bf u}_{I(\xi\xi')}$. 

We can now repeat the arguments above almost verbatim. The main difference is that one has to divide real and imaginary contributions to eq.~\eqref{zeroruleseven2}. Suppose now that ${\rm Re}\,{\bf v}$ is such that  $\langle{\rm Re}\,{\bf v},{\bf u}_{I(\xi\xi')}\rangle\geq 0$ for all  ${\bf u}_{I(\xi\xi')}$'s.
In more formal terms, assume that ${\rm Re}\,{\bf v}\in {\cal C}^*$, where ${\cal C}^*$ is the dual cone to ${\cal C}$. 
In such a case, ${\rm Re}\langle {\bf v},\sigma(s)\rangle\geq 0$ and then
from eq.~\eqref{zeroruleseven2} we get the positivity constraints
\be\label{pcRe}
{\rm Re}\langle {\bf v},a^{(2k)}\rangle\geq 0\quad~~~~\text{for}\quad {\bf v}\in {\cal V}_+\quad\text{and}\quad  {\rm Re}\,{\bf v}\in{\cal C}^* \, .
\ee
Analogously, for the imaginary component we get
\be\label{pcIm}
{\rm Im}\langle {\bf v},a^{(2k)}\rangle\geq 0\quad~~~~\text{for}\quad {\bf v}\in {\cal V}_+\quad\text{and}\quad  {\rm Im}\,{\bf v}\in{\cal C}^* \, .
\ee
One could then proceede as described is subsection \ref{sec:evenpos} to extract a minimal set of independent positivity constraints from eqs.~\eqref{pcRe} and \eqref{pcIm}.

\section{Examples}\label{examples}

So far we have been completely general and did not restrict to any specific symmetry group $H$. 
Let us now summarize the algorithm to extract the sum rules for the scattering of two (identical and real) representations $\mathbf{r}$ of $H$:
\begin{itemize}
\item do the CG decomposition $\r\otimes\r=\bigoplus_{I,\xi} \mathbf{r}_{{I(\xi)}} $ into irreps  $\mathbf{r}_{I(\xi)}$, and calculate the crossing matrix $X$ that acts on the independent eigen-amplitudes $\Amp_{I(\xi\xi')}$, by using e.g.~the expressions \eqref{Qdef} and \eqref{matrixXfromQ}; 
\item diagonalize  $X$ with a non-singular matrix $M$ that brings it to a canonical form
\be
MXM^{-1}=\left(\begin{array}{cc} -\mathbbm{1}_{m_-} & 0 \\ 0 & \mathbbm{1}_{m_+} \end{array} \right)\,;
\ee
\item read off the sum rules' coefficients from the rows of $M$.  
In particular, for massless  particles, when no degenerate irreps occur in the CG decomposition, the sum rules for one and two subtractions are
\begin{subequations}
\label{ex:zeromassrules1}
\begin{align}
 [M a^{(1)}]_\alpha &=\frac{2}{\PI}
 \int_{0}^{\infty} \frac{\di s}{s} [M\sigma^{\text{tot}}(s)]_\alpha\,, && \hspace{-1cm}\alpha=1,\ldots,m_-\, ,\hspace{1cm} \\
 \label{ex:zeromassrules1B}
 [M a^{(2)}]_a &=\frac{2}{\PI}
 \int_{0}^{\infty} \frac{\di s}{s^2} [M\sigma^{\text{tot}}(s)]_a\,, && \hspace{-1cm} a=1,\ldots,m_+\,,\hspace{1cm}\\
   [M a^{(1)}]_a &= 0 && \hspace{-1cm} a=1,\ldots,m_{+}\,,\hspace{1cm}\\
  [M a^{(2)}]_\alpha &= 0 && \hspace{-1cm} \alpha=1,\ldots,m_{-}\,,\hspace{1cm}
  \end{align}
 \end{subequations}
 where $a^{(n)}=\Amp^{(n)}(0)$ are the expansion coefficients around $s=0$. When instead degenerate irreps appear in the CG decompositions, one should work as described in section \ref{sumrules}. The last two eqs.~\eqref{ex:zeromassrules1} actually represent a constraint that follows directly from the symmetry structure of the theory and crossing symmetry, without relying on unitarity. They imply that not all the amplitudes coefficients are linearly independent. 
 \item derive the positivity constraints  that follow from eq.~\eqref{ex:zeromassrules1B} by taking linear combinations of the last $m_+$ rows of $M$ that return only non-negative entries. The strongest positivity constraints obtained in this way take the form $\sum_{I}\mathbf{v}^A_I \mathrm{dim}\,\mathbf{r}_I a^{(2)}_I\geq 0$, and can be derived by following the algorithm outlined at the end of subsection~\ref{sec:evenpos} that finds the edge generators of a convex polyhedral cone $\mathbf{v}^A$ that belong to the $+1$-eigenspace and have $m_+-1$ vanishing entries (with the remaining ones being strictly positive). 
 \end{itemize}
 
 For massive particles  it is useful to study the behavior of the eigen-amplitudes at scales $s\approx \mu^2$ through the expansion
 \be
 \Amp(s)=\Amp^{(0)}(\mu^2)+\Amp^{(1)}(\mu^2)(s-\mu^2)+\Amp^{(2)}(\mu^2)(s-\mu^2)^2+\ldots
 \ee
 with $\mu^2$ larger than the squared masses and any other IR structure such as extra light poles or small widths: $m_{\mathrm{IR}}^2 \ll \mathrm{Re}\mu^2\sim \mathrm{Im}\,\mu^2\ll \Lambda^2_{\mathrm{IR}}$. For example, within this approximation, the sum rules and the constraints from diagonalizing eq.~\eqref{exact_sumrulen2} are
\bes
\begin{align}
 [M \Amp^{(1)}(\mu^2)]_\alpha &= \frac{2}{\PI}
\int_{4m^2}^{\infty} \di s\,  \frac{(s^2+\mu^4)s}{(s^2-\mu^4)^2}  \sqrt{1-\f{4m^{2}}{s}} [M \sigma^{\text{tot}}(s)]_\alpha \,,   && \alpha=1,\ldots,m_-\,,\\
\label{constrmixsumrule1sub}
[M \Amp^{(1)}(\mu^2)]_a &= \frac{2}{\PI}
\int_{4m^2}^{\infty} \di s\,  \frac{2\mu^2 s^2}{(s^2-\mu^4)^2}  \sqrt{1-\f{4m^{2}}{s}} [M \sigma^{\text{tot}}(s)]_a \,,   && a=1,\ldots,m_+\,,\\
 [M \Amp^{(2)}(\mu^2)]_a & = \frac{2}{\PI}
 \int_{4m^2}^{\infty} \di s \frac{(s^2+3\mu^4)s^2}{(s^2-\mu^4)^3}  \sqrt{1-\f{4m^{2}}{s}} [M\sigma^{\text{tot}}(s)]_a\,,  && a=1,\ldots,m_+\,,\\
 [M \Amp^{(2)}(\mu^2)]_\alpha & = \frac{2}{\PI}
 \int_{4m^2}^{\infty} \di s \frac{(3s^2+\mu^4)s\mu^2}{(s^2-\mu^4)^3}  \sqrt{1-\f{4m^{2}}{s}} [M\sigma^{\text{tot}}(s)]_\alpha\,,  && \alpha=1,\ldots,m_-\, .  
\end{align}
\ees
They reproduce  eqs.~\eqref{ex:zeromassrules1} in the limit $\mu^2, m^2 \rightarrow 0$ with $|\mu|^2\gg m^2$ if this limit exists. If it does not, one should collect all the IR divergent terms on one side and take the limit afterwards so that the convergence of one side enforces the convergence of the other.
Alternatively, one can work with $\Amp^{(n)}(\mu^2)$ with a finite and real $\mu^2$ smaller than $4m^{2}$, e.g.~setting it at the crossing symmetric point $\mu^2=2m^2$. In such a regime the sum rules are given by eqs.~\eqref{ex:zeromassrules1} up to the replacement \eqref{substitutionmusymmetric}. They are guaranteed to be real and IR convergent as in eq.~\eqref{massivesumruleforpositivity}. One can thus extract positivity constraints from the even-subtracted sum rules as discussed in section~\ref{sec:positivity}. 

The sum rules with two or more subtractions are UV convergent. Once-subtracted sum rules are also UV convergent under the general assumptions discussed in subsection~\ref{sec:convergence} about the universality behavior of the amplitudes saturating the Froissart bound.

In this section we go through detailed examples and show concretely the powerful information carried by the sum rules. 
For simplicity we assume that the expansion around $\mu^2= 0$ does not give rise to any IR singularity and work directly with the expansion coefficients $a^{(n)}$ (at least for $n=1,2$), bearing in mind the simple modifications $a^{(n)}\rightarrow \Amp^{(n)}(\mu^2)$ for the sum rules and the positivity constraints at finite $\mu^2$. In case IR residues are present and cannot be neglected with respect to $\mu^{2}$, they should also be included in the left-hand side of eqs.~\eqref{ex:zeromassrules1}, see e.g.~eq.~\eqref{exact_sumrulen2}.

\subsection{Fundamentals of $SO(N)$}
\label{SO4example}

We consider first the case of the forward elastic scattering of two particles transforming as fundamental representations of SO$(N)$ with $N\neq4$. 
The case $N=4$ is discussed in subsection~\ref{subsec:SO4}.

The tensor product of fundamental representations decomposes as ${\bf N}\otimes {\bf N}={\bf 1}\oplus {\bf S}\oplus{\bf A}$ into a singlet, the symmetric and anti-symmetric representations, whose dimensions are
\be
\Delta_{\bf 1}=1\,,\qquad \Delta_{\bf S}=\frac12N(N+1)-1\,,\qquad  \Delta_{\bf A}=\frac12N(N-1)\,.
\ee
The crossing matrix is given by (see appendix \ref{XforSON} for details about the construction of $X$)
\be
\label{Xson}
X=\left(\begin{array}{ccc}\frac{1}{N} & \frac{\Delta_{\bf S}}{N} & -\frac{\Delta_{\bf A}}{N} \\
 \frac{1}{N} & \frac{\alpha}{\Delta_{\rm S}} & \frac{\Delta_{\rm A}}{\Delta_{\rm S}}(\frac12+\frac1N)\\
 -\frac{1}{N} & \frac12+\frac1N &\frac12
 \end{array}
\right)\,,
\ee 
with $\alpha=N(N-1)/4-1+1/N$. The states are ordered as ${\bf 1}$, ${\bf S}$, and ${\bf A}$.  One can simply verify that this matrix satisfies all the properties discussed in appendix \ref{app:crossingM} and in particular
$X^2=\mathbbm{1}$, $\det X=-1$, and $\tr X=m_+-m_-=1$. Since $m=m_{+} + m_{-} =3$, we get  that the number of independent once-subtracted and twice-subtracted sum rules are respectively $m_-=1$ and $m_+=2$, consistently with the general discussion of appendix \ref{app:crossingM}, which relates these numbers to the number of (anti)-symmetric representations appearing in the decomposition.
As expected by eq.~\eqref{onevector}, the $+1$-eigenspace contains the vector $(1,1,1)^T$ and the columns sum up to $1$ for each row.

We diagonalize $X$ with a non-singular matrix $M$
\be 
\label{MSON}
M=\left(\begin{array}{ccc}
\frac{1}{2N}  & -\frac{N+2}{4N} & \frac{1}{4}  \\
-\frac{1}{2N} & \frac{N+2}{4N} & \frac{3}{4} \\
\frac{1}{2N} & \frac{6N-4}{8N} & \frac{1}{4}  
 \end{array}
\right)\,,\qquad M X M^{-1}=\left(\begin{array}{cc} -\mathbbm{1}_{1} & 0 \\ 0 & \mathbbm{1}_{2} \end{array} \right)\,.
\ee 
Its first row gives the coefficients of the sum rule with an odd number of subtractions\footnote{We have implicitly taken all masses to zero at the end of the computation, as we are allowed to do if no massless mode propagates in the $t$-channel. This is the case for the theory of GBs discussed below where no IR divergence arises with one subtraction.}, e.g.
\be 
\label{sumruleodd}
 2a^{(1)}_{\mathbf{1}}-(N+2)a^{(1)}_{\mathbf{S}}+N a^{(1)}_{\mathbf{A}}
= \frac{2}{\PI}\int_0^\infty \frac{\di s}{s}\left[2\sigma^{\mathrm{tot}}_\mathbf{1}(s)-(N+2)\sigma^{\mathrm{tot}}_{\mathbf{S}}(s)+N\sigma^{\mathrm{tot}}_{\mathbf{A}}(s)\right]\,.
\ee
The convergence for one subtraction is guaranteed  by the fact that the coefficients in front of the cross-sections add up to zero being orthogonal to the vector $(1,1,1)^T$.
The other two rows set instead the constraints on the $a^{(1)}_I$'s that we can write as
\begin{equation}
\label{costr1SOn}
a^{(1)}_{\mathbf{S}}=-a^{(1)}_{\mathbf{A}}=-\frac{1}{N-1}a^{(1)}_{\mathbf{1}}\,,
 \end{equation}
 and apply to the once-subtracted sum rule~\eqref{sumruleodd} that in terms of just one eigen-amplitude, e.g.~$a^{(1)}_{\mathbf{A}}$, takes the form 
 \be 
a^{(1)}_{\mathbf{A}}
= \frac{1}{2\PI N}\int_0^\infty \frac{\di s}{s}\left[2\sigma^{\mathrm{tot}}_\mathbf{1}(s)-(N+2)\sigma^{\mathrm{tot}}_{\mathbf{S}}(s)+N\sigma^{\mathrm{tot}}_{\mathbf{A}}(s)\right]\,.
\ee

Let us pass to  two subtractions. The first row of $M$ tells us which combination of even scattering amplitude coefficients must vanish, e.g.
\begin{equation}
\label{planeofAs}
2a^{(2)}_{\mathbf{1}}-(N+2)a^{(2)}_{\mathbf{S}}+N a^{(2)}_{\mathbf{A}}=0\,.
\end{equation}
Any (linearly independent) combination of the other two rows gives sum rules with an even number of subtractions. If the coefficients are arranged to be positive these sum rules imply inequalities because of the positivity of the total cross-sections. For example,  summing the second and third row of the matrix \eqref{MSON} we get that $a^{(2)}_{\mathbf{S}}+a^{(2)}_{\mathbf{A}}$ equals an integral over positive combination of cross-sections, hence $a^{(2)}_{\mathbf{S}}+a^{(2)}_{\mathbf{A}}\geq 0$.

\begin{figure}[t!]
\vspace{2mm}\begin{center}
\includegraphics[scale=0.35]{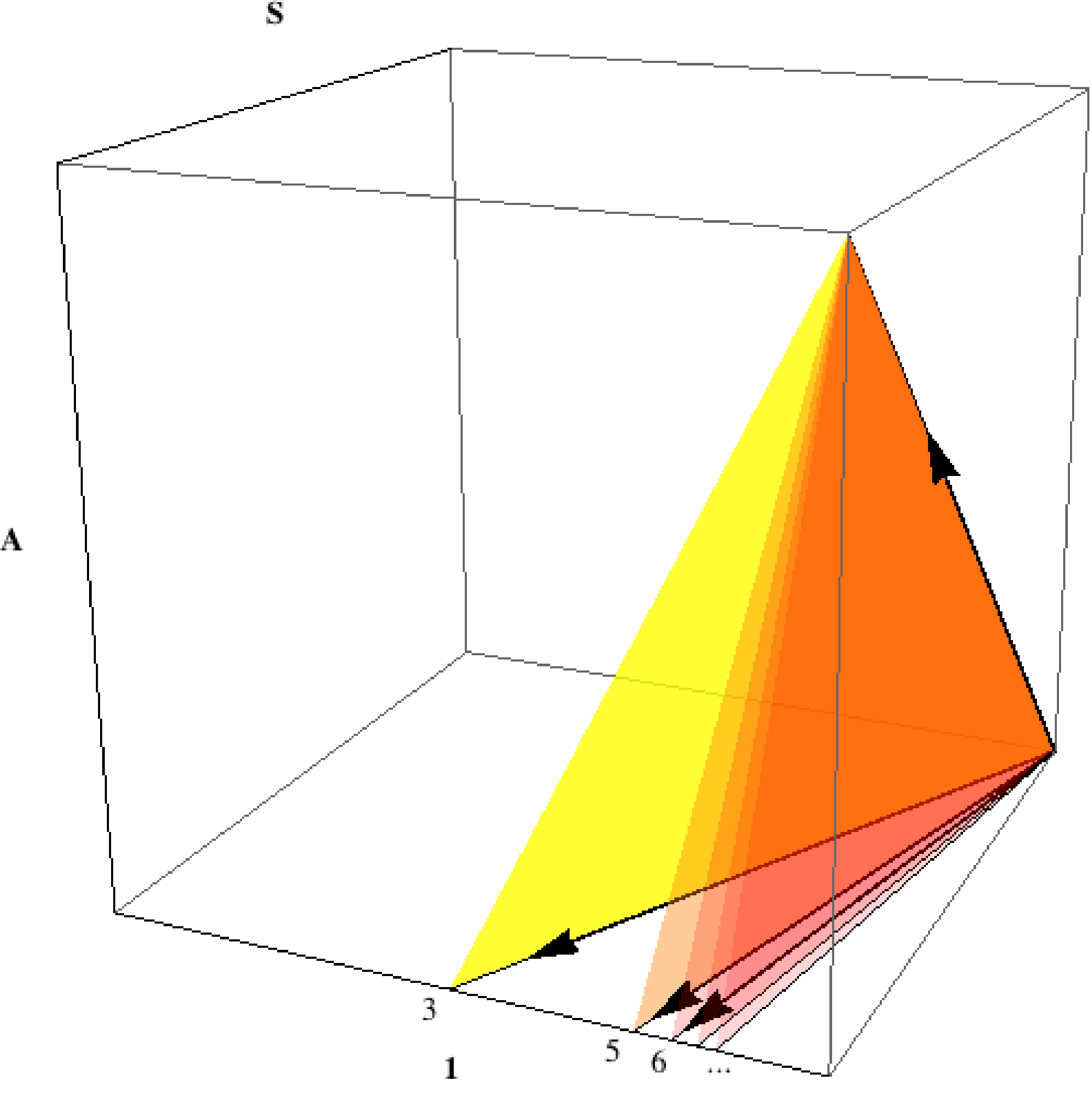}
\caption{Two dimensional convex polyhedral cone in three dimensions (i.e.~planar cone or sector of a plane) generated by $\sum_I \mathbf{v}^A_I \mathcal{G}_{IJ}$ which are the coefficients of the strongest positivity constraints for $SO(N\neq4)$ for $N=3,5,6,\ldots$}
\label{polyconeSON}
\end{center}
\end{figure}

\noindent More systematically, we can apply the prescription of section \ref{sec:positivity} where the edge generators $\mathbf{v}$'s of the positivity convex polyhedral cone satisfy $2\mathbf{v}_{\mathbf{1}}-(N+2)\mathbf{v}_{\mathbf{S}}+N \mathbf{v}_{\mathbf{A}}=0$, and have $(m_{+}-1)$ vanishing components, while the remaining ones are positive. In this way we find two edges in a three dimensional space generated by $\mathbf{v}^1=(0,N,N+2)^T$ and $\mathbf{v}^2=(N+2,2,0)^T$. The associated cone is depicted in figure \ref{polyconeSON}. 
We can thus determine the coefficients that set the strongest  positivity constraints  $\sum_{I,J}\mathbf{v}^A_I \mathcal{G}_{IJ}a^{(2)}_J\geq 0$ with $A=1,\,2$ by contracting with the metric $\mathcal{G}=\mathrm{diag}(\Delta_{\bf 1},\Delta_{\bf S}, \Delta_{\bf A})$, 
\bes
\l{conditionsSON}
\begin{align}
a^{(2)}_{\mathbf{S}}+a^{(2)}_{\mathbf{A}} & \geq 0\,,\\
a^{(2)}_{\mathbf{1}}+(N-1)a^{(2)}_{\mathbf{S}} & \geq 0\,.
\end{align}
\ees
Notice that the constraint \eqref{planeofAs} allows one to single out two independent amplitude coefficients, for instance $a^{(2)}_{\mathbf{S,A}}$, and to recast the associated positivity constraints in the form
\bes
\l{conditionsSONbis}
\begin{align}
a^{(2)}_{\mathbf{S}}+a^{(2)}_{\mathbf{A}} & \geq 0\,,\\
3a^{(2)}_{\mathbf{S}}-a^{(2)}_{\mathbf{A}} & \geq 0\,,
\end{align}
\ees
that immediately imply $a^{(2)}_{\mathbf{S}}\geq0$.  
Equality in the expressions \eqref{conditionsSON} and \eqref{conditionsSONbis} is reached only for trivial non-interacting theories where the cross-sections are vanishing. Analogous positivity constraints from twice subtracted sum rules in the specific case of $SO(3)\sim SU(2)$ have been studied also in refs.~\cite{Pennington:1994do,Ananthanarayan:1995ks,Distler:2006if,Manohar:2008tc}.

\subsubsection{Goldstone bosons from $SO(N,1)/SO(N)$ and $SO(N+1)/SO(N)$}

As we stressed in the Introduction, the sum rules become useful when the IR side can be calculated using the LECs of an EFT. 
We consider now the theory of GBs emerging from the spontaneous breaking patterns $SO(N+1)\rightarrow SO(N)$ or $SO(N,1)\rightarrow SO(N)$. The Lagrangian at $O(p^2)$ is given by 
\be 
\label{cosetGBint}
\mathcal{L}=\frac{1}{2}\partial_\mu\pi^a \partial^\mu \pi^a \mp\frac{1}{6 f_\pi^2}\left[(\pi^b\pi^b)(\partial_\mu\pi^a \partial^\mu \pi^a)-(\partial_\mu\pi^a \pi^a) (\partial^\mu\pi^b \pi^b)\right]\,,
\ee
where the GBs live on a $N$-dimensional sphere (upper sign) or hyperboloid (lower sign) respectively.
We also add two light states, $h\in \mathbf{1}$ and $h_{ab}\in \mathbf{S}$, that are coupled as
\be 
\label{interactions_scalar_symm}
\f{1}{f_\pi}\left(a h \delta_{ab}+b h_{ab}\right)\partial_\mu\pi^a \partial^\mu\pi^b \,. 
\ee
We can think of them as Higgs-like states. The LECs are the decay constant $f_\pi$, and the couplings $a$ and $b$.
With these ingredients we can calculate the amplitudes for the scattering at low-energy
\begin{equation}
\label{AmpforsoN}
\Amp(\pi^a\pi^b\rightarrow \pi^c \pi^d)=(\pm 1-a^2+\frac{N+2}{2N} b^2 )\frac{s}{f_\pi^2}\left(\delta^{ab}\delta^{cd}-\delta^{cb}\delta^{ad}\right)\,,
\end{equation}
and the corresponding eigen-amplitude coefficients
\be
\bry{l}
\dst a^{(1)}_{\mathbf{1}}=\frac{(N-1)}{f_\pi^2}\left(\pm1 -a^2 + \frac{N+2}{2N} b^2 \right)\,, \vspace{2mm}\\  
\dst a^{(1)}_{\mathbf{S}}=  -\frac{1}{f_\pi^2}\left(\pm 1 -a^2 + \frac{N+2}{2N} b^2 \right)\,, \vspace{2mm}\\
\dst a^{(1)}_{\mathbf{A}}= \frac{1}{f_\pi^2}\left(\pm1 -a^2 + \frac{N+2}{2N} b^2 \right) \,,
\ery
\ee
which, as expected, satisfy the constraints~\eqref{costr1SOn}.
Substituting these coefficients into eq.~\eqref{sumruleodd} the once-subtracted sum rule takes now an explicit expression in terms of the LECs of the EFT:
\be 
\label{SO(N)coset_example}
\left(\pm 1-a^2+\frac{N+2}{2N}b^2\right)=\frac{f_\pi^2}{2\PI N}\int_0^\infty \frac{\di s}{s}\left[2\sigma^{\mathrm{tot}}_\mathbf{1}(s)+N\sigma^{\mathrm{tot}}_{\mathbf{A}}(s)-(N+2)\sigma^{\mathrm{tot}}_{\mathbf{S}}(s)\right]\,.
\ee
The two signs $+$ and $-$ correspond to the sphere and the hyperboloid respectively.
For \mbox{$SO(4)/SO(3)\sim SU(2)_L\times SU(2)_R/SU(2)_V$} we recover the sum rule of ref.~\cite{Falkowski:2012bu} for $b=0$, and the original Olsson sum rule of QCD for $a=b=0$.
The sum rule for $SO(3)$ improves the one proposed for $U(1)\sim SO(2)$ in ref.~\cite{Low:2009di} in the context of composite Higgs models since the deep UV contribution $c^{\infty(1)}$ from the big circle is projected out by $P_{-}$.

More generally, in a non-linear sigma model defined by the constraint $\sum_{i=1}^N \phi^2_i+c \phi^2_{N+1}=f_\pi^2$ (ellipsoid) where $H=SO(N)$ is unbroken,  one needs just to rescale the $+1$ of the sphere in eq.~\eqref{SO(N)coset_example} by a factor $1/c$. This shows that the sum rule with no Higgses  is insensitive to the geometric structure of the coset as long as we deform it in a way that respects $H$ and rescale $f_\pi$.

\subsection{Composite Higgs models and $SO(4)$}
\label{subsec:SO4}

We now move on to consider the case $H=SO(4)\sim SU(2)_L \times SU(2)_R$ that is important for custodially symmetric composite Higgs models, see e.g.~ref.~\cite{Bellazzini:2014yua} for a recent comprehensive review. 
The scattering of two $\mathbf{4}\in SO(4)$ is special because the anti-symmetric $\mathbf{6}$ is further reducible into $(\mathbf{1},\mathbf{3})\oplus(\mathbf{3},\mathbf{1})$ of $SU(2)_L \times SU(2)_R$.
An immediate consequence is that there are two sum rules for an odd number of subtractions, rather than just one like for $SO(N\neq 4)$, and two sum rules for an even number of subtractions.

Let us work directly with $SU(2)_L\times SU(2)_R$ where every irrep carries pairs of indices in the irreps of $SU(2)$. In particular, we have $(\mathbf{2},\mathbf{2})\otimes (\mathbf{2},\mathbf{2})=(\mathbf{1},\mathbf{1})\oplus (\mathbf{1},\mathbf{3})\oplus(\mathbf{3},\mathbf{1})\oplus (\mathbf{3},\mathbf{3})$. The CG coefficients are thus the product of the well known CG coefficients for $3$-dimensional rotations.  We can therefore calculate the crossing matrix $X$ directly from its definition \eqref{Qdef} and get
 \be 
 X=\frac{1}{4}
 \left(\begin{array}{cccc}
1 & -3 & -3 & 9\\
-1 & -1 & 3 & 3\\ 
-1 & 3 & -1 & 3\\
1 & 1 & 1 & 1
 \end{array}\right)\,,\qquad 
 M=\frac{1}{8} \left(\begin{array}{cccc}
-1 & -1 & -1 & 3\\
1 & -3 & 5 & -3\\ 
1 & 1 & 1 & 5\\
-1 & 3 & 3 & 3
 \end{array}\right)\,,
 \ee
where $M$ is the matrix that diagonalizes $X$ as $M X M^{-1}=\mathrm{diag}\left(-1,-1,1,1\right)$.
The crossing matrix has two linearly independent $-1$-eigenvectors, and therefore there are two independent sum rules with an odd number of subtractions. Taking e.g.~linear combinations of the first two rows of $M$ we get the sum rules
\bes
\begin{align}
\label{firstSO4}
&a^{(1)}_{(\mathbf{1},\mathbf{1})}+a^{(1)}_{(\mathbf{1},\mathbf{3})}+a^{(1)}_{(\mathbf{3},\mathbf{1})}-3 a^{(1)}_{(\mathbf{3},\mathbf{3})}= \frac{2}{\PI}\int_0^\infty \frac{\di s}{s}\left[\sigma^{\mathrm{tot}}_{(\mathbf{1},\mathbf{1})}(s)+\sigma^{\mathrm{tot}}_{(\mathbf{1},\mathbf{3})}(s)+\sigma^{\mathrm{tot}}_{(\mathbf{3},\mathbf{1})}(s)-3 \sigma^{\mathrm{tot}}_{(\mathbf{3},\mathbf{3})}(s)\right]\,,\\
\label{secondSO4}
&a^{(1)}_{(\mathbf{1},\mathbf{3})}- a^{(1)}_{(\mathbf{3},\mathbf{1})}=\frac{2}{\PI}\int_0^\infty \frac{\di s}{s}\left[\sigma^{\mathrm{tot}}_{(\mathbf{1},\mathbf{3})}(s)-\sigma^{\mathrm{tot}}_{(\mathbf{3},\mathbf{1})}(s)\right]\,.
\end{align}
\ees
The last two rows of $M$ provide the constraints on the $a^{(1)}_I$ that we can write as
 \be
 \label{so4constr1sub}
 a^{(1)}_{(\mathbf{3},\mathbf{3})}=- \frac{1}{3}a^{(1)}_{(\mathbf{1},\mathbf{1})}=-\frac{1}{2}( a^{(1)}_{(\mathbf{1},\mathbf{3})}+ a^{(1)}_{(\mathbf{3},\mathbf{1})})\,,
 \ee
and use to recast eq.~\eqref{firstSO4} in terms of a single eigen-amplitude, e.g.~$a^{(1)}_{(\mathbf{3},\mathbf{3})}$:
\begin{equation}
a^{(1)}_{(\mathbf{3},\mathbf{3})}= -\frac{1}{4\PI}\int_0^\infty \frac{\di s}{s}\left[\sigma^{\mathrm{tot}}_{(\mathbf{1},\mathbf{1})}(s)+\sigma^{\mathrm{tot}}_{(\mathbf{1},\mathbf{3})}(s)+\sigma^{\mathrm{tot}}_{(\mathbf{3},\mathbf{1})}(s)-3 \sigma^{\mathrm{tot}}_{(\mathbf{3},\mathbf{3})}(s)\right]\,.
\end{equation}

The first two rows of $M$ provide instead the constraints for $a^{(2)}_I$:
 \bes
 \label{neweqso4}
 \begin{align}
 a^{(2)}_{(\mathbf{1},\mathbf{1})}+a^{(2)}_{(\mathbf{1},\mathbf{3})}+a^{(2)}_{(\mathbf{3},\mathbf{1})}-3 a^{(2)}_{(\mathbf{3},\mathbf{3})}& =0\,, \\
a^{(2)}_{(\mathbf{1},\mathbf{3})}- a^{(2)}_{(\mathbf{3},\mathbf{1})}& =0\,.
 \end{align}
 \ees
Following the prescription of section \ref{sec:positivity}, we can derive the strongest positivity conditions for an even number of subtractions, e.g.
 \bes
 \begin{align}
 \label{firstSO4even}
 a^{(2)}_{(\mathbf{1},\mathbf{3})}+a^{(2)}_{(\mathbf{3},\mathbf{1})}+2 a^{(2)}_{(\mathbf{3},\mathbf{3})} & \geq 0\,,\\
\label{secondSO4even}
 a^{(2)}_{(\mathbf{1},\mathbf{1})}+3 a^{(2)}_{(\mathbf{3},\mathbf{3})} & \geq 0\,.
 \end{align}
 \ees
Solving the constraints \eqref{neweqso4}, e.g.~for $a^{(2)}_{(\mathbf{3},\mathbf{1})}$ and $a^{(2)}_{(\mathbf{1},\mathbf{1})}$, one can obtain the two strongest positivity constraints for the two remaining independent amplitudes which immediately imply $a^{(2)}_{(\mathbf{3},\mathbf{3})}\geq 0$.

\subsubsection{Goldstone bosons and composite Higgs}

 Let us now assume that the IR theory is well described by GBs from the cosets $SO(5)/SO(4)$ or $SO(4,1)/SO(4)$, or more generally by a non-linear sigma model 
 \begin{equation}
 \sum_{i=1}^4 \phi^2_i + \frac{1}{c_H} \phi^2_{5}=f_\pi^2\,.
 \end{equation}
 For the Higgs boson at $O(p^2)$ it accounts for the deformation
\begin{equation}
\mathcal{O}_H= \frac{c_H}{2f^2_\pi} (\partial_\mu |H|^2)^2\,,
\end{equation}
which is the leading one in custodially symmetric composite Higgs models \cite{Giudice:1024017}.
We may also add a light $SO(4)$ singlet and a light symmetric (traceless) scalar coupled to the GBs with couplings $a$ and $b$ as in eq.~\eqref{interactions_scalar_symm}.
The resulting eigen-amplitudes
\be
\bry{l}
\dst a^{(1)}_{(\mathbf{1},\mathbf{1})}=\frac{3}{f_\pi^2}\left(c_H -a^2 + \frac{3}{4} b^2 \right)\,,\vspace{2mm}\\  
\dst a^{(1)}_{(\mathbf{1},\mathbf{3})}= a^{(1)}_{(\mathbf{3},\mathbf{1})} =\frac{1}{f_\pi^2}\left(c_H -a^2 + \frac{3}{4} b^2 \right)\,, \vspace{2mm}\\
\dst a^{(1)}_{(\mathbf{3},\mathbf{3})}= -\frac{1}{f_\pi^2}\left(c_H -a^2 + \frac{3}{4} b^2 \right)\,,
\ery
\ee
satisfy the constraints \eqref{so4constr1sub} and allow us to evaluate the left-hand side of the sum rule \eqref{firstSO4} that now reads
   \begin{align}
 \label{SO4GB1}
\left(c_H -a^2 + \frac{3}{4} b^2 \right)= \frac{f^2_\pi}{4\PI}\int_0^\infty \frac{\di s}{s}\left[\sigma^{\mathrm{tot}}_{(\mathbf{1},\mathbf{1})}(s)+\sigma^{\mathrm{tot}}_{(\mathbf{1},\mathbf{3})}(s)+\sigma^{\mathrm{tot}}_{(\mathbf{3},\mathbf{1})}(s)-3 \sigma^{\mathrm{tot}}_{(\mathbf{3},\mathbf{3})}(s)\right]\,.
\end{align}
The case with $a=b=0$ was originally found in ref.~\cite{Urbano:2013ty} that, however, missed the second sum rule \eqref{secondSO4}.
Our construction instead systematically allows one to find all the independent sum rules.

 The second sum rule with one subtraction shows an interesting feature.  The IR theory  of GBs has an accidental discrete $P_{LR}$ symmetry at $O(p^2)$ that exchanges $SU(2)_L\leftrightarrow SU(2)_R$ and therefore sets $a_{(\mathbf{1},\mathbf{3})}=a_{(\mathbf{3},\mathbf{1})}$. Higher dimensional operators spoil this symmetry. Yet the sum rule enforces an \textit{averaged} $P_{LR}$ 
 \be
 \label{integratedPLR}
\int_0^\infty \frac{\di s}{s} \sigma^{\mathrm{tot}}_{(\mathbf{1},\mathbf{3})}(s)=\int_0^\infty \frac{\di s}{s}  \sigma^{\mathrm{tot}}_{(\mathbf{3},\mathbf{1})}(s)\,,
 \ee
 on top of the asymptotic equality 
  \be 
\sigma^{\mathrm{tot}}_{(\mathbf{1},\mathbf{3})}(s\rightarrow\infty) =\sigma^{\mathrm{tot}}_{(\mathbf{3},\mathbf{1})}(s\rightarrow\infty)\,,
 \ee
analogous to the Pomeranchuk's theorem or, similarly, to the condition of eq.~\eqref{eq:Asymptotic}. The averaged $P_{LR}$ relation is a surprising result where the IR/UV connection is clearly at work: an IR accidental symmetry puts constraints on the theory at all energies.
  
\subsection{Adjoints of $SU(N)$ and chiral QCD}\l{SUN}
 
We now consider the $2\to2$ forward scattering between particles transforming in the adjoint representation of $SU(N)$ for $N\geq 4$. 
The simpler cases $N=2$ and $N=3$, relevant for chiral QCD, are discussed below in a separate subsection.

We have the following decomposition of ${\bf Adj}\otimes {\bf Adj}$:
\be\l{AdjxAdj}
{\bf Adj}\otimes {\bf Adj}={\bf 1}^{\text{s}}\oplus{\bf D}^{\text{s}}\oplus {\bf F}^{\text{a}}\oplus{\bf Y}^{\text{s}}\oplus{\bf T}^{\text{a}}\oplus{\bf \ovl{T}}^{\text{a}}\oplus{\bf X}^{\text{s}}\,,
\ee
where the s and a labels stand for symmetric and antisymmetric with respect to the two original adjoints.\footnote{We adopt for $SU(N)$ the same conventions as in ref.~\cite{1992hep.ph....6222S}.}
The dimensions of the irreps appearing in the decomposition are
\be
\bry{lll}
\dst \Delta_{\bf 1}=1\,,\quad &\Delta_{\bf D}=N^{2}-1\,,\quad &\Delta_{\bf F}=N^{2}-1\,,\vspace{2mm}\\
\dst \Delta_{\bf Y}=\f{N^{2}(N+1)(N-3)}{4}\,,\quad &\dst \Delta_{\bf T}=\Delta_{\bf \ovl{T}}=\f{(N^{2}-4)(N^{2}-1)}{4}\,,\quad &\dst \Delta_{\bf X}=\f{N^{2}(N-1)(N+3)}{4}\,.
\ery
\ee
The crossing matrix $X$ can be computed as shown in appendix \ref{XforSUN}. In this case only $\hat{X}$ is relevant\footnote{The matrix $X$ is block diagonal in the mixed and non-mixed indexes. Moreover, the amplitude corresponding to the mixed entry $\mathbf{FD}$ (or $\mathbf{DF}$) vanishes due to conservation of angular momentum in the forward limit.} and is given by 
\be\l{matrixXSUNreduced}
\hat{X}=\(
\begin{array}{cccccc}
 \f{1}{N^2-1} & 1 & -1 & \f{(N-3) N^2}{4 (N-1)} & 2-\f{N^2}{2} & \f{N^2 (N+3)}{4 (N+1)} \vspace{2mm}\\
 \f{1}{N^2-1} & \f{N^2-12}{2 \(N^2-4\)} & -\f{1}{2} & -\f{(N-3) N^2}{4 (N-2) (N-1)} & 1 & \f{N^2 (N+3)}{4 (N+1) (N+2)} \vspace{2mm}\\
 \f{1}{1-N^2} & -\f{1}{2} & \f{1}{2} & -\f{(N-3) N}{4 (N-1)} & 0 & \f{N (N+3)}{4 (N+1)} \vspace{2mm}\\
 \f{1}{N^2-1} & \f{1}{2-N} & -\f{1}{N} & \f{1}{N-2}+\f{1}{4}+\f{1}{2-2 N} & \f{N+2}{2 N} & \f{N+3}{4 N+4} \vspace{2mm}\\
 \f{1}{1-N^2} & \f{2}{N^2-4} & 0 & \f{(N-3) N}{4 \(N^2-3 N+2\)} & \f{1}{2} & \f{N (N+3)}{4 \(N^2+3 N+2\)} \vspace{2mm}\\
 \f{1}{N^2-1} & \f{1}{N+2} & \f{1}{N} & \f{N-3}{4 (N-1)} & \f{N-2}{2 N} & \f{N^2+N+2}{4N^2+12 N+8}
\end{array}
\)\,,
\ee
where the entries are ordered as ${\bf 1},{\bf D},{\bf F},{\bf Y},{\bf T},{\bf X}$.
One can simply verify that this matrix satisfies all the properties discussed in appendix \ref{app:crossingM}. The matrix $M$ that diagonalizes $\hat{X}$ as
\be
M \hat{X} M^{-1}=\text{diag}(-1,-1,1,1,1,1)\,,
\ee
is given by
\be\l{matrixUSUNreduced}
M=
\(
\begin{array}{cccccc}
 \frac{1}{2-2 N^2} & -\frac{1}{2 N+4} & -\frac{1}{2 N} & \frac{3-N}{8 (N-1)} & \frac{2-N}{4 N} & \frac{(N+3) (3 N+2)}{8 (N+1)
   (N+2)} \vspace{2mm}\\
 \frac{1}{2 \left(N^2-1\right)} & \frac{1}{4-N^2} & 0 & -\frac{(N-3) N}{8 \left(N^2-3 N+2\right)} & \frac{1}{4} & -\frac{N (N+3)}{8
   \left(N^2+3 N+2\right)} \vspace{2mm}\\
 \frac{1}{2 \left(N^2-1\right)} & \frac{1}{2 N+4} & \frac{1}{2 N} & \frac{N-3}{8 (N-1)} & \frac{N-2}{4 N} & \frac{1}{8}
   \left(-\frac{4}{N+2}+5+\frac{2}{N+1}\right) \vspace{2mm} \\
 \frac{1}{2-2 N^2} & \frac{1}{N^2-4} & 0 & \frac{(N-3) N}{8 \left(N^2-3 N+2\right)} & \frac{3}{4} & \frac{N (N+3)}{8 \left(N^2+3
   N+2\right)} \vspace{2mm}\\
 \frac{1}{2 \left(N^2-1\right)} & \frac{1}{4-2 N} & -\frac{1}{2 N} & \frac{1}{8} \left(-\frac{2}{N-1}+5+\frac{4}{N-2}\right) &
   \frac{N+2}{4 N} & \frac{N+3}{8 N+8} \vspace{2mm}\\
 \frac{1}{2-2 N^2} & -\frac{1}{4} & \frac{3}{4} & -\frac{(N-3) N}{8 (N-1)} & 0 & \frac{N (N+3)}{8 (N+1)} 
\end{array}
\)\,.
\ee
We see that the $-1$-eigenspace has dimension two, leading to two once-subtracted sum rules and $\dim\hat{X}-2=4$ twice-subtracted sum rules. These numbers match the number of (anti)-symmetric irreps appearing in the matrix $\hat{X}$. The first two rows of $M$ in eq.~\eqref{matrixUSUNreduced} allow us to read the two independent once-subtracted sum rules
\bes\l{sumrumesSUN}
\begin{align}
& \bry{l}
 \dst \frac{4 a^{(1)}_{\mathbf{1}}}{N^2-1}+\frac{4 a^{(1)}_{\mathbf{D}}}{N+2}+\frac{4 a^{(1)}_{\mathbf{F}}}{N}+\frac{(N-3) a^{(1)}_{\mathbf{Y}}}{N-1}+\frac{2 (N-2) a^{(1)}_{\mathbf{T}}}{N}-\frac{(N+3) (3 N+2)a^{(1)}_{\mathbf{X}}}{(N+1) (N+2)}\vspace{2mm}\\
\dst = \frac{2}{\PI}\int_0^\infty \frac{\di s}{s}\left[\frac{4 \sigma^{\text{tot}}_{\mathbf{1}}}{N^2-1}+\frac{4 \sigma^{\text{tot}}_{\mathbf{D}}}{N+2}+\frac{4 \sigma^{\text{tot}}_{\mathbf{F}}}{N}+\frac{(N-3) \sigma^{\text{tot}}_{\mathbf{Y}}}{N-1}+\frac{2 (N-2) \sigma^{\text{tot}}_{\mathbf{T}}}{N}-\frac{(N+3) (3 N+2)\sigma^{\text{tot}}_{\mathbf{X}}}{(N+1) (N+2)}\right]\,,
\ery\\
& \bry{l}
 \dst \frac{4 a^{(1)}_{\mathbf{1}}}{N^2-1}-\frac{8 a^{(1)}_{\mathbf{D}}}{N^2-4}-\frac{(N-3) N a^{(1)}_{\mathbf{Y}}}{N^2-3N+2}+2 a^{(1)}_{\mathbf{T}}-\frac{N (N+3) a^{(1)}_{\mathbf{X}}}{N^2+3 N+2}\vspace{2mm}\\
\dst = \frac{2}{\PI}\int_0^\infty \frac{\di s}{s}\left[\frac{4 \sigma^{\text{tot}}_{\mathbf{1}}}{N^2-1}-\frac{8 \sigma^{\text{tot}}_{\mathbf{D}}}{N^2-4}-\frac{(N-3) N \sigma^{\text{tot}}_{\mathbf{Y}}}{N^2-3N+2}+2 \sigma^{\text{tot}}_{\mathbf{T}}-\frac{N (N+3) \sigma^{\text{tot}}_{\mathbf{X}}}{N^2+3 N+2}\right]\,.
\ery
\end{align}
\ees
The coefficients $a_{I}$ in these sum rules are not all independent since they satisfy the constraints set by the last four rows of $M$ that, taking linear combinations, 
can be written e.g.~as
\bes
\label{exSU3:constr1}
\begin{align}
a^{(1)}_{\mathbf{1}}- N a^{(1)}_{\mathbf{Y}}-(N^2+N-2)a^{(1)}_{\mathbf{T}}= &0 \,,\\
2a^{(1)}_{\mathbf{D}}-(N-2)a^{(1)}_{\mathbf{T}}-N a^{(1)}_{\mathbf{Y}} = & 0 \,, \\
2a^{(1)}_{\mathbf{F}}-Na^{(1)}_{\mathbf{Y}}-(N+2)a^{(1)}_{\mathbf{T}} = & 0\,, \\
a^{(1)}_{\mathbf{X}}+2 a^{(1)}_{\mathbf{T}}+a^{(1)}_{\mathbf{Y}}= & 0\,.
\end{align}
\ees
Solving these constraints, e.g.~for $a^{(1)}_{\mathbf{1,D,F,X}}$, the sum rules \eqref{sumrumesSUN} read
\bes
\begin{align}
& \bry{l}
2a^{(1)}_{\mathbf{T}}+a^{(1)}_{\mathbf{Y}} \vspace{2mm}\\
\dst =-\frac{2}{\PI}\int_0^\infty \frac{\di s}{s}\left[\frac{4 \sigma^{\text{tot}}_{\mathbf{1}}}{N^2-1}+\frac{4 \sigma^{\text{tot}}_{\mathbf{D}}}{N+2}+\frac{4 \sigma^{\text{tot}}_{\mathbf{F}}}{N}+\frac{(N-3) \sigma^{\text{tot}}_{\mathbf{Y}}}{N-1}+\frac{2 (N-2) \sigma^{\text{tot}}_{\mathbf{T}}}{N}-\frac{(N+3) (3 N+2)\sigma^{\text{tot}}_{\mathbf{X}}}{(N+1) (N+2)}\right]\,,\ery\\
& a^{(1)}_{\mathbf{T}} = \frac{2}{\PI}\int_0^\infty \frac{\di s}{s}\left[\frac{4 \sigma^{\text{tot}}_{\mathbf{1}}}{N^2-1}-\frac{8 \sigma^{\text{tot}}_{\mathbf{D}}}{N^2-4}-\frac{(N-3) N \sigma^{\text{tot}}_{\mathbf{Y}}}{N^2-3N+2}+2 \sigma^{\text{tot}}_{\mathbf{T}}-\frac{N (N+3) \sigma^{\text{tot}}_{\mathbf{X}}}{N^2+3 N+2}\right]\,.
\end{align}
\ees
In the next subsection we  relate these $a^{(1)}_I$ to the LECs of the $SU(N)_{L}\times SU(N)_{R}/ SU(N)_{V}$ non-linear sigma model.

The first two rows give rise also to the following constraints for the second derivatives
\bes\l{constraintsSUN2}
\begin{align}
\frac{4 a^{(2)}_{\mathbf{1}}}{N^2-1}+\frac{4 a^{(2)}_{\mathbf{D}}}{N+2}+\frac{4 a^{(2)}_{\mathbf{F}}}{N}+\frac{(N-3) a^{(2)}_{\mathbf{Y}}}{N-1}+\frac{2 (N-2) a^{(2)}_{\mathbf{T}}}{N}-\frac{(N+3) (3 N+2)a^{(2)}_{\mathbf{X}}}{(N+1) (N+2)}&=0\,,\\
\frac{4 a^{(2)}_{\mathbf{1}}}{N^2-1}-\frac{8 a^{(2)}_{\mathbf{D}}}{N^2-4}-\frac{(N-3) N a^{(2)}_{\mathbf{Y}}}{N^2-3N+2}+2 a^{(2)}_{\mathbf{T}}-\frac{N (N+3) a^{(2)}_{\mathbf{X}}}{N^2+3 N+2}&=0\,.
\end{align}
\ees
The positivity conditions corresponding to the crossing matrix \eqref{matrixXSUNreduced} are computed as prescribed in section \ref{sec:positivity}. In this case the procedure is the following: we take linear combinations with free coefficients of the last four rows of $M$ in eq.~\eqref{matrixUSUNreduced}. We obtain a $6$-vector depending on four free coefficients. We set three entries at a time to zero, and express three of the coefficients as functions of the remaining one. We then substitute their expression into the 6-vector linear combination and we check if the three non-zero entries are all positive (or negative, since they still depend on one free parameter). We repeat the procedure for all the combinations of three entries of the 6-vector linear combination. In this way we get the five strongest positivity constraints
 \bes \label{SUNeven}
 \begin{align}
 \label{firstSUNeven}
 2a^{(2)}_{\mathbf{F}}+(N-2)a^{(2)}_{\mathbf{T}}+N a^{(2)}_{\mathbf{X}} & \geq 0\,,\\
  \label{secondSUNeven}
 2 a^{(2)}_{\mathbf{D}}+(N+2) a^{(2)}_{\mathbf{T}}+N a^{(2)}_{\mathbf{X}} & \geq 0\,,\\
  \label{thirdSUNeven}
 2 a^{(2)}_{\mathbf{1}}+(N-2)(N+1)a^{(2)}_{\mathbf{Y}}+(N-1)(N+2) a^{(2)}_{\mathbf{X}} & \geq 0\,,\\
  \label{fourthSUNeven}
 a^{(2)}_{\mathbf{1}}+2(N+1)a^{(2)}_{\mathbf{F}}+N (N+2) a^{(2)}_{\mathbf{X}} & \geq 0\,,\\
\label{fifthSUNeven}
 (N+2)a^{(2)}_{\mathbf{1}}+2(N-2)(N+1)a^{(2)}_{\mathbf{D}}+N^{3} a^{(2)}_{\mathbf{X}} & \geq 0\,.
 \end{align}
 \ees
These equations generate a 4-dimensional convex polyhedral cone with five edges in a 5-dimensional space. Therefore the positivity conditions are not all linearly independent. However, they are the minimal set necessary to construct all possible positivity constraints through linear combinations with only positive coefficients (see figures \eqref{polycone} and \eqref{polyconeSON} for illustration). Notice that using the constraints \eqref{constraintsSUN2} solved for $a^{(2)}_{\mathbf{1}}$ and $a^{(2)}_{\mathbf{D}}$, eq.~\eqref{SUNeven} implies the simple positivity constraints $a^{(2)}_{\mathbf{F}},a^{(2)}_{\mathbf{Y}},a^{(2)}_{\mathbf{T}},a^{(2)}_{\mathbf{X}}\geq 0$.

\subsubsection{Goldstone bosons from $SU(N)_{L}\times SU(N)_{R}/ SU(N)_{V}$}

The sum rules \eqref{sumrumesSUN} are completely general and independent of the structure of the IR effective theory. An interesting case corresponds to the IR effective theory being given by the non-linear sigma model for the coset $SU(N)_{L}\times SU(N)_{R}/ SU(N)_{V}$. The $O(p^{2})$ effective Lagrangian can be written as
\be\l{NSMSUN}
\La_{\text{eff}}^{(2)}=\f{f_{\pi}^{2}}{4}\tr\left[(\demub \Sigma)^{\dag} \demua \Sigma\right]\,,
\ee
where the non-linear $\Sigma$ field is defined has
\be
\Sigma=\e^{\f{2\i\pi^{a}T^{a}}{f_{\pi}}}\,,\qquad \Sigma \to g_{R}\Sigma g_{L}^{\dag}\,,\quad g_{L,R}\in SU(N)_{L,R}\,,\ee
with the $SU(N)$ generators $T^{a}$ defined according to eq.~\eqref{SUNGener1}. 
Expanding the Lagrangian \eqref{NSMSUN} in the number of fields up to four we get
\be\l{SUNLag}
\La_{\text{eff}}^{(2)}=\dst\f{\delta^{ab}}{2}\demub \pi^{a}\demua \pi^{b}-\f{1}{6f_{\pi}^{2}}f^{abe}f^{cde}\pi^{a}\pi^{c}\demub \pi^{b}\demua\pi^{d}\,.
\ee
From this effective Lagrangian we get the four Goldstone bosons scattering amplitude
\be
\Amp\left(\pi^a \pi^b\rightarrow \pi^c\pi^d\right)(s,t=0)=\f{s}{f_{\pi}^{2}}f^{ace}f^{bde}\,.
\ee
Projecting this amplitude into the space of irreps, by using the projectors given in appendix \ref{AppProjectors}, we obtain  
\be\l{NLSMSUN}
\bry{lll}
\dst \Amp_{\mathbf{1}}=\f{N s}{f_{\pi}^{2}}\,,\qquad &\dst \Amp_{\mathbf{D}}=\f{N s}{2f_{\pi}^{2}}\,,\qquad &\dst\Amp_{\mathbf{F}}=\f{Ns}{2f_{\pi}^{2}}\,,\vspace{2mm}\\
\dst \Amp_{\mathbf{Y}}=\f{s}{f_{\pi}^{2}}\,,\qquad &\dst \Amp_{\mathbf{T}}=\Amp_{\ovl{\mathbf{T}}}=0\,,\qquad &\dst \Amp_{\mathbf{X}}=-\f{s}{f_{\pi}^{2}}\,,
\ery
\ee
which nicely satisfy the constraints \eqref{exSU3:constr1}.
This last equation allows us to write down the contribution of the non-linear sigma model \mbox{$SU(N)_{L}\times SU(N)_{R}/ SU(N)_{V}$} to the sum rules \eqref{sumrumesSUN}:
\bes\l{sumrumesSUNsigmamodel}
\begin{align}
& \bry{l}
\dst \int_0^\infty \frac{\di s}{s}\left[\frac{4 \sigma^{\text{tot}}_{\mathbf{1}}}{N^2-1}+\frac{4 \sigma^{\text{tot}}_{\mathbf{D}}}{N+2}+\frac{4 \sigma^{\text{tot}}_{\mathbf{F}}}{N}+\frac{(N-3) \sigma^{\text{tot}}_{\mathbf{Y}}}{N-1}+\frac{2 (N-2) \sigma^{\text{tot}}_{\mathbf{T}}}{N}-\frac{(N+3) (3 N+2)\sigma^{\text{tot}}_{\mathbf{X}}}{(N+1) (N+2)}\right]=-\frac{\PI}{2f_{\pi}^{2}}\,,
\ery\\
& \bry{l}
\dst \int_0^\infty \frac{\di s}{s}\left[\frac{4 \sigma^{\text{tot}}_{\mathbf{1}}}{N^2-1}-\frac{8 \sigma^{\text{tot}}_{\mathbf{D}}}{N^2-4}-\frac{(N-3) N \sigma^{\text{tot}}_{\mathbf{Y}}}{N^2-3N+2}+2 \sigma^{\text{tot}}_{\mathbf{T}}-\frac{N (N+3) \sigma^{\text{tot}}_{\mathbf{X}}}{N^2+3 N+2}\right]=0\,.
\ery
\end{align}
\ees

\subsubsection{Chiral $SU(2)$ and $SU(3)$}

The scattering of two adjoints of $SU(2)\sim SO(3)$ is covered by the discussion in  subsection~\ref{SO4example} in the case $N=3$. 
Let us move to the scattering of two $\mathbf{8}\in SU(3)$. In this case, \mbox{$\mathbf{8}\otimes\mathbf{8}=\mathbf{1}^{\text{s}} \oplus \mathbf{8}_1^{\text{s}} \oplus \mathbf{8}_2^{\text{a}} \oplus \mathbf{10}^{\text{a}} \oplus \ovl{\mathbf{10}}^{\text{a}} \oplus \mathbf{27}^{\text{s}}$} where we have renamed the irreps with their dimension. Note that the representation $\mathbf{Y}$ of the general decomposition of eq.~\eqref{AdjxAdj} does not appear. Therefore, after dropping this irrep in the matrix \eqref{matrixXSUNreduced} we get the following crossing matrix
\be\l{matrixXSU3reduced}
\hat{X}=\(
\begin{array}{ccccccccc}
 \f{1}{8} & 1 & -1 &  -\f{5}{2} & \f{27}{8} \vspace{2mm}\\
 \f{1}{8} & -\f{3}{10} & -\f{1}{2} & 1 & \f{27}{40} \vspace{2mm}\\
 -\f{1}{8} & -\f{1}{2} & \f{1}{2} & 0 & \f{9}{8} \vspace{2mm}\\
 -\f{1}{8} & \f{2}{5} & 0 &  \f{1}{2} & \f{9}{40} \vspace{2mm}\\
 \f{1}{8} & \f{1}{5} & \f{1}{3} & \f{1}{6} & \f{7}{40}
\end{array}
\)\,,
\ee
which  agrees with the crossing matrix used in ref.~\cite{Mateu:2008gv}. The matrix $M$ that diagonalizes $\hat{X}$ as
\be
M \hat{X} M^{-1}=\text{diag}(-1,-1,1,1,1)
\ee
is given by
\be\l{matrixUSU3reduced}
M=
\(
\bry{ccccc}
 -\f{1}{16} & -\f{1}{10} & -\f{1}{6} & -\f{1}{12} & \f{33}{80} \\
 \f{1}{16} & -\f{1}{5} & 0 & \f{1}{4} & -\f{9}{80} \\
 \f{1}{16} & \f{1}{10} & \f{1}{6} & \f{1}{12} & \f{47}{80} \\
 -\f{1}{16} & \f{1}{5} & 0 & \f{3}{4} & \f{9}{80} \\
 -\f{1}{16} & -\f{1}{4} & \f{3}{4} & 0 & \f{9}{16} \\
\ery
\)\,.
\ee
This is nothing but the matrix \eqref{matrixUSUNreduced} for $N=3$ where the entry corresponding to the representation $\mathbf{Y}$ has been dropped.
The once-subtracted sum rules can therefore be read off the first two rows of this matrix that give
\bes\l{sumrumesSUNsigmamodel}
\begin{align}
& \bry{l}
\dst a^{(1)}_{\mathbf{27}}=-\f{1}{120\PI}\int_0^\infty \frac{\di s}{s}\left[15\sigma^{\text{tot}}_{\mathbf{1}}+24 \sigma^{\text{tot}}_{\mathbf{8_{1}}}+40 \sigma^{\text{tot}}_{\mathbf{8_{2}}}+20\sigma^{\text{tot}}_{\mathbf{10}}-99\sigma^{\text{tot}}_{\mathbf{27}}\right] \,,
\ery\\
& \bry{l}
\dst a^{(1)}_{\mathbf{10}}=\f{1}{40 \PI }\int_0^\infty \frac{\di s}{s}\left[5\sigma^{\text{tot}}_{\mathbf{1}}-16 \sigma^{\text{tot}}_{\mathbf{8_{1}}}+20 \sigma^{\text{tot}}_{\mathbf{10}}-9 \sigma^{\text{tot}}_{\mathbf{27}}\right]\,.
\ery
\end{align}
\ees
On the right-hand side we have solved the constraints set on the $a^{(1)}_I$'s by the last three rows of $M$ for $a^{(2)}_{\mathbf{1}},\,a^{(2)}_{\mathbf{8_{1}}},\,a^{(2)}_{\mathbf{8_{2}}}$. The non-linear sigma model $SU(3)_{L}\times SU(3)_{R}/ SU(3)_{V}$ gives $a^{(1)}_{\mathbf{10}}=0$ and $a^{(1)}_{\mathbf{27}}=-1/f_\pi^2$, see eq.~\eqref{NLSMSUN}.

The first two rows of $M$ give rise also to
\bes
\begin{align} 
\label{constsu3ex}
15a^{(2)}_{\mathbf{1}}+24a^{(2)}_{\mathbf{8_{1}}}+40 a^{(2)}_{\mathbf{8_{2}}}+20a^{(2)}_{\mathbf{10}}-99a^{(2)}_{\mathbf{27}} &=0 \,, \\
5a^{(2)}_{\mathbf{1}}-16 a^{(2)}_{\mathbf{8_{1}}}+20 a^{(2)}_{\mathbf{10}}-9 a^{(2)}_{\mathbf{27}} &=0\,.
\end{align}
\ees
The positivity conditions are obtained with the usual procedure from the last three rows of $M$ and read
 \bes
 \begin{align}
 \label{firstSU3even}
 2a^{(2)}_{\mathbf{8_{2}}}+a^{(2)}_{\mathbf{10}}+3 a^{(2)}_{\mathbf{27}} & \geq 0\,,\\
  \label{secondSU3even}
 2a^{(2)}_{\mathbf{8_{1}}}+5 a^{(2)}_{\mathbf{10}}+3 a^{(2)}_{\mathbf{27}} & \geq 0\,,\\
  \label{fourthSU3even}
 a^{(2)}_{\mathbf{1}}+8a^{(2)}_{\mathbf{8_{2}}}+15 a^{(2)}_{\mathbf{27}} & \geq 0\,,\\
\label{fifthSU3even}
 5a^{(2)}_{\mathbf{1}}+8a^{(2)}_{\mathbf{8_{1}}}+27 a^{(2)}_{\mathbf{27}} & \geq 0\,.
 \end{align}
 \ees
These are exactly the first two and the last two conditions \eqref{SUNeven} for $N=3$. They can be seen as the generating vectors of a 4-edged 3-dimensional convex polyhedral cone in five dimensions.
Notice that solving the constraints \eqref{constsu3ex} for $a^{(2)}_{\mathbf{1}}$ and $a^{(2)}_{\mathbf{8}_1}$ the positivity constraints imply $a^{(2)}_{\mathbf{8}_2},\, a^{(2)}_{\mathbf{10}},\,a^{(2)}_{\mathbf{27}}\geq 0$.
 Certain positivity constraints for the LECs appearing in twice-subtracted sum rules of chiral $SU(3)$ have been discussed in refs.~\cite{Pham:1985cr,Mateu:2008gv}. 

\section{Longitudinal $WW$ scattering}
\label{sectWWscattering}
 
Some of the once-subtracted sum rules presented in the previous sections for GBs in $SO(4)/SO(3)$ and $SO(5)/SO(4)$ have been interpreted in the context of the EW chiral Lagrangian and Composite Higgs models in refs.~\cite{Low:2009di,Falkowski:2012bu,Urbano:2013ty} by means of the Equivalence Theorem (ET). 
There are however three caveats:
\begin{itemize}
\item As recently noticed in ref.~\cite{Espriu:2014jya} the application of the ET in the forward limit $t=0$ or $t\ll m_W^2$ is questionable since large corrections of the order of $m_{W}^{2}/t$ can be expected. 
\item Theories with massive gauge bosons require particular care since they may affect the sum rules with one subtraction by a finite $\delta c^{\infty(1)}$ coming from the deep UV, as we discussed already in section~\ref{sec:convergence}.
\item A propagating photon in the $t$-channel in the forward scattering, $t=0$, gives rise to a Coulomb singularity so that one may wonder whether  $g^\prime=0$ and $g^{\prime\,2}\ll 1$ yield different sum rules.
\end{itemize}
In this section we address in steps each of these subtle points, and derive a robust sum rule for $W_L W_L$ scattering with arbitrary $g$ and small but finite $g^{\prime\,2}\ll 1$.
\subsection{Longitudinal $WW$ scattering and the equivalence theorem}

Let us start  with the first point and compare $W_L W_L$ scattering with $\pi\pi$ scattering. 
We focus first on $SU(2)_{L}$ broken completely with $g\neq 0$, and $g'=0$ strictly so that the photon is not included. We come back to the case of finite $g^\prime$ in subsection \ref{sec:photonincluded}. 
We also include a propagating singlet Higgs-like state $h$  coupled to the longitudinal components of the gauge bosons with a strength $a$ in units of the SM coupling. The relevant part of the Lagrangian in a $R_{\xi}$-gauge is given by
\be
\La=-\f{1}{4}W_{\mu\nu}^{a}W^{\mu\nu\,a}-\f{1}{2\xi}\(\demub W^{\mu\,a}+m_{W}\xi\pi^{a}\)^{2}+\f{v^{2}}{4}\Tr\left[(D_{\mu}\Sigma)^{\dag}D^{\mu}\Sigma\right](1+2a\f{h}{v})+\f{1}{2}\demub h \demua h\,,
\ee
where $\Sigma=\e^{\i\pi^{a}\sigma^{a}/v}$ and $m_{W}=gv/2$. 
 
The amplitude for the process $W^{a}_L W^{b}_L\to W^{c}_L W^{d}_L$ receives contributions from the quadrilinear $W$ interaction, the $s,t,u$-channel $W$ exchange, and $s,t,u$-channel $h$ singlet exchange. There is no $\pi WW$ vertex, so that there is no contribution from Goldstone exchange. The exact form of the eigen-amplitudes is given in appendix \ref{WWscattering}. Here we are interested in the limit $s\gg m_{\mathrm{IR}}^2 \gg t$ with $m_{\mathrm{IR}}^2=m_W^2\,,\,m_h^2$ since we eventually need the forward amplitude.
From the first row of the matrix $M$ in eq.~\eqref{MSON} with $N=3$ we extract the left-hand side of the sum rule
\be
\l{WWright}
\lim_{\mu^2\gg m_{\mathrm{IR}}^2 \gg t}[M\Amp^{(1)}(\mu^2)]_1
=\f{\(3-a^{2}\)}{v^{2}}\,.
\ee
This result should be contrasted with the usual limit $s,t\gg m_{\mathrm{IR}}^2$ that gives instead
\be
\l{WWwrong}
\lim_{s,t\gg m_{\mathrm{IR}}^2 }[M\Amp^{(1)}(s,t)]_1 =\f{\(1-a^{2}\)}{v^{2}}\,.
\ee
The latter result clearly agrees with the prediction of the ET for the $\pi\pi$ scattering in that kinematical region. 
But in fact, we want to emphasize that even eq.~\eqref{WWright} agrees with the prediction of the ET at $t\ll m_{\mathrm{IR}}^2$.
Indeed, the diagrams contributing to the $\pi^{a}\pi^{b}\to \pi^{c}\pi^{d}$ scattering are exactly the same as in the previous case with all the external $W_{L}$ legs replaced by the corresponding Goldstone bosons. In particular, they include the contribution of a $t$-channel exchange of a $W$ boson which has a pole of the form $g^2 (4m_W^2-2s-t)/(t-m_W^2)$, where we cannot neglect $m_W^2$ compared to $t$ in the forward limit.  In other words, at $t=0$, the $\pi\pi$ scattering in a gauge theory with $g\neq 0$ is different from the $\pi\pi$ scattering in the gauge-less limit $g=0$. The latter does not include the diagram with the $t$-channel $W$ boson exchange which, for $t=0$, contributes instead to the scattering amplitude $\Amp^{(1)}$ of the former by an extra $2/v^2$ factor explaining the mismatch between eq.~\eqref{WWwrong} and eq.~\eqref{WWright}.
More explicitly, using the scattering amplitude computed with GBs as external legs given by eq.~\eqref{pipiWcontrib} we get 
\be
\lim_{s\gg m_{\mathrm{IR}}^2 }[M\Amp^{(1)}(s,t)]_1= \frac{(1-a^2)}{v^2}-\frac{g^2}{2}\frac{1}{t-m_W^2}\,.
\ee
This expression reduces to the correct limits \eqref{WWright} and \eqref{WWwrong} for $t\ll m_{W}^2$ and $t \gg m_W^2 $ respectively. 
The bottom line is that the ET does provide the correct answer when handled properly and when all relevant contributions are taken into account.

\subsection{The sum rule for $g^\prime=0$}

From the matrix $M$ that diagonalizes the crossing matrix for $SO(3)\sim SU(2)$, see eqs.~\eqref{Xson} and \eqref{MSON}, we can read off the sum rule with one subtraction\footnote{There are in fact two additional dispersion relations that we could write, see eq.~\eqref{constrmixsumrule1sub}. However, we are eventually interested in the case $\mu^2=2m_W^2$, see eq.~\eqref{finalsumrule1subso3}, where these extra equations are nothing but the trivial constraints set by crossing symmetry, $\Amp^{(1)}_{\mathbf{5}}(2m_W^2)=-\Amp^{(1)}_{\mathbf{3}}(2m_W^2)=-\Amp^{(1)}_{\mathbf{1}}(2m_W^2)/2$.} at $s=\mu^2$ with $\text{Re}\,\mu^{2}\gg m_{W}^{2}$
\begin{equation}
\label{sumrulebeforephotonsubtraction}
\frac{\left(3-a^2 \right)}{v^2}=\frac{1}{6\PI}\int_{4m_W^2}^{\infty} \di s\,  \frac{(s^2+\mu^4)s}{(s^2-\mu^4)^2}\sqrt{1-\f{4m_W^{2}}{s}} \left[2\sigma^{\mathrm{tot}}_\mathbf{1}(s)+3\sigma^{\mathrm{tot}}_{\mathbf{3}}(s)-5\sigma^{\mathrm{tot}}_{\mathbf{5}}(s)\right]+[M\delta c^{\infty(1)}]_1\,.
\end{equation}
 The left-hand side supports the claim of ref.~\cite{Espriu:2014jya}. However, the right-hand side contains a finite contribution coming from the massive gauge boson exchange, if they are still propagating degrees of freedom in the deep UV.\footnote{For the other contributions to $c^{\infty(1)}$ we assume that eq.~\eqref{eq:Asymptotic} holds and hence they are projected out.} In this case, the very same terms that are responsible for the mismatch between \eqref{WWright} and \eqref{WWwrong} 
also affect the contribution from the big circle $c^{\infty (1)}$, and by exactly the same amount
\begin{equation}
\delta c^{\infty(1)}_{\mathbf{1}}=\frac{4}{v^2}\,,\qquad \delta c^{\infty(1)}_{\mathbf{3},\mathbf{5}}=\pm \frac{2}{v^2}\qquad\Longrightarrow\qquad  [M \delta c^{\infty(1)}]_1=\frac{2}{v^2}\,.
\end{equation}
We therefore recover the original sum rule
\begin{equation}
\label{sumrulegprime=0}
\left(1-a^2 \right)=\frac{v^2}{6\PI}\int_{4m_W^2}^{\infty} \di s\,  \frac{(s^2+\mu^4)s}{(s^2-\mu^4)^2}\sqrt{1-\f{4m_W^{2}}{s}} \left[2\sigma^{\mathrm{tot}}_\mathbf{1}(s)+3\sigma^{\mathrm{tot}}_{\mathbf{3}}(s)-5\sigma^{\mathrm{tot}}_{\mathbf{5}}(s)\right]\,,
\end{equation}
up to the finite mass terms, and the $\mu^2$ factor. Notice that $\mu^2$ can not be generically sent to zero while keeping $m_W$ finite. Moreover, $\mu^2$ regularizes the otherwise divergent integral in the IR in the general formula \eqref{exact_sumrulen2}.\footnote{\label{footIRdiv}The integral on the right-hand side of eq.~\eqref{exact_sumrulen2} is indeed generically IR divergent for massive gauge bosons as $\mu\rightarrow 0$  since the longitudinal  polarisations $\epsilon^{\mathrm{L}}_\mu(k)$ do not vanish as $s\rightarrow 4m_W^2$.  This should be contrasted with the case of GBs where the $\epsilon^{\mathrm{L}}_\mu(k)$ of the gauge bosons is replaced by the GB momentum $k_\mu$ which does instead go to zero at the IR boundary $s=0$.}

Notice that we used the same gauge coupling on the IR side and on the big circle (where \mbox{$s=\Lambda^2\rightarrow\infty$}) because $t=0$ and thus the exchanged momentum and the scattering angle are zero. 
More concretely, the eikonal approximation that resums all the ladder diagrams (including the crossed ones that enforce crossing symmetry) \cite{Abarbanel:1969ek,Levy:1969cr,Giudice:2001ce}
\begin{equation}
\delta\Amp_{I}= -2i \int \di^2b_\perp e^{\i q_\perp b_\perp}\left(e^{i \chi_I}-1\right)\,,\qquad \chi_{I}(b_\perp)=\frac{1}{2s}\int \frac{\di^2 q_\perp}{(2\PI)^2} e^{-\i q_\perp b_\perp}\delta\Amp_{I}^{\mathrm{Born}}(q_\perp)\,,
\end{equation}
returns for $s\rightarrow \infty$ and $t=0$ the Born amplitude $\Amp=\Amp^{\text{Born}}$, as can be explicitly checked with the extra gauge contribution to the full tree-level amplitude given in eq.~\eqref{pipiWcontrib}.  
Since the extra contribution to the amplitude is the same in the UV and in the IR, the $\delta c^{\infty(1)}$ from the big circle is the same, by analyticity, of the $\delta\Amp^{(1)}$ returned by the contour integral along any $\mathcal{C}$ in the IR.

The punch line is that the sum rule  for $WW$ scattering at $g^\prime=0$ agrees with the one for GBs scattering because the extra gauge boson contributions are the same on both sides of the sum rule.
  
\subsection{The sum rule for $g^{\prime\,2}\ll 1$}
\label{sec:photonincluded}
 
We now extend the discussion of the previous subsection and allow for a small electric-charge and a propagating photon while we still neglect the mass splitting among the massive gauge bosons. 
In other words, we now focus on the limit $g^{\prime\,2}\ll 1$.

First, we regulate the IR with a finite  $t$ and a finite photon mass $m_\gamma^2$, to be sent to zero later. The sum rules at finite $t$ are discussed in appendix~\ref{beyondforward}.
Analogously to the case of massive $W$'s the contribution from the photon exchange gives an extra IR term to the left-hand side of the sum rule of the form
\be 
 \delta\Amp(s,t) =\frac{c_*s+\tilde{c}_* t}{t-m_\gamma^2}\longrightarrow \delta\Amp^{(1)}=\frac{c_* }{t-m_\gamma^2}
\ee
where $c_*$ and $\tilde{c}_*$ are proportional to the square of the electric charge.
The photon also generates an extra contribution to $c^{\Lambda(1)}$ that can be computed by expanding in the size of the radius of the big circle: since $\e^{\i\theta}$ always multiplies such a radius, apart from the first term that is finite, the others average to zero (see eq.~\eqref{cinfty})
\be 
\delta c^{\Lambda(1)}
=\frac{c_* }{t-m_\gamma^2}\,.
\ee
As expected, the IR contribution and the one from the big circle are equal, $\delta c^{\infty(1)}= \delta\Amp^{(1)}$, and thus they cancel, disappearing from the sum rules. Again, $c_*$ is the same coefficient on both sides since we are working at finite but small $t$ (in fact, we send $t\rightarrow0$ at the end) so that the exchanged momentum seen from the photon is always small, even though the center of mass energy for the big circle is large. Therefore, analyticity ensures that $\delta c^{\infty(1)}= \delta\Amp^{(1)}$.

After removing these terms we can take the limits $m_\gamma\rightarrow 0$ and $t\rightarrow 0$, in any order, getting the same expression as for the gauge-less limit $g^{\prime}=0$. In particular, the limit $t\rightarrow 0$ allows us to link $\mathrm{Im} \Amp$ in \eqref{sumrule_tfinite} to the total cross-section so that the sum rules essentially reduce to those of GBs discussed in the previous sections.
 For example, in the SM  with a light Higgs-like singlet coupled to the $W$ bosons with a coupling constant rescaled by a factor $a$ we recover again eq.~\eqref{sumrulegprime=0}.
Adding a quintuplet coupled to $W$'s as in eq.~\eqref{interactions_scalar_symm} changes the left-hand side as $\left(1-a^2\right)\rightarrow \left(1-a^2+5 b^2/6 \right)$.

This result agrees with the sum rule for the GBs living in $SO(4)/SO(3)$ found in ref.~\cite{Falkowski:2012bu} with $b, \mu^2\,, m^2_W\rightarrow 0$. The main difference is that our version for $W_L W_L$-scattering has $g\neq 0$, $g^{\prime\,2}\ll 1$ and finite masses, see also footnote~\ref{footIRdiv} for the IR convergence. Actually,  we checked in appendix~\ref{WWscattering} that keeping all the residues, the left-hand side of eq.~\eqref{sumrulegprime=0} does not explicitly depend on $\mu^2$ at tree-level.
The dependence on $\mu^2$ on the right-hand side thus captures the radiative corrections such as the running of the coupling constants. For $\mu^2$ real and below the cut (but finite and away from the poles) one should also reintroduce the full dependence on $m_W^2$ according to eq.~\eqref{exact_sumrulen2}, that is made by the replacement 
\be
\frac{(s^2+\mu^4)s}{(s^2-\mu^4)^2} \rightarrow  \frac{(s^2+\mu^4-4m_W^2 s-4m_W^2\mu^2+8m_W^4)s}{(s-\mu^2)^{2}(s+\mu^2-4m_W^2)^{2}}
\ee
in eq.~\eqref{sumrulegprime=0}. 
As long as $\mu^2$ is away from the poles and the IR singularity at $\mu^2=0$ there is very little sensitivity to its actual value. Choosing for convenience the crossing symmetric point $\mu^2=2m_W^2$ one gets
 \be 
 \label{finalsumrule1subso3}
\left(1-a^2\right)=\frac{v^2}{6\PI}\int_{4 m_W^2}^\infty \di s  \frac{s}{(s-2m_W^2)^2}\sqrt{1-\frac{4m_W^2}{s}}\left[2\sigma^{\mathrm{tot}}_\mathbf{1}(s)+3\sigma^{\mathrm{tot}}_{\mathbf{3}}(s)-5\sigma^{\mathrm{tot}}_{\mathbf{5}}(s)\right]\,,
\ee 
which can be used, thanks to the reality of $\mu^2$, to argue that a Higgs coupling $a$ bigger than one would require sizable contributions to longitudinal $WW$-scattering from quintuplets \cite{Falkowski:2012bu} that contain doubly charged states \cite{Low:2009di}. 

Summarizing the result, we have proved that the sum rule for the GBs survives after gauging. In particular, the sum rule for $SO(4)/SO(3)$ carries over $W_L W_L$ scattering with finite $g$ and small $g^{\prime\,2}\ll 1$. This is a non-trivial result since the theory contains gauge bosons that contribute to $c^{\infty(1)}$, as well as a photon exchanged in the $t$-channel at or near the forward limit. While the forward amplitudes are not continuous in the gauge couplings at $g=0$ or $g^\prime=0$ (as opposed to the continuity in the non-forward scattering \cite{Weinberg:1996kr}), the resulting sum rules are actually continuous.  

\section{Conclusions and Discussion}
\label{conclusions}

We derived dispersion relations that provide universal sum rules for the $2\rightarrow 2$ forward scattering amplitudes of real particles transforming in a unitary representation $\mathbf{r}=\ovl{\mathbf{r}}$ of an arbitrary internal symmetry group $H$. The sum rules represent identities between an IR side where the amplitudes are presumably calculable, e.g.~within an EFT, and a UV side that encodes information about the asymptotic behavior of the amplitudes at very high energy. 
The theory of GBs living in a coset space $G/H$ represents the typical system where our sum rules can be used to set non-trivial constraints on the low-energy coupling constants in addition to the usual symmetry requirements. But in fact our approach applies also to more general systems, e.g.~with massive and spinning particles, as long as $H$ is a good symmetry linearly  realized on the states. 

The sum rules, aside from the usual ingredients of unitarity, analyticity and crossing symmetry are crucially based on two general properties of the $s\leftrightarrow u$ crossing matrix $X$ that acts on the space of the eigen-amplitudes $\Amp(s)$.
First, the crossing matrix is involutory, $X^2=\mathbbm{1}$. Second, it admits (at least) one  $+1$-eigenvector $v$, $Xv=v$, that for non-degenerate irreps has all identical entries.
Since $X^2=\mathbbm{1}$, one can construct two projectors $P_{\pm}=(\mathbbm{1} \pm X)/2$.
 We showed that the eigen-amplitudes projected on the $+1$-eigenspace admit dispersion relations that can be regarded as a multidimensional generalization of the ordinary dispersion relations for non-symmetric theories (where $X$ is trivial). In practice, along certain directions provided by the eigenvectors of $X$, we are able to recast the usual dispersion relation arguments to prove, e.g., positivity constraints that generalize those found in ref.~\cite{Pham:1985cr,Adams:2006ch,Distler:2006if}.
We provided a systematic and simple way to construct all such positivity constraints for the coefficients $a^{(n)}_I$ with even $n$ of the low-energy expansion of the eigen-amplitudes.

Projecting instead on the $-1$-eigenspace with $P_-$ we studied once-subtracted dispersion relations. The resulting sum rules are very interesting since they can be used to put constraints on the low-energy coupling constants of EFTs at $O(p^2)$.
Under very general assumptions discussed in section~\ref{sec:convergence}, and summarized by the universality condition \eqref{eq:Asymptotic} on the saturation of the Froissart bound, we showed that the sum rules based on once-subtracted dispersion relations are UV convergent.
Indeed, amplitudes that grow maximally fast as $\Amp(s) \sim s\log^2 s$ turn out to be proportional to $v$, i.e.~a $+1$-eigenvector of $X$, and are thus projected out by $P_-$. 
Again, our method allows us to systematically find all the sum rules and the associated constraints on the LECs.

We discussed  several illustrative examples that are relevant for theories of GBs such as $SO(N+1)/SO(N)$ and $SO(N,1)/SO(N)$ that appear in composite Higgs models, and \sloppy\mbox{$SU(N)_L\times SU(N)_R/SU(N)_{L+R}$}  for e.g.~chiral QCD. 
In the context of composite Higgs models respecting the custodial $SO(4)$ symmetry, we obtain two once-subtracted sum rules, see eqs.~\eqref{SO4GB1} and \eqref{integratedPLR} in section~\ref{subsec:SO4}, one of which constrains the operator $\mathcal{O}_H$ of the SILH Lagrangian \cite{Giudice:1024017}.

Finally, we discussed the once-subtracted sum rule for longitudinal $WW$-scattering with finite $g$ and small $g^\prime$, that is in the custodial $SO(3)$ limit of the SM.
We carefully compared the resulting sum rule to the one obtained for GBs of $SO(4)/SO(3)$ in the gauge-less limit. We showed that even though the amplitudes in the forward limit are not continuous in the gauge couplings at $g=g^\prime=0$, the resulting sum rule for GBs does actually carry over to the scattering of longitudinal $W$'s. While the same conclusion could be reached with a naive use of the equivalence theorem, we emphasized that in fact a non-trivial cancellation between two extra contributions on both sides of the sum rule takes place.

There are various directions that are worth exploring further. One immediate option would be  to use the positivity conditions  on the $a^{(2)}_I$ coefficients to set constraints on the dimension six operators that deform the SM but respect custodial symmetry. More speculative directions involve extensions of the space-time symmetry. For example, it would be interesting to look more carefully at the way our arguments adapt to higher or lower dimensions, as well as to curved space-times.
Finally, even though we have restricted our analysis to internal symmetries, it would be very interesting to extend our approach to space-time symmetries such as supersymmetry.

\section*{Acknowledgements}

We thank Massimo Passera, Andrea Wulzer, Anatoly Dymarsky and Roberto Contino for valuable discussions.
B.B thanks Slava Rychkov, Andi Weiler, Alfredo Urbano, Filippo Sala and Borut Bajc  for useful discussions, and Flip Tanedo for reading the manuscript. B.B. would like to express special thanks to the Mainz Institute for Theoretical Physics (MITP) for its hospitality and support. R.T. thanks Adam Falkowski for valuable comments and Christoffer Petersson and Andrea Thamm for fruitful conversations.
B.B. is supported in part by the MIUR-FIRB grant RBFR12H1MW, and by the Agence Nationale de la Recherche under contract ANR 2010 BLANC 0413 01. The work of R.T. was supported by the ERC Advanced Grant no.~267985 {\it DaMeSyFla} and by the Research Executive Agency (REA) of the European Union under the Grant Agreement number PITN-GA- 2010-264564 {\it LHCPhenoNet}. R.T. finally thanks the ``Centro de Ciencias de Benasque Pedro Pascual'' for hospitality during the completion of this work.

\appendix 

\setcounter{tocdepth}{1}

\section{Crossing matrix and its properties}
\label{app:crossingM}
 
In this appendix we explain how to construct the crossing matrix $X$ in eq.~\eqref{eq:crossEigenA} and derive some of its properties.
We start by using the CG coefficients in eq.~\eqref{CGcoeff} to construct the matrix $Q$ with elements
\be 
\label{Qdef}
Q_{I(\xi\xi^\prime)J(\zeta\zeta^\prime)}= \frac{1}{\mathrm{dim}\,\r_I}\sum_{abcd} P_{I(\xi\xi')}^{ab,cd}\,P_{J(\zeta'\zeta)}^{cb,ad}\,,
\ee
where
\be
P_{I(\xi\xi')}^{ab,cd}=\sum_iC^{ab}_{I(\xi) i} \bar C^{cd}_{I(\xi^\prime) i}\,.
\ee
with $\bar C^{ab}_{I(\xi) i}\equiv ( C^{ab}_{I(\xi) i})^*\equiv C^{ab}_{\bar I(\xi) i}$. 
We recall that the CG coefficients are unitary matrices
\be
\sum_{a,b} C^{ab}_{I(\xi)i} \bar{C}^{ab}_{J(\zeta)j}=\delta_{IJ}\delta_{ij}\delta_{\xi\zeta}\,,\qquad \sum_{I,\xi,i}\bar{C}^{ab}_{I(\xi)i} C^{cd}_{I(\xi)i}=\delta^{ac}\delta^{bc}\,.
\ee
In particular $P_{I(\xi\xi)}^{ab,cd}$ can be regarded as the projector onto the subspace ${\bf r}_{I(\xi)}$, expressed in the basis $|ab\rangle$.  
Furthermore, notice that, by definition, 
\be\label{ccQ}
\bar Q_{I(\xi\xi^\prime)J(\zeta\zeta^\prime)}=Q_{I(\xi'\xi)J(\zeta'\zeta)}=Q_{\bar I(\xi\xi')\bar J(\zeta\zeta')}\,.
\ee

\tocless\subsection{Crossing symmetry in non-minimal notation}

The crossing symmetry of eq.~\eqref{eq:crossA} can be written in terms of the eigen-amplitudes ${\cal A}_{I(\xi\xi')}(s)$ appearing in \eqref{WE} as
\be\label{extcross}
{\cal A}_{I(\xi\xi')}(u)=\sum_{J\zeta\zeta^\prime}Q_{I(\xi\xi^\prime)J(\zeta\zeta^\prime)}{\cal A}_{J(\zeta\zeta')}(s)\,.
\ee
Let us organize the eigen-amplitudes ${\cal A}_{I(\xi\xi')}(s)$ in a vector
\be\l{unconstreigamp}
\tilde {\cal A}(s)=\left(\begin{array}{c}\vdots \\
\Amp_{I(\xi\xi')}(s)\\
\vdots
\end{array}\right)
\ee
where the index $I(\xi\xi')$ takes {\em all} possible values. Hence, the vector $\tilde {\cal A}(s)$ differs from the vector ${\cal A}(s)$ introduced in eq.~\eqref{indfree}, since the latter is restricted to eigen-amplitudes which are not related by eq.~\eqref{eq:Amp_real_irrep}. 

The crossing relation \eqref{extcross} can then be written in the compact form
\be\label{excross}
\tilde {\cal A}(u)=Q\tilde {\cal A}(s)\,.
\ee
This equation shows that $Q_{I(\xi\xi^\prime)J(\zeta\zeta^\prime)}$ are the crossing matrix elements for the unconstrained eigen-amplitudes \eqref{unconstreigamp}.
The  matrix $Q$  has the following important properties that follow directly from unitarity of the CG coefficients
\begin{enumerate}

\item $Q$ is involutory:
\be
\label{Qinv}
 Q^2=\mathbbm{1}\,,
 \ee
 and then it has only $\pm1$ eigenvalues;

\item The $+1$-eigenspace of $Q$ contains the vector $\tilde v$ whose components are
\be\label{onevectorext}
\tilde v_{I(\zeta\zeta')}=\left\{\begin{array}{l} 1\quad~~~ \text{if }\zeta=\zeta'\\
0\quad~~~ \text{if }\zeta\neq\zeta'\end{array}\right. \,;
\ee

\item $Q$ is unitary with respect to the diagonal metric $\Delta$ with entries $\Delta_{I(\xi\xi')}=\dim{\bf r}_{I}$:
\be\label{Qunitary}
Q^\dagger\Delta Q=\Delta\,.
\ee
This condition follows from eq.~\eqref{Qinv} and the symmetry $\dim{\bf r}_{I} \bar{Q}_{I(\xi\xi^\prime)J(\zeta\zeta^\prime)}=\dim{\bf r}_{J} Q_{J(\zeta\zeta^\prime)I(\xi\xi^\prime)}$ implied by the definition \eqref{Qdef}.
\end{enumerate}

\noindent The vector $\tilde v$ is related to the vector $v$ defined in eq.~\eqref{onevector}, the only difference being that, as for $\tilde {\cal A}$, its indices $I(\xi\xi')$ are unrestricted. Other properties of $Q$ have been studied in detail in ref.~\cite{1968AnPhy..48..541F}.

In general, we can assign to each irrep ${\bf r}_{I(\xi)}$ appearing in eq.~\eqref{CGseries} a definite $\pm1$ parity under the exchange $|ab\rangle \rightarrow |ba\rangle$ in ${\bf r}\otimes{\bf r}$. We indicate such parity by $(-)^{{\cal I}(\xi)}$, with ${\cal I}(\xi)\in \mathbb{Z}\ ({\rm mod}\ 2)$, and we explicitly identify it by the property: 
\be\l{CGdef}
C^{ab}_{I(\xi)i}=(-)^{{\cal I}(\xi)}C^{ba}_{I(\xi)i}\,.
\ee
The rows and columns of the matrix $Q$ have $q\equiv \dim Q$ indices, labelled by $I(\xi\xi')$. We say that an index $I(\xi\xi')$ is {\em even} if $(-)^{{\cal I}(\xi)}=1$, while an index $I(\xi\xi')$ is {\em odd} if $(-)^{{\cal I}(\xi)}=-1$. We can write $q=q_{+}+q_-$, where $q_\pm$ denote the number of even/odd indices $I(\xi\xi')$.
Reference \cite{1968AnPhy..48..541F} proved that 
\be\label{trace}
\tr\, Q\equiv \sum_{I(\xi\xi')} Q_{I(\xi\xi')I(\xi\xi')}=q_+-q_-\,.
\ee 
On the other hand $\tr\, Q$ is also equal to the difference between the number of $+1$ and $-1$  eigenvalues of $Q$. Hence, $q_\pm$ exactly give the number of $\pm1$ eigenvalues of $Q$. 

\tocless\subsection{Getting rid of redundancies}\l{reductionQX}

The matrix $Q$ and eq.~\eqref{excross} could directly be used to derive dispersion relations and sum rules for the eigen-amplitudes, along the lines followed in the main part of the present paper. On the other hand, the eigen-amplitudes ${\cal A}_{I(\xi\xi')}(s)$ are not all independent and they are related by eq.~\eqref{eq:Amp_real_irrep}. Hence, in general, there is some redundancy in eq.~\eqref{excross} and it is then convenient to eliminate it.

In order to do that, let us first explicitly distinguish the irreps ${\bf r}_{I(\xi)}$ appearing in eq.~\eqref{CGseries} in three independent sets,  ${\bf r}_{I_{\rm r}(\xi)}$, ${\bf r}_{I_{\rm c}(\xi)}$ and ${\bf r}_{\bar I_{\rm c}(\xi)}$, where $I_{\rm r}$ and $I_{\rm c}$ label real and complex representations respectively. (The choice of the separation of the complex representations into ${\bf r}_{I_{\rm c}(\xi)}$ and ${\bf r}_{\bar I_{\rm c}(\xi)}$ is of course arbitrary.) Then, we can correspondingly group the indices $I(\xi\xi')$ in three sets. The first set contains the indices $I_{\rm r}(\xi\xi)$. The second set contains the indices $I_{\rm r}(\xi\xi')$ with $\xi<\xi'$ and the indices $I_{\rm c}(\xi\xi')$. The third set contains the indices  $I_{\rm r}(\xi'\xi)$  and the indices $\bar I_{\rm c}(\xi'\xi)$, by using the same ordering of $\xi$ and $\xi'$ used for the second set. 
Using this subdivision the vector $\tilde{\cal A}(s)$ splits in three subvectors ${\cal A}^1(s),{\cal A}^2(s),{\cal A}^3(s)$ as follows
\be
\tilde {\cal A}(s)=\left(\begin{array}{c}{\cal A}^1(s) \\
{\cal A}^2(s)\\
{\cal A}^3(s)
\end{array}\right)\,.
\ee
Correspondingly, we can write the block-decomposition of the matrices $Q$ and $\Delta$ according to the above index subdivision
\be\label{Qdec}
Q=\left(\begin{array}{rrr}Q^{11}& Q^{12}& Q^{13} \\
Q^{21} & Q^{22}& Q^{23}\\
Q^{31}& Q^{32} & Q^{33} 
\end{array}\right)\,,\quad~~~~\Delta=\left(\begin{array}{rrr}\Delta^1 & 0&0 \\
0 & \Delta^2 & 0\\
0&  0 &\Delta^3
\end{array}\right)\,.
\ee
Notice that $\Delta^2=\Delta^3$ and that, by eq.~\eqref{ccQ}, 
\be\label{Qrel1}
Q^{12}=Q^{13}\,,\quad Q^{21}=Q^{31}\,,\quad Q^{23}=Q^{32}\,,\quad Q^{22}=Q^{33}\,.
\ee

The relation \eqref{eq:Amp_real_irrep} now reads ${\cal A}^2(s)= {\cal A}^3(s)$. 
We can therefore consider the minimal amplitude vector
\be
{\cal A}(s)=\left(\begin{array}{c}{\cal A}^1(s) \\
{\cal A}^2(s)
\end{array}\right)\,,
\ee
already introduced in eq.~\eqref{indfree}.  The crossing relation \eqref{excross} can be written in terms of the vector ${\cal A}(s)$ as
\be\label{excross3}
{\cal A}(u)=X {\cal A}(s)\,,
\ee
where $X$ is the matrix
\be\l{matrixXfromQ}
X=\left(\begin{array}{cc}Q^{11}& Q^{12}+Q^{13} \\
Q^{21} & Q^{22}+Q^{23}
\end{array}\right)\,.
\ee
This is the crossing matrix introduced in section \ref{sumrules}, which plays a crucial role in the present paper. 
The matrix $X$ inherits some properties from those of $Q$, as one can check by direct inspection. They have been already mentioned in section \ref{sumrules}, see eqs.~\eqref{eq:crossEigenA} and \eqref{onevector}. Furthermore $X$ is unitary with respect to the diagonal metric
\be\label{Gmetric}
{\cal G}=\left(\begin{array}{cc}\Delta^1 & 0 \\
0 & 2\Delta^2 
\end{array}\right)\,,
\ee
which means
 \be\label{Gunitary}
 X^\dagger {\cal G}X={\cal G}\,.
 \ee

\tocless\subsection{Reducibility of the crossing matrix $X$}
In this subsection we discuss some other properties of the matrix $X$, useful to determine whether it takes the block-diagonal form \eqref{blockdiag}.

From the definition of $Q$ in eq.~\eqref{Qdef} and the property \eqref{CGdef}, it immediately follows that
\be
Q_{I(\xi\xi')J(\zeta\zeta')}=(-)^{{\cal I}(\xi)+{\cal I}(\xi')+{\cal J}(\zeta)+{\cal J}(\zeta')}Q_{I(\xi\xi')\bar J(\zeta'\zeta)}\,,
\ee  
where $(-)^{{\cal I}(\xi)}$ denotes the parity of ${\bf r}_{I(\xi)}$ defined above eq.~\eqref{CGdef}. This in turn implies that
\be
X_{I(\xi\xi')J(\zeta\zeta')}=(-)^{{\cal I}(\xi)+{\cal I}(\xi')+{\cal J}(\zeta)+{\cal J}(\zeta')} X_{I(\xi\xi')J(\zeta\zeta')}\,.
\ee 
From this identity  we  see  that $X_{I(\xi\xi')J(\zeta\zeta')}=0$ if $(-)^{{\cal I}(\xi)+{\cal I}(\xi')+{\cal J}(\zeta)+{\cal J}(\zeta')}=-1$ and, in particular, we have
\be\label{Xparity}
X_{I(\xi\xi')J(\zeta\zeta)}=X_{J(\zeta\zeta)I(\xi\xi')}=0 \quad~~~~~~~~~\text{if}\quad~~~~~(-)^{{\cal I}(\xi)}=-(-)^{{\cal I}(\xi')}\,.
\ee
This provides a useful restriction on the components of $X$. For instance, it often happens that degenerate irreps appear only in pairs of opposite parity (see, e.g., the $\mathbf{D}$ and $\mathbf{F}$ representations in the decomposition of the product of adjoints of $SU(N)$ in eq.~\eqref{AdjxAdj}). This immediately implies that $X$ has the block-diagonal structure \eqref{blockdiag}.

Another possible restriction on the structure of $Q$, and then of $X$, could come from an additional (possibly discrete) symmetry group $K$ which commutes with the symmetry group $H$. 
To make this argument more concrete, suppose that $K$ is a $U(1)$ symmetry that acts as $|I(\xi),i\rangle \to\e^{\i q_{I(\xi)}\theta}|I(\xi),i\rangle$, where $\theta$ is the $U(1)$ angle. Then from the definition of $Q$ in eq.~\eqref{Qdef} we get
\be
Q_{I(\xi\xi')J(\zeta\zeta')}=\e^{\i(q_{I(\xi)}-q_{I(\xi')}+q_{J(\zeta')}-q_{J(\zeta)})}Q_{I(\xi\xi')J(\zeta\zeta')}\,.
\ee
Hence, if for instance $q_{I(\xi)}\neq q_{I(\xi')}$ for $\xi\neq \xi'$, then $Q_{I(\xi\xi')J(\zeta\zeta)}=Q_{J(\zeta\zeta)I(\xi\xi')}=0$. This clearly implies that $X_{I(\xi\xi')J(\zeta\zeta)}=X_{J(\zeta\zeta)I(\xi\xi')}=0$ too. In such a case  $X$ has the block diagonal structure \eqref{blockdiag}.

\vspace{7mm}
\section{Analytic structure with light unstable resonances}
\label{AnalyticLR}

\begin{figure}[t!]
\begin{center}
\includegraphics[scale=0.5]{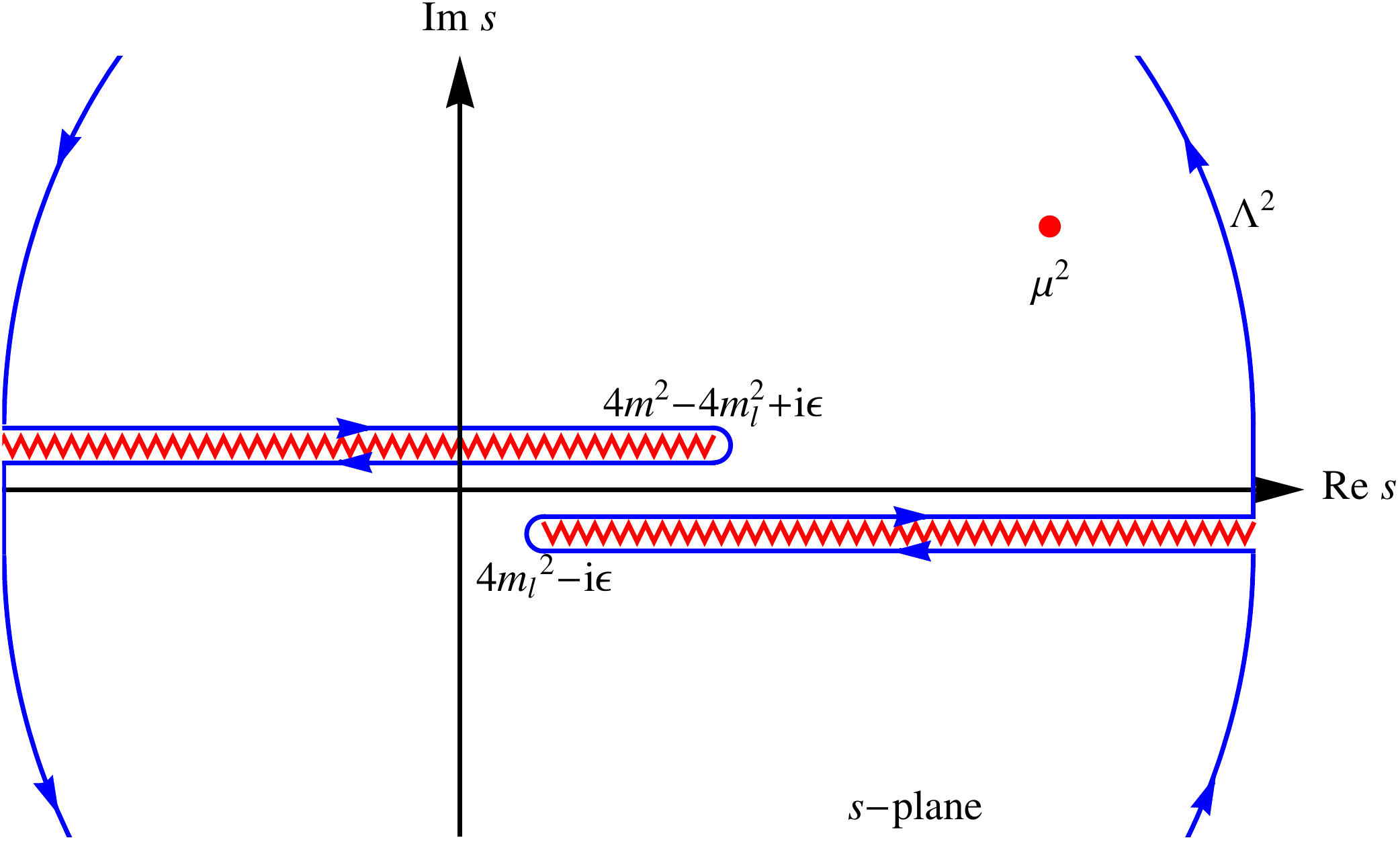}
\caption{Analytic structure in the presence of light particles of mass $m_{l}$.}\label{fig:path3}
\end{center}
\end{figure}

Light poles can turn into unstable resonances in presence of extra light states of mass $m_{\ell}^2$ in which they can decay. When this happens the poles move to the unphysical Riemann and the branch cut extends down to the masses of the light states $4m_{\ell}^2$.  The analytic structure of the amplitude at $t=0$ in this case is depicted in figure \ref{fig:path3}, where the cuts extend from $s=-\infty+\i\epsilon$ to $s=4m^2-4m_\ell^2+\i\epsilon$, and from $s=4m_{\ell}^2-\i\epsilon$ to $s=+\infty-\i\epsilon$. Notice that here the $\i \epsilon$ prescription is important to ensure the correct cross symmetric structure of the amplitude. The dispersion relation involves now an extra unphysical region of integration, namely from $4m_{\ell}^{2}$ to $4m^{2}$:  
\begin{align}
 \Amp^{(n)}(\mu^2)=
c^{\Lambda\,(n)}+ \int_{4m_\ell^2}^{\Lambda^2+2m^2} \frac{\di s}{2\PI \i} \left[\frac{1}{(s-\mu^2)^{n+1}}+(-1)^n \frac{X}{(s-4m^2+\mu^2)^{n+1}} \right]\left[\Amp(s+\i\epsilon)-\Amp(s-\i\epsilon) \right]\,.
\label{eq:dispersionApp}
\end{align}

Moreover, when massless particles are present, e.g.~when $m_\ell=0$, the dispersion relations for finite $t\neq0$ may be needed, as discussed in appendix~\ref{beyondforward}. For $m_\ell=0$ and finite $t$ the cuts go from $s=0-\i\epsilon$ to $s=+\infty-\i\epsilon$, and from $s=-\infty+\i\epsilon$ to $s=4m^2-t+\i\epsilon$.

In all such cases, we can discard the unphysical region of integration $4m_\ell^2< s< 4m^2$ as long as the widths are narrow compared to $m_i$, $\mu$ and $\Lambda$. This is equivalent to working at leading order in the couplings that make the resonance unstable. For example, the SM Higgs has a tiny width dominated by the small bottom Yukawa coupling, and we can basically approximate the analytic structure with a pole on the real axis at $s=m_h^2$, which anyway gives a negligible contribution to the sum rules  when $\mu^2\gg m_i^2$.

\section{Beyond the forward limit}
\label{beyondforward}

When massless particles can be exchanged in the $t$-channel one cannot strictly consider the forward limit $t=0$ because of the Coulomb singularity.
While this problem does not arise for GBs, it may be relevant e.g.~for massless gauge bosons such as the photon.
In such a case one can work at fixed and finite $t\neq0$ and/or add an IR regulator like a mass term.

Consider first the contour integral at fixed and finite $t$
\be 
\label{eq:Cauchy2}
\frac{1}{2\PI \i}\oint _\mathcal{C} \frac{\Amp(s,t)}{(s-\mu^2)^{n+1}}= \sum_{s_{i}}\mathrm{Res}\left[\frac{\Amp(s,t)}{(s-\mu^2)^{n+1}}\right]+ \Amp^{(n)}(\mu^2,t)\,,
\ee
around the cuts running from $s=4m^2$ to $+\infty$, and from $s=-\infty$ to $-t$, by $s\leftrightarrow u=-s-t+4m^2$ crossing.
The case with unstable resonances below the $4m^2$ slightly changes the analytic structure as we discussed in appendix \ref{AnalyticLR}. 
Adapting the arguments presented for $t=0$, we get the dispersion relations 
\begin{subequations}
\begin{align}\l{sumrulesfinitet}
P_{-}\sum \mbox{ (residues)}^{(n)}=& 
 \int_{4m^2}^{\infty} \frac{\di s}{\PI} \left[\frac{1}{(s-\mu^2)^{n+1}}-(-1)^n \frac{1}{(s+t-4m^2+\mu^2)^{n+1}} \right]P_{-}\mathrm{Im}\Amp(s+\i\epsilon,t)\\
P_{+}\sum \mbox{ (residues)}^{(n)}=& 
 \int_{4m^2}^{\infty} \frac{\di s}{\PI} \left[\frac{1}{(s-\mu^2)^{n+1}}+(-1)^n \frac{1}{(s+t-4m^2+ \mu^2)^{n+1}} \right]P_{+}\mathrm{Im}\Amp(s+\i\epsilon,t) 
\end{align}
\end{subequations}
where the convergence for $n>1$ is guaranteed by the Froissart bound for $t\neq0$  \cite{Froissart:1961ux}
\be 
|\Amp(s,t\neq0)|\leq \text{const}\times\,\f{s\log^{\frac{3}{2}} s}{(-t)^{\f{1}{4}}}\,,\qquad \text{for }s\to\infty\,.
\ee
For $n=1$ the convergence could be spoilt by amplitudes that grow maximally fast. However, as we discussed in section~\ref{sec:convergence}  for $t=0$,  if the amplitudes  grow maximally fast in a universal way, that is as 
\begin{equation}
\Amp(s,t\neq0)\sim \mathrm{const} \times s \log^{\frac{3}{2}} s\,  \delta_{\xi\xi^\prime} 
\end{equation}
with $\mathrm{const}$ independent of $I(\xi\xi')$, then the integral in eq.~\eqref{sumrulesfinitet} is UV convergent, and the contribution $c^{\infty(1)}$ from the big circle is projected out by $P_{-}$. Thus the once-subtracted sum rule  holds too.

We can also take the limit of $\mu^2$ much larger than all IR scales, including $\mu^2\gg t$, so that, e.g., the first sum rule in eq.~\eqref{sumrulesfinitet} for $n=1$ can be expressed as
 \be 
  \label{sumrule_tfinite}
 P_{-} \Amp^{(1)}(s=\mu^2,t)=\frac{2}{\PI}
\int_{4m^2}^{\infty} \di s\,  \frac{(s^2+\mu^4)}{(s^2-\mu^4)^2} P_{-}\mathrm{Im}\Amp(s+\i\epsilon,t)+O(\frac{m^2}{\mu^2}, \frac{m_i^2}{\mu^2},\frac{t}{\mu^2})\,.
 \ee

In order to relate the non-forward imaginary amplitude to the physical cross-sections one make an expansion in partial waves that, e.g.~for spin-0 particles, reads 
\begin{equation}
\Amp(t,s)=\sum_\ell (2\ell+1)P_\ell(1+2t/(s-4m^2))\Amp^{\ell}(s)\,,
\end{equation}
 where $P_\ell(\cos\theta)$ are the Legendre polynomials and $\mathrm{Im}\Amp^{\ell}(s)=s \sigma^\ell(s)\sqrt{1-\f{4m^{2}}{s}}  $.

\section{Projectors and crossing matrix for $SO(N)$ and $SU(N)$}\l{AppProjectors}

In this appendix we construct the matrices  $Q$, $X$ and $\hat{X}$ for the product of fundamentals of $SO(N\neq 4)$ and $SU(N\geq4)$.

\tocless\subsection{$SO(N)$}\l{XforSON}
In the  product of two fundamentals of $SO(N)$, ${\bf N}\otimes {\bf N}={\bf 1}\oplus {\bf S}\oplus {\bf A}$, no degenerate irreps appear and then $Q=X=\hat{X}$, which can be constructed by eq.~\eqref{Qdef} with the projectors 
\be\l{projSON}
\begin{aligned}
P_{\bf 1}^{ab,cd}&=\frac1N\,\delta^{ab}\delta^{cd}\,,\\
P_{\bf S}^{ab,cd}&=\frac12(\delta^{ac}\delta^{bd}+\delta^{ad}\delta^{bc})-\frac1N\,\delta^{ab}\delta^{cd}\,,\\
P_{\bf A}^{ab,cd}&=\frac12(\delta^{ac}\delta^{bd}-\delta^{ad}\delta^{bc})\,.
\end{aligned}
\ee 
The result is given in eq.~\eqref{Xson}.

\tocless\subsection{$SU(N)$}\l{XforSUN}
In the product \eqref{AdjxAdj} of two adjoint representations of $SU(N)$ there appear degenerate irreps and the matrix $X$ is given by the consistent reduction of the matrix $Q$ discussed in appendix \ref{reductionQX}. In order to compute the matrices $P_{I(\xi\xi')}^{ab,cd}$ appearing in eq.~\eqref{Qdef} we follow the conventions
\be\l{SUNGener1}
[T^{a},T^{b}]=if^{abc}T^{c}\,,\quad \tr(T^{a}T^{b})=\f{\delta^{ab}}{2}\,,\quad \{T^{a},T^{b}\}=\f{1}{n}\delta^{ab}\mathbbm{1}_{n}+d^{abc}T^{c}\,,\quad d^{ace}d^{bce}=\f{N^{2}-4}{N}\delta^{ab}\,,
\ee
which imply
\be\l{SUNGener2}
f^{abc}=-2i\tr([T^{a},T^{b}]T^{c})\,,\qquad d^{abc}=2\tr(\{T^{a},T^{b}\}T^{c})\,,
\ee
for the $SU(N)$ generators in the fundamental representation $T^{a}$. Other useful relations are
\be\l{trace4SUN}
\bry{l}
\dst \tr\(T^{a}T^{b}T^{c}\)=\f{1}{4}(\i f^{abc}+d^{abc})\,,\vspace{2mm}\\
\dst \tr\(T^{a}T^{b}T^{c}T^{d}\)= \f{1}{4N}\delta^{ab}\delta^{cd}+\f{1}{8}(\i f^{abe}+d^{abe})(\i f^{cde}+d^{cde})\,,\vspace{2mm}\\
\dst f^{abe}f^{cde}=d^{ace}d^{bde}-d^{ade}d^{bce}+\f{2}{N}\(\delta^{ac}\delta^{bd}-\delta^{ad}\delta^{bc}\)\,,\vspace{2mm}\\
\dst d^{abe}f^{bce}+d^{ace}f^{bde}+d^{cde}f^{bae}=0\,.
\ery
\ee
The matrices $P_{I(\xi\xi')}^{ab,cd}$ are then given by\footnote{The last term in the projector $P_{\bf T}^{ab,cd}$ differs from the one of ref.~\cite{1992hep.ph....6222S}, which does not square to one (presumably due to a typo in their equation).}
\bes\l{projSUN}
\small\begin{align}
P_{\bf 1}^{ab,cd}&=\frac{\delta^{ab}\delta^{cd}}{N^{2}-1}\,,\\
P_{\bf D}^{ab,cd}&=\frac{N}{N^2-4} d^{abe}d^{cde}\,,\\
P_{\bf DF}^{ab,cd}&=-\frac{1}{\sqrt{N^2-4}} d^{abe}f^{cde}\,,\\
P_{\bf FD}^{ab,cd}&=-\frac{1}{\sqrt{N^2-4}} f^{abe}d^{cde}\,,\\
P_{\bf F}^{ab,cd}&=\frac{f^{abe}f^{cde}}{N}\,,\\
P_{\bf Y}^{ab,cd}&=\f{N-2}{4N}\(\delta^{ac}\delta^{bd}+\delta^{ad}\delta^{bc}\)+\f{N-2}{2N(N-1)}\delta^{ab}\delta^{cd}-\f{1}{4}\(d^{ace}d^{bde}+d^{ade}d^{bce}\)+\f{N-4}{4(N-2)}d^{abe}d^{cde}\,,\\
P_{\bf T}^{ab,cd}&=\f{N^{2}-4}{4N^{2}}\(\delta^{ac} \delta^{bd}-\delta^{ad}\delta^{bc}\)-\f{1}{2N}\(d^{ace}d^{bde}-d^{ade}d^{bce}\)-\f{\i}{4}\(d^{bce} f_{ade}+ d^{ade} f_{bce}\) \,,\\
P_{\bf \ovl{T}}^{ab,cd}&=(P_{\bf T}^{ab,cd})^{*}\,,\\
P_{\bf X}^{ab,cd}&=\f{N+2}{4N}\(\delta^{ac}\delta^{bd}+\delta^{ad}\delta^{bc}\)-\f{N+2}{2N(N+1)}\delta^{ab}\delta^{cd}+\f{1}{4}\(d^{ace}d^{bde}+d^{ade}d^{bce}\)-\f{N+4}{4(N+2)}d^{abe}d^{cde}\,.
\end{align}\normalsize
\ees
Using these matrices we can construct the matrix $Q$ as in eq.~\eqref{Qdef} getting $Q=$
\be\l{matrixXSUN}
\hspace{-4mm}
\scriptsize\(
\begin{array}{ccccccccc}
 \f{1}{N^2-1} & 1 & 0 & 0 & -1 & \f{(N-3) N^2}{4 (N-1)} & 1-\f{N^2}{4} & 1-\f{N^2}{4} & \f{N^2 (N+3)}{4 (N+1)} \\
 \f{1}{N^2-1} & \f{N^2-12}{2 \(N^2-4\)} & 0 & 0 & -\f{1}{2} & -\f{(N-3) N^2}{4 (N-2) (N-1)} & \f{1}{2} &
   \f{1}{2} & \f{N^2 (N+3)}{4 (N+1) (N+2)} \\
 0 & 0 & -\f{1}{2} & -\f{1}{2} & 0 & 0 & \f{1}{4} \i \sqrt{N^2-4} & -\f{1}{4} \i \sqrt{N^2-4} & 0 \\
 0 & 0 & -\f{1}{2} & -\f{1}{2} & 0 & 0 & -\f{1}{4} \i \sqrt{N^2-4} & \f{1}{4} \i \sqrt{N^2-4} & 0 \\
 \f{1}{1-N^2} & -\f{1}{2} & 0 & 0 & \f{1}{2} & -\f{(N-3) N}{4 (N-1)} & 0 & 0 & \f{N (N+3)}{4 (N+1)} \\
 \f{1}{N^2-1} & \f{1}{2-N} & 0 & 0 & -\f{1}{N} & \f{1}{N-2}+\f{1}{4}+\f{1}{2-2 N} & \f{N+2}{4 N} & \f{N+2}{4
   N} & \f{N+3}{4 N+4} \\
 \f{1}{1-N^2} & \f{2}{N^2-4} & -\f{\i}{\sqrt{N^2-4}} & \f{\i}{\sqrt{N^2-4}} & 0 & \f{(N-3) N}{4 \(N^2-3 N+2\)} &
   \f{1}{4} & \f{1}{4} & \f{N (N+3)}{4 \(N^2+3 N+2\)} \\
 \f{1}{1-N^2} & \f{2}{N^2-4} & \f{\i}{\sqrt{N^2-4}} & -\f{\i}{\sqrt{N^2-4}} & 0 & \f{(N-3) N}{4 \(N^2-3 N+2\)} &
   \f{1}{4} & \f{1}{4} & \f{N (N+3)}{4 \(N^2+3 N+2\)} \\
 \f{1}{N^2-1} & \f{1}{N+2} & 0 & 0 & \f{1}{N} & \f{N-3}{4 (N-1)} & \f{N-2}{4 N} & \f{N-2}{4 N} & \f{N^2+N+2}{4
   N^2+12 N+8}
\end{array}
\)\,,\normalsize
\ee
where the representations are ordered as they appear in eqs.~\eqref{projSUN}. This $9\times 9$ matrix can now be consistently reduced to a $7\times 7$ block diagonal matrix $X$ with the prescriptions of appendix \ref{reductionQX}. These prescriptions here can be simply implemented by summing the two columns corresponding to the $\mathbf{DF}$ and $\mathbf{FD}$ entries (columns $3$ and $4$) and the two columns corresponding to the $\mathbf{T}$ and $\mathbf{\ovl{T}}$ entries (columns $7$ and $8$) and by removing one of each of the two equal rows so obtained. In this way one gets the block diagonal $X$ matrix
\be\l{matrixXSUNreduced7x7}
X=\(
\begin{array}{ccccccc}
 -1 & 0 & 0 & 0 & 0 & 0 & 0\vspace{2mm}\\
 0 &\f{1}{N^2-1} & 1 & -1 & \f{(N-3) N^2}{4 (N-1)} & 2-\f{N^2}{2} & \f{N^2 (N+3)}{4 (N+1)} \vspace{2mm}\\
 0 &\f{1}{N^2-1} & \f{N^2-12}{2 \(N^2-4\)} & -\f{1}{2} & -\f{(N-3) N^2}{4 (N-2) (N-1)} & 1 & \f{N^2 (N+3)}{4 (N+1) (N+2)} \vspace{2mm}\\
 0 &\f{1}{1-N^2} & -\f{1}{2} & \f{1}{2} & -\f{(N-3) N}{4 (N-1)} & 0 & \f{N (N+3)}{4 (N+1)} \vspace{2mm}\\
 0 &\f{1}{N^2-1} & \f{1}{2-N} & -\f{1}{N} & \f{1}{N-2}+\f{1}{4}+\f{1}{2-2 N} & \f{N+2}{2 N} & \f{N+3}{4 N+4} \vspace{2mm}\\
 0 &\f{1}{1-N^2} & \f{2}{N^2-4} & 0 & \f{(N-3) N}{4 \(N^2-3 N+2\)} & \f{1}{2} & \f{N (N+3)}{4 \(N^2+3 N+2\)} \vspace{2mm}\\
 0 &\f{1}{N^2-1} & \f{1}{N+2} & \f{1}{N} & \f{N-3}{4 (N-1)} & \f{N-2}{2 N} & \f{N^2+N+2}{4N^2+12 N+8}
\end{array}
\)\,,
\ee
where the first block containing only the $-1$ entry corresponds to the $\mathbf{DF}$ (or equivalently $\mathbf{FD}$) mixed entry while the block $6\times 6$ corresponds to the non-degenerate representations in the order $\mathbf{1},\mathbf{D},\mathbf{F},\mathbf{Y},\mathbf{T},\mathbf{X}$.

\section{$W_{L}W_{L}\to W_{L}W_{L}$ scattering amplitude}\l{WWscattering}

The scattering of two longitudinal $W\in \mathbf{3}$ of $SO(3)$ can be written as
\be
\label{WWAmp}
\mathcal{A}\(W_{L}^{a}W_{L}^{b}\to W_{L}^{c}W_{L}^{d}\)=A_s\(s,t,u\)\delta^{ab}\delta^{dc}+A_t\(t,s,u\)\delta^{ac}\delta^{bd}+A_u\(u,t,s\)\delta^{ad}\delta^{bc}\,,
\ee
where $u=4m_W^2-s-t$. The functions $A_{s,t,u}$ are related by crossing symmetry. For $t=0$, crossing symmetry simply acts as $s\leftrightarrow u$ and $b\leftrightarrow d$, that is $A_s\(s,0,u\)=A_u\(s,0,u\)$ and \mbox{$A_t\(0,s,u\)=A_t\(0,u,s\)$}.
The full tree-level amplitude appearing in eq.~\eqref{WWAmp} can be written in the c.o.m. frame as
\bes
\l{allchan}
\begin{align}
& \begin{array}{lll}\l{schan}
\dst A_s(s,t,u)&=&\dst -\f{a^2 \(s-2m_{W}^2\)^2}{v^{2}\(s-m_{h}^2\)}+\f{1}{v^2\(s-4 m_{W}^2\)^2 \(m_{W}^2-t\) \(s+t-3 m_{W}^2\)}\vspace{2mm}\\
&&\dst \times\bigg[768m_{W}^{10}-128 m_{W}^8 (5 s+4 t)+32 m_{W}^6\(7 s^2+8 s t+4 t^2\)\vspace{1mm}\\
&&\dst -8 m_{W}^4 s \(5 s^2+11s t+4 t^2\)+m_{W}^2 s^2 \(3 s^2+18 s t+14 t^2\)-s^3 t (s+t)\bigg]\,,
\end{array}\\
& \begin{array}{lll}\l{tchan}
\dst A_t(t,s,u)&=&\dst \f{a^{2} \(s t+2 m_{W}^2 s-8 m_{W}^4\)^2}{v^{2}\(m_{h}^2-t\)\(s-4 m_{W}^2\)^2}+\f{1}{v^{2} \(s-4 m_{W}^2\)^2\(s-m_{W}^2\) \(s+t-3 m_{W}^2\)}\vspace{2mm}\\
&&\dst\times \bigg[-768 m_{W}^{10}+64 m_{W}^8 (4 s+9 t)+16m_{W}^6 \(3 s^2+3 s t-8 t^2\)\vspace{1mm}\\
&&\dst -8 m_{W}^4 s (s+t) (3 s+4 t)+m_{W}^2 s^2 \(2 s^2-2 s t-3 t^2\)+s^3 t(s+t)\bigg]\,,
\end{array}
\end{align}
\ees
\bes
\begin{align}
& \begin{array}{lll}\l{uchan}
\dst A_u(u,t,s)&=&\dst \f{a^{2} \(8 m_{W}^4-6 m_{W}^2 s+s (s+t)\)^2}{\(s+t-4 m_{W}^2+m_{h}^2\)v^2 \(s-4 m_{W}^2\)^2}-\f{1}{v^2 \(s-4 m_{W}^2\)^2\(s-m_{W}^2\) \(m_{W}^2-t\)}\vspace{2mm}\\
&&\dst\times \bigg[512 m_{W}^{10}-64 m_{W}^8 (6 s+7 t)+16 m_{W}^6 \(9 s^2+3 s t+8 t^2\)\vspace{1mm}\\
&&\dst-16 m_{W}^4 s \(2 s^2+s t-2 t^2\)+3 m_{W}^2 s^2 \(s^2+4 s t+t^2\)-s^3 t (s+t)\bigg]\,.
\end{array}
\end{align}
\ees
The first one agrees with the one computed in ref.~\cite{Espriu:2014jya} but for  the sign of the last two terms in the last line (presumably due to a typo in their equation).

 The function $A_s(s,t,u=-s-t+4m_W^2)$ at fixed $t$ has poles at $s=m_{h}^{2}$, $s=3m_{W}^{2}-t$ and $s=4m_{W}^{2}$, while $A_u(u=-s-t+4m_W^2,t,s)$ has poles at $s=-t+4m_{W}^{2}-m_{h}^{2}$, $s=m_{W}^{2}$ and $s=4m_{W}^{2}$. 
 
 The amplitudes can now be decomposed in eigen-amplitudes of $\mathbf{1}$, $\mathbf{3}$, and $\mathbf{5}$ as follows
\be
\label{eigendecWW}
\Amp_\mathbf{1}=3A_s\(s,t,u\)+A_t\(t,s,u\)+A_u\(u,t,s\)\,,\qquad \Amp_{\mathbf{3},\mathbf{5}}=A_t\(t,s,u\)\mp A_u\(u,t,s\)\,.
\ee
From these eigen-amplitudes and the first row of the matrix $M$ in eq.~\eqref{MSON}
 with $N=3$ we see that the combination of amplitudes and residues that enter on the left-hand side of the sum rules is given by 
\be\l{lefthandside}
[M\Amp^{(1)}(\mu^2,t=0)]_1+\sum_{s_{i}}\mathrm{Res}\left[\frac{[M\Amp^{(1)}(s,t=0)]_1}{(s-\mu^2)^{2}}\right]\,,
\ee
where $\Amp=(\Amp_\mathbf{1},\Amp_{\mathbf{5}},\Amp_\mathbf{3})^T$ and 
\begin{equation}
[M\Amp(s,t=0)]=\left(\begin{array}{c}
\frac{1}{2}\(A_s(s,0,u)-A_u(u,0,s)\)\\
A_t(0,s,u)- \frac{1}{2}\(A_s(s,0,u)+A_u(u,0,s)\)\\
A_t(0,s,u)+\frac{1}{2}\(A_s(s,0,u)+A_u(u,0,s)\)
\end{array}
\right)\,.
\end{equation}
Notice that at $t=0$ the residues in $s=4m_{W}^{2}$ vanish. The other residues  in eq.~\eqref{lefthandside} give, for $t=0$,
\be\l{residues}
\bry{l}
\dst \text{Res}_{s=m_{W}^{2}}=-\frac{27 m_{W}^{4}}{2v^{2} \(\mu^{2}-m_{W}^{2}\)^2}\vspace{2mm}\\
\dst \text{Res}_{s=m_{h}^{2}}=\frac{a^{2} \(m_{h}^{2}-2 m_{W}^{2}\)^{2}}{2v^{2}\(\mu^{2}-m_{h}^{2}\)^2}\vspace{2mm}\\
\dst \text{Res}_{s=4m_{W}^{2}-m_{h}^{2}}=-\f{a^2 \(m_{h}^2-2 m_{W}^2\)^2}{2v^2 \(\mu^2-4 m_{W}^2+m_{h}^2\)^{2}}\vspace{2mm}\\
\dst \text{Res}_{s=3m_{W}^{2}}=\f{27 m_{W}^{4}}{2v^{2}\(\mu^{2}-3 m_{W}^{2}\)^{2}}\,,
\ery
\ee
while the derivative of the amplitude computed at $s=\mu^{2}$ and $t=0$ gives
\be\l{derivativepole}
\bry{lll}
[M\Amp^{(1)}(\mu^2,t=0)]_1 &=&\dst \f{4 a^2 \(\mu ^2-2 m_{W}^2\)^2 \(-\mu ^4+3m_{h}^4-12 m_{h}^2 m_{W}^2+8 m_{W}^4+4 \mu ^2m_{W}^2\)}{4 v^2\(m_{h}^2-\mu ^2\)^2 \(\mu ^2+m_{h}^2-4m_{W}^2\)^2}\vspace{2mm}\\
&&\dst -\f{12 \(-\mu ^8+36 m_{W}^8-12 \mu ^2m_{W}^6-13 \mu ^4 m_{W}^4+8 \mu ^6 m_{W}^2\)}{4 v^2\(\mu ^4+3m_{W}^4-4 \mu ^2 m_{W}^2\)^2}\,.
\ery
\ee
By expanding eqs.~\eqref{residues} and \eqref{derivativepole} we get
\be
\bry{rcl}
\dst \lim_{\mu\gg m_{W}^{2},m_{h}^{2}}  [M\Amp^{(1)}(\mu^2,t=0)]_1
&=&\dst \f{3-a^{2}}{v^{2}}+O\(\f{m_{h}^{2}}{\mu^{2}},\f{m_{W}^{2}}{\mu^{2}}\)\,,\vspace{2mm}\\
\dst \lim_{\mu\gg m_{W}^{2},m_{h}^{2}} \sum_{s_{i}}\mathrm{Res}\left[\frac{ [M\Amp(s,t=0)]_1}{(s-\mu^2)^{2}}\right] &=&\dst O\(\f{m_{h}^{2}}{\mu^{2}},\f{m_{W}^{2}}{\mu^{2}}\)\,.
\ery
\ee
Intriguingly, these corrections $O(m_{h}^{2}/\mu^{2})$ and $O(m_{W}^{2}/\mu^{2})$ actually cancel in the sum \eqref{lefthandside} yielding 
\be\l{lefthandsidefinal}
 [M\Amp^{(1)}(\mu^2,t=0)]_1+\sum_{s_{i}}\mathrm{Res}\left[\frac{ [M\Amp(s,t=0)]_1}{(s-\mu^2)^{2}}\right]=\f{3-a^{2}}{v^{2}}
\ee
as exact result.
From eq.~\eqref{lefthandsidefinal} one obtains the sum rule \eqref{sumrulebeforephotonsubtraction} that, after subtraction of the contributions form the big circle at infinity for finite $g\ll 1$, gives eq.~\eqref{sumrulegprime=0}. 
Notice that eq.~\eqref{lefthandsidefinal}, i.e.~the left-hand side of the sum rule \eqref{sumrulebeforephotonsubtraction}, does not depend explicitly on $\mu^{2}$, whereas the right-hand side does.  Therefore, our sum rule \eqref{sumrulegprime=0}  captures information about the radiative corrections, i.e.~about the running of the couplings and their $\beta$-functions.

Analogously, one can consider the other two once-subtracted dispersion relations~\eqref{sr1subtrconstr}
\be\label{seconandthirdsumruleforSO3}
\bry{lll}
0&=&\dst [M\Amp^{(1)}(\mu^2,t=0)]_{2,3}+\sum_{s_{i}}\mathrm{Res}\left[\frac{ [M\Amp(s,t=0)]_{2,3}}{(s-\mu^2)^{2}}\right]\vspace{2mm}\\
&=&\dst  \frac{2}{\PI}
\int_{4m_{W}^{2}}^{\infty} \di s\,  \frac{2(\mu^2-2m_W^2)(s-2m_W^2)s}{(s-\mu^2)^2(s-4m_W^2 +\mu^2)^2}\sqrt{1-\frac{4m_W^2}{s}}  \,[M \sigma^{\mathrm{tot}}(s)]_{2,3}\,.
\ery
\ee
As for the previous sum rule, the left-hand side turns out to be $\mu^2$ independent within our tree-level calculation, and vanishing. At the crossing symmetric point $\mu^2=2m_W^2$, the residues and $[M\Amp^{(1)}]_{2,3}$ on the left-hand side are separately vanishing 
\begin{equation}
\Amp^{(1)}_{\mathbf{5}}(2m_W^2)+\Amp^{(1)}_{\mathbf{3}}(2m_W^2)=0\,,\qquad \Amp^{(1)}_{\mathbf{1}}(2m_W^2)+2\Amp^{(1)}_{\mathbf{5}}(2m_W^2)=0\,,
\end{equation}
as expected by crossing symmetry, and confirmed by eq.~\eqref{seconandthirdsumruleforSO3}. For the other values of $\mu^2$, the amplitudes in eqs.\eqref{allchan} still nicely combine in  simple expressions:
\bes
\begin{align}
\Amp^{(1)}_{\mathbf{5}}(\mu^2)+\Amp^{(1)}_{\mathbf{3}}(\mu^2)=\frac{216 m_W^6}{v^2\mu^2}\frac{(\mu^2-2m_W^2)}{(\mu^4-4m_W^2\mu^2+3m_W^4)^2}\,,\\
\Amp^{(1)}_{\mathbf{1}}(\mu^2)+2\Amp^{(1)}_{\mathbf{5}}(\mu^2)=\frac{12a^2}{v^2}\frac{(\mu^2-2m_W^2)(m_h^2-2m_W^2)^3 }{(m_h^2-\mu^2)^2(m_h^2-4m_W^2+\mu^2)^2}\,.
\end{align}
\ees
%
 
The calculation performed with the GBs $\pi$ on the external legs is totally analogous, up to the replacement $A_s(s,t,u)\rightarrow A^\pi(s,t,u)$ and $A_u(u,t,s)\rightarrow A^\pi(u,t,s)$ where
\be
\l{pipiWcontrib}
A^{\pi}(s,t,u)=\f{s}{v^{2}}-\f{s^{2}a^{2}}{v^{2}(s-m_{h}^{2})}+\f{g^{2}}{4}\f{u-s}{t-m_{W}^{2}}-\f{g^{2}}{4}\f{s-t}{u-m_{W}^{2}}\,.
\ee

\bibliographystyle{mine}
\bibliography{paper}

\end{document}